\documentclass[12pt]{article}
\pdfoutput=1
\usepackage{jheppub}
\usepackage{amsmath}
\usepackage{amsfonts}
\usepackage{amssymb}
\usepackage{latexsym}
\usepackage{simplewick}
\usepackage{colonequals}
\RequirePackage{amssymb,amsmath,amsthm,amsfonts}
\RequirePackage{bbm}
\numberwithin{equation}{section}

\usepackage{mathtools}
\usepackage{simpler-wick}

\usepackage{extarrows}

\usepackage{epsf}
\usepackage{color}
\usepackage{graphicx}
\usepackage{subfig}
\usepackage{dsfont}

\usepackage[export]{adjustbox}
\usepackage{hyperref}
\definecolor{darkred}{rgb}{0.8,0.1,0.1}
\hypersetup{colorlinks=true, linkcolor=darkred, citecolor=blue, linktoc=page}

\graphicspath{{figures/}}
\usepackage[utf8]{inputenc} 

\AtBeginDocument{}

\def\tr{{\rm tr}}
\newcommand{\hq}{\hat q}
\newcommand{\HHsymb}{\spadesuit}

\newcommand{\cJ}{\mathcal{J}}

\newcommand{\bpm}{\ensuremath{\begin{pmatrix}}}
\newcommand{\epm}{\ensuremath{\end{pmatrix}}}

\DeclareMathSymbol{:}{\mathord}{operators}{"3A}

\newcommand*\pPqskip{8mu}
\catcode`,\active
\newcommand*\pPq{\begingroup
        \catcode`\,\active
        \def ,{\mskip\pPqskip\relax}%
        \dopPq
}
\catcode`\,12
\def\dopPq#1#2#3#4#5{%
        {}_{#1}\Phi_{#2}\biggl[\genfrac..{0pt}{}{#3}{#4};#5\biggr]%
        \endgroup
}

\usepackage{comment}

\title{Quantum groups, non-commutative $AdS_2$, and chords in the double-scaled SYK model}

\author[a]{Micha Berkooz}
\author[b]{, Misha Isachenkov}
\author[c]{, Prithvi Narayan}
\author[d]{and Vladimir Narovlansky}
\emailAdd{micha.berkooz@weizmann.ac.il}
\emailAdd{m.isachenkov@uva.nl}
\emailAdd{prithvi.narayan@gmail.com}
\emailAdd{narovlansky@princeton.edu}
\affiliation[a]{Department of Particle Physics and Astrophysics, \\ Weizmann Institute of Science, Rehovot 7610001, Israel}
\affiliation[\,b]{
Institute of Physics, University of Amsterdam, 1098 XH Amsterdam, the Netherlands \\
Korteweg-de Vries Institute for Mathematics, University of Amsterdam, \\ 1098 XG \mbox{Amsterdam}, the Netherlands
}
\affiliation[\,c]{
\it{Department of Physics, Indian Institute of Technology, Palakkad 678557, India}}
\affiliation[d]{Department of Physics,  Princeton University, Princeton, NJ 08544, USA}

\abstract{
We study the double-scaling limit of SYK (DS-SYK) model and elucidate the underlying quantum group symmetry. The DS-SYK model is characterized by a parameter $q$, and in the $q\rightarrow 1$ and low-energy limit it goes over to the familiar Schwarzian theory. We relate the chord and transfer-matrix picture to the motion of a ``boundary particle" on the Euclidean Poincar{\'e} disk, which underlies the single-sided Schwarzian model. $AdS_2$ carries an action of $\mathfrak{s}\mathfrak{l}(2,{\mathbb R}) \simeq \mathfrak{s}\mathfrak{u}(1,1)$, and
we argue that the symmetry of the full DS-SYK model is a certain $q$-deformation of the latter, namely $\mathcal{U}_{\sqrt q}(\mathfrak{s}\mathfrak{u}(1,1))$. We do this by obtaining the effective Hamiltonian of the DS-SYK as a (reduction of) particle moving on a lattice deformation of $AdS_2$, which has this $\mathcal{U}_{\sqrt q}(\mathfrak{s}\mathfrak{u}(1,1))$ algebra as its symmetry. We also exhibit the connection to non-commutative geometry of $q$-homogeneous spaces,
by obtaining the effective Hamiltonian of the DS-SYK as a (reduction of) particle moving on a non-commutative deformation of $AdS_3$. There are families of possibly distinct $q$-deformed $AdS_2$ spaces, and we point out which are relevant for the DS-SYK model.

}

\begin{document}

\setcounter{footnote}{0}

\maketitle

\setcounter{equation}{0}
\setcounter{footnote}{0}

\section{Introduction}\label{sec:intro}

The Sachdev-Ye-Kitaev (SYK) model is a model of $N$ Majorana fermions with random all-to-all interactions \cite{Sachdev_1993,Kitaev_talk}, which is important as a simple toy model that is both solvable and maximally chaotic \cite{Maldacena:2016hyu,Polchinski_2016,ABK1,ABK2}. Its solvability is both due to large $N$ Schwinger-Dyson (SD) techniques and due to the fact that it is nearly conformal in the IR, with low-energy fluctuations described by a Schwarzian effective action. The latter is the same dynamics that describes Jackiw-Teitelboim (JT) gravity on $AdS_2$ \cite{Maldacena:2016upp,Almheiri2015,Stanford_Witten2017,Sachdev_2019}, and the SYK model has emerged as a tractable example of $AdS_2/CFT_1$ holography \cite{Kitaev_talk,Kitaev:2018wpr}. One then uses it to study the issues of quantum gravity, including black hole thermodynamics and the information paradox \cite{Jensen_2016,Kitaev:2018wpr,Cotler:2016fpe,Davison_2017,Garc_a_Garc_a_2016,Lam_2018}.

Focusing on the universal infrared behaviour means focusing on JT gravity and its description via the Schwarzian action \cite{Sachdev_1993,Kitaev_talk,Kitaev:2017awl} and the boundary particle \cite{Maldacena:2016upp,Kitaev:2018wpr}. (Note that JT gravity is related to the Schwarzian for large cutoff surfaces, while finite cutoff was studied in \cite{Iliesiu:2020zld,Stanford:2020qhm}.)  This however changes the problem significantly since that theory by itself is dual to a $\beta$-ensemble type RMT model \cite{Saad:2019lba} and not to the SYK model. The models behave the same at long time, but they differ significantly at short times (for a radical example see \cite{Berkooz:2020fvm}). An example of a short time process that we might be interested in is fluctuation of the horizon at finite times\footnote{For example, the recent discussions of the structure of algebras in black holes backgrounds is at such short time scales \cite{Chandrasekaran:2022eqq,Witten:2021unn,Leutheusser:2021frk}.} (i.e., such that do not scale with $N$).

The SYK model was also studied in the {\it double-scaled limit} (DS-SYK), where the interaction size goes as $\sqrt{N}$ \cite{Cotler:2016fpe,Berkooz:2018qkz,Berkooz:2018jqr}. In this limit, the model has a known asymptotic density of states \cite{Erdos14}, which can be calculated through combinatorial tools \cite{feng2018spectrum,Erdos14,Berkooz:2018qkz}.  Correlation functions have been calculated in this limit using the technique of chord diagrams \cite{Berkooz:2018qkz,Berkooz:2018jqr}. This limit has been connected to $q$-Brownian motion processes in \cite{Speicher2019}.
It was also suggested that the model might be related to gravity in de Sitter space \cite{HermanDeSitter,Susskind:2021esx,Susskind:2022dfz,Lin:2022nss,Susskind:2022bia,Rahman:2022jsf}.
In addition to the version of the model with real fermions, similar techniques were applied to models with $U(1)$ symmetry and to the supersymmetric SYK model \cite{Berkooz:2020xne,Berkooz:2020uly}.

\begin{figure}[h]
\centering
\includegraphics[width=0.5\textwidth]{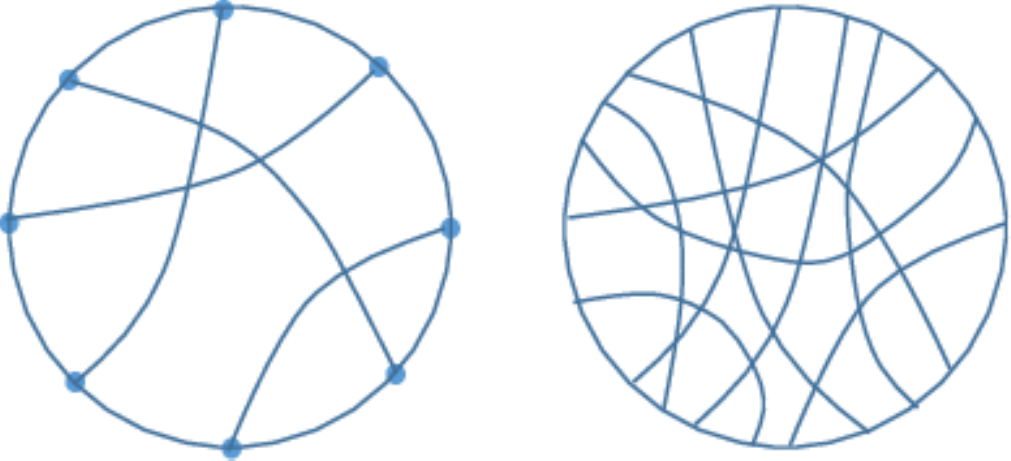}
\caption{An example of a chord diagram, and the emergence of a bulk at large chord number.}
\label{ChordIntro}
\end{figure}

The solution to the DS-SYK model relies on combinatorial objects called chord diagrams, which is very different from the SD technique, being applicable at all energy scales and not just at the IR. In fact, the techniques used are particularly geared towards computing the short time behavior of the theory (at any average energy scale), i.e. at an energy range which is not captured well by the standard $\beta$-ensemble RMT techniques. A chord diagram is a set of $k$ points on a circle which are connected in pairs as in the left diagram in figure \ref{ChordIntro}. In the limit of a large number of points and a dense set of chords this just 'becomes' $AdS_2$ and its boundary -- see the right side of figure \ref{ChordIntro}. One of our objectives here is to make this pictorial intuition more precise in terms of a non-commutative $AdS_2$ space.  

In section \ref{sec:CrdABK} we review the $q\rightarrow 1$ and low-energy limit, and see how we restore the usual Schwarzian description. But DS-SYK does more than that. It solves the model at all energy scales,
which makes a question about the possible role of the parameter $q$ very interesting. In section \ref{sec:hands-on} and onward we will argue that a general $q$ away from 1 explores strong quantum gravity effects. In this paper we will see how it turns spacetime into a non-commutative geometry (or discretizes it, which is essentially an equivalent description). This applies to $AdS_2$ turning it into what we will call $AdS_{2,\hq}$, but since $AdS_2$ is the non-trivial part of near-horizon geometries of many extremal black holes in higher dimension, it is tempting to think that this is a way in which quantum effects modify the near-horizon dynamics for any black hole.

The outline and main results of the paper are the following:
\begin{itemize}

\item In section \ref{sec:transfer_matrix} we recall how the transfer-matrix method works in double-scaled SYK model.

\item In section \ref{sec:CrdABK}, we carry out the $q\rightarrow 1$ limit and identify the role of chord number as lengths in spacetime. We argue that the bulk of the chord diagram corresponds to the region of the Poincar{\'e} disk carved out by the boundary particle. This is implied by the fact that the solution yields the Liouville quantum mechanics in this limit \cite{Berkooz:2018qkz}, and the latter has a clear spacetime interpretation \cite{ABK1,ABK2,Mertens:2017mtv,Maldacena:2016upp}. We will be discussing the single-sided Euclidean $AdS_2$ disk, rather then the two-sided Minkowski interpretation given in \cite{Lin:2022rbf}

\item In section \ref{sec:hands-on} we discuss one approach to the main abjective of the paper, which is the non-commutative $AdS_2$. We construct some hand-on models of the latter using lattice discretizations of $AdS_2$ on which $\mathcal{U}_{\sqrt q}(\mathfrak{s}\mathfrak{u}(1,1))$ acts, and show how to obtain the DS-SYK transfer-matrix out of them.

Part of the discussion focuses on $q$-Fourier transforms for such lattices: a specific class of generalizations of the usual Fourier transform to the non-commutative geometry (NCG) setting that we consider. We survey what is known about these transforms, and adapt one of them to our purposes.

\item In the rest of the paper we will need the full power of quantum groups and $q$-homogeneous spaces. In \ref{sec:q-groups-intro} we provide what some of us consider to be a user-friendly introduction to quantum groups and the specific details that we will require.

\item In \ref{sec:Lobachevsky} we discuss a plethora of $q$-deformations of the Lobachevsky space $EAdS_3$, following (with minor corrections) \cite{Olshanetsky:1993sw,Olshanetsky:2001}.

\item Finally, in section \ref{sec:down-to-AdS2} we carry out reductions from the $q$-deformed $EAdS_{3,\hq}$ to a q-deformed $AdS_{2,\hq}$ to obtain a family of models in rough correspondence with the hands-on models in section \ref{sec:hands-on}.  
\end{itemize}

\section{The transfer-matrix method for double-scaled SYK}
\label{sec:transfer_matrix}

The Sachdev-Ye-Kitaev (SYK) model \cite{Sachdev_1993,Kitaev_talk} is a quantum-mechanical theory of $N$ Majorana fermions $\psi_i$, $i=1,\cdots ,N$, satisfying $ \{ \psi_i,\psi_j\}=2\delta _{ij}$. The Hamiltonian gives rise to random all-to-all interactions
\begin{equation}
    H = i^{p/2} \sum _{i_1<\cdots <i_p} J_{i_1 \cdots i_p} \psi_{i_1} \cdots \psi_{i_p}\,.
\end{equation}
$J_{i_1\cdots i_p}$ are Gaussian random couplings, with zero mean, and variance
\begin{equation}\label{eq:JVar}
    \langle J_{i_1 \cdots i_p}^2 \rangle _J = \binom{N}{p} ^{-1} \cJ^2\,.
\end{equation}
We denote by angular brackets the average over the couplings.\footnote{In the von Neumann algebra language,
one can think of this model as a von Neumann algebra generated by a single operator-valued random variable $H$. It carries a natural state defined by averaging over all the independent Gaussian distributions. While $N$ is finite, the algebra is
just a finite-dimensional matrix von Neumann algebra. As we go to the double-scaled limit, the limiting hyperfinite von Neumann algebra is of type $II_1$. Adding matter operators to the set of generators for this von Neumann algebra turns out not to change the type, see \cite{sniady2004factoriality}, \cite{Ricard2005}.} In most places we will set $\cJ=1$, but when we will need to match dimensionful objects we will restore it.

In the double-scaling limit, we take $N,p \to \infty $ keeping fixed
\begin{equation}
    \lambda \colonequals \frac{2p^2}{N}\,.
\end{equation}

In order to describe the transfer-matrix method, it is easiest to consider the moments
\begin{equation}
    m_k \colonequals \langle \tr H^k \rangle _J .
\end{equation}
The partition function $ \langle \tr e^{-\beta H}\rangle _J$ can be obtained by summing the moments with appropriate coefficients, corresponding to powers of the inverse temperature. We normalize the trace such that $\tr 1=1$.

\begin{figure}[h]
\centering
\includegraphics[width=0.5\textwidth]{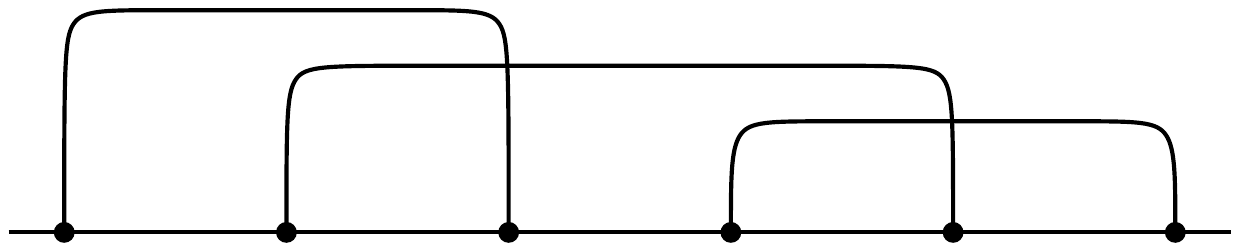}
\caption{An example of a chord diagram contributing to $m_k$ with $k=6$.}
\label{fig:chord_diagram_example}
\end{figure}

The moments can be represented using chord diagrams.\footnote{In mathematical literature, this type of chord diagrams is called linear chord diagrams. That is, if we draw the chords' endpoints on a circle, we don't necessarily identify the chord diagrams related to each other by rotation. We will always abbreviate linear chord diagrams as 'chord diagrams' to save space.} We start with the circular chord diagram in figure \ref{ChordIntro}, which represent the cyclic trace, and then cut it at some point to make it into a line.    That is, we draw a line with $k$ nodes marked on it, each one corresponding to one Hamiltonian. In every chord diagram, we connect by chords pairs of nodes, where each node has exactly one chord attached to it. An example of a chord diagram is shown in Fig.\ \ref{fig:chord_diagram_example}. The origin of these chords is in the average over the couplings, giving rise to Wick's theorem when applied to contractions of the random $J$'s. We are therefore instructed to go over all contractions. As a result, in order to obtain the moment, we should sum over all chord diagrams with $k$ nodes.

As explained in \cite{Erdos14,Berkooz:2018qkz,Berkooz:2018jqr}, the value one should assign to a given chord diagram is simply determined by the number of intersections. More concretely, defining
\begin{equation}
    q \colonequals e^{-\lambda }\,,
\end{equation}
each intersection of two chords is assigned a factor of $q$. I.e., the value of the moment is then given by
\begin{equation}\label{eq:MomentsChords}
    m_k = \sum_{\text{chord diagrams on $k$ nodes}} q^{\# \text{ intersections}}\,.
\end{equation}

While these chord diagrams may naively be thought of as merely providing a combinatorial tool to evaluate the moments, they in fact give rise to a much richer structure, in the form of a Hilbert space and an effective Hamiltonian for double-scaled SYK model. In fact they explicitly implement the duality which takes us from the field theory degrees of freedom to the gravitational ones.  In the next subsection we review the origin and structure of this effective Hamiltonian and the Hilbert space of chords that it acts on.

\subsection{A crash review of the derivation}

For completeness, we provide a crash derivation of \eqref{eq:MomentsChords}.
First, it is convenient to abbreviate sets of indices $i_1<i_2< \cdots <i_p$ by capital multi-indices $I$, and strings of fermions by $\psi_I=\psi_{i_1}\cdots \psi_{i_p}$. When performing the average over the coupling, we get a factor of $\binom{N}{p} ^{-k/2}$ from the variance of the random couplings \eqref{eq:JVar}. Then, every chord diagram reduces after the ensemble average to an expression of the form
\begin{equation}
    i^{kp/2} \binom{N}{p}^{-k/2} \sum _{I_1,\cdots ,I_{k/2}} \tr \left( \psi_{I_1} \cdots \psi_{I_1} \cdots \right) \,,
\end{equation}
where every two $\psi_I$'s connected by a chord appear with the same index.

Next we notice that if a pair of fermions with the same indices appear next to each other, their product is simply proportional to the identity from the fermionic algebra. Therefore, our goal is to disentangle the chords by commuting $\psi_I$'s in order to get rid of intersections. For example, in Fig.\ \ref{fig:chord_diagram_example} we should exchange the fourth node with the fifth node. By the fermionic algebra, commuting $\psi_I$ with $\psi_{I'}$ gives a sign $(-1)^{|I \cap I'|}$, where $|I \cap I'|$ is the number of sites in the intersection of $I$ with $I'$. This number is a random variable, since we sum over the site indices. The prefactor $\binom{N}{p} ^{-k/2}$ precisely turns the summation into an average. For $p \ll N$, when we form $I'$, we make $p$ choices out of $N$, and each one is in the intersection with $I$ with probability $p/N$. Therefore, $|I \cap I'|$ is Poisson distributed with mean value of $p^2/N$. Finally, each intersection is given by an average of signs $(-1)^m$ with $m$ Poisson distributed, so that the average is $e^{-2p^2/N}=e^{-\lambda}=q$. This is the result we mentioned: each intersection is assigned a factor of $q$. Note that for $p \ll N$, the different intersections are statistically independent. These values assigned to the intersections are all there is, since after commuting the fermions to be next to each other, we remain with trace of the identity operator.

\subsection{Evaluating the moments: the transfer-matrix and the chord Hilbert space}
\label{subsec:transfer-matrix}

\subsubsection{The transfer-matrix}

Fortunately there are closed formulas for the chord partition functions. One method to evaluate them relies on a transfer-matrix approach \cite{Berkooz:2018qkz}. Before we arranged the Hamiltonians as nodes on a circle, or a line (choosing a point on the circle and starting and ending on it). Next we associate a state from a Hilbert space to each interval between consecutive nodes -- in each such interval, there is a particular number of propagating chords, and the Hilbert space is then built using these states. For every non-negative integer $l=0,1,2,\cdots $ we define $|l\rangle $ to be the state of $l$ chords. This set of states constitutes a basis for the Hilbert space $\mathcal{H}$.

\begin{figure}[h]
\centering
\includegraphics[width=0.7\textwidth]{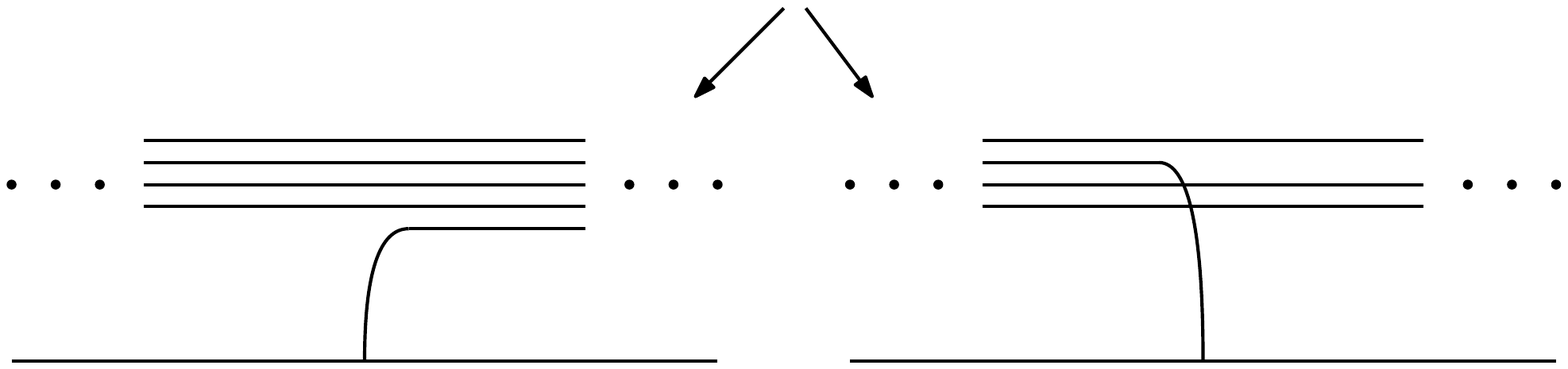}
\caption{The possibilites at each node.}
\label{fig:cases_at_chords}
\end{figure}

Next, when crossing a node, the state of the system changes by going from a particular number of chords to a different number. So there is an effective Hamiltonian which acts in the Hilbert space of chords. In fact, since each node has a single chord attached, the number of chords changes by one at each node -- this is depicted in figure \ref{fig:cases_at_chords}. Thinking about time evolution as going from, say, left to right along the chord diagram, there are two options for what can happen in a node. One possibility is that a new chord opens, hereby increasing the number of open chords by one. The other option is that one chord closes at the given node. In fact, in this case, there are several possibilities for which chord closes. If the lowest chord closes, it just reduces the number of chords by one. However, if the second chord from the bottom is closed, in addition to reducing the number of chords, it also intersects the chord below it, and should therefore give a factor of $q$. Similarly, any of the chords can close, and by doing so it would give a factor of $q$ for every chord it crosses.

Such a transition can be represented by an operator that is referred to as the transfer-matrix. The description above corresponds to the action
\begin{equation}
    T | l\rangle  = |l+1\rangle +\left(1+q+\cdots +q^{l-1}\right) |l-1\rangle =|l+1\rangle + \frac{1-q^l}{1-q}|l-1\rangle\,.
\end{equation}
By acting $k$ times with the transfer-matrix, all the different chord diagrams are generated, each one appearing with the correct power of $q$. In other words,
\begin{equation}
    m_k = \langle 0|T^k|0\rangle \,,
\end{equation}
since we start with no chords, i.e., the state $|0\rangle $, and at the end we should remain with no chords as well.

The operator $T$ can be represented by a matrix using the basis $|l\rangle $. The action above corresponds to the matrix
\begin{equation}
\label{eq:T Matrix form}
T \colonequals \begin{bmatrix}
0 & {1-q \over 1-q}&  0  &  0 & 0 &  \dots   \\
1 & 0 &{1-q^2 \over 1-q} &  0 & 0 &  \dots  \\
0 &  1 & 0 & {1-q^3 \over 1-q} & 0 & \dots \\
\vdots  &  \ddots & \ddots & \ddots &  \ddots & \ddots \\
\end{bmatrix}\, .
\end{equation}

The matrix $T$ in \eqref{eq:T Matrix form} naively appears to be non-Hermitian, but one can conjugate it to a symmetric version\footnote{Or, in other words, the matrix $T$ is self-adjoint with respect to another measure which differs from the simple counting measure by a square root of the factor in the similarity transformation.} by defining a new matrix:
\begin{equation}\label{eq:Def hat T from T}
\hat T \colonequals P T P^{-1} \,,
\end{equation}
where $P$ is a diagonal matrix with entries $(P_0,P_1,P_2\dots )$. Let us define
\begin{equation}
    \label{eq:def eta}\eta_l \colonequals 1+q+...+q^l= {1-q^{l+1}\over 1-q}
\end{equation}
and
\begin{equation}
P_l \colonequals \prod_{i=0}^{l-1} \sqrt{\eta_i} ={ \sqrt{ (q;q)_l } \over (1-q)^{l \over 2}}\,, \quad l\neq 0 \quad \quad P_0=1\,,
\end{equation}
where $(a;q)_{l}$ is the $q$-Pochhammer symbol defined by
\begin{equation} \label{eq:q-poch}
    (a;q)_l \colonequals \prod _{i=0}^{l-1} (1-aq^i)\, .
\end{equation}
$\hat T$ has matrix elements
\begin{equation}
  (\hat T)^{\ l_2}_{l_1} = \sqrt{\eta_{l_2}}\delta^{l_2}_{l_1-1} + {\sqrt{\eta_{l_1}}} \delta^{l_2}_{l_1+1}\,,
\label{eq:sym-T}
\end{equation}
thus, it is manifestly symmetric:
\begin{equation}\label{eq:The Sym Transfer Matrix}
\hat T = \begin{bmatrix}
0 & 1 &  0  &  0 & 0 &  \dots   \\
1 & 0 & \sqrt{\eta_1}&  0 & 0 &  \dots  \\
0 &  \sqrt{\eta_1} & 0 & \sqrt{\eta_2}& 0 & \dots \\
\vdots  &  \ddots & \ddots & \ddots &  \ddots & \ddots \\
\end{bmatrix}\,.
\end{equation}

So we have a rule in which we replace $H^k$, inside a correlation function, by ${\hat T}^k$, or the time evolution operator $e^{-iHt}$ by $e^{-i{\hat T}t}$. This means that $\hat T$ is an effective Hamitonian after averaging over the $J$'s. This is exactly what the gravitational Hamiltonian does, and in fact the latter will just be $\hat T$. Furthermore, it was understood in \cite{Berkooz:2018jqr} that the model is governed by a quantum group symmetry, which is a $q$-deformation of (the universal enveloping algebra of) $\mathfrak{s}\mathfrak{l}(2)$. In particular, the 4-point function was calculated and is given via the 6j-symbol of this quantum group. So, we will essentially be looking for a $q$-deformation of JT gravity.

\subsubsection{The Hilbert space}

\begin{figure}[h]
\centering
\includegraphics[width=0.7\textwidth]{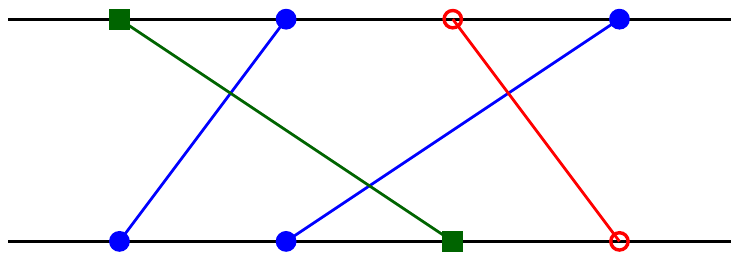}
\caption{A contribution to the overlap of $|C_1C_0C_2C_0\rangle $ and $|C_0C_0C_1C_2\rangle$.}
\label{fig:Inner_Product}
\end{figure}

The Hilbert space of chords has a natural inner product. As we will discuss in the next section, the chord number is associated with the boundary particle degrees of freedom, but it is simple, in this language, to also introduce additional particles in the bulk beyond the gravitational degrees of freedom. In this case additional particles are naturally associated with additional chords of different types \cite{Berkooz:2018qkz,Berkooz:2018jqr}. We will denote the types of chords by $C_0,C_1,\cdots,C_L$ where $C_0$ are chords associated with the Hamiltonian. For each pair of chords there is a weight $q_{ij},\ i,j=0, \cdots, L$ associated with their intersection (determined by the mass of the particle in the bulk). A basis for states in the total generalized Hilbert space is an ordered set of finite length of chords
\begin{equation}
|C_{i_1}C_{i_2}... C_{i_n}\rangle,\ \ n=0,1,\cdots,\ \ i_j\in\{0,\cdots,L\}\,.
\end{equation}
The inner product $\langle C_{j_1}...C_{j_n}| C_{i_1}..C_{i_n}\rangle$ is then defined\footnote{In the $\mathcal{N}=2$ SUSY case there is a variant with fermionic chords and a more intricate structure of the Hilbert space \cite{Berkooz:2020xne}.} \cite{Speicher1991} pictorially with the help of figure \ref{fig:Inner_Product}:
\begin{itemize}
\item We sum over all the pairings of identical types of chords between the set $C_j$ and the set $C_i$ (so they have to have the same overall set of chords, but perhaps in different orderings).
\item Each pairing defines a different set of intersections of chords going from one state to the other, and each intersection between chord $i$ and chord $j$ carries the weight $q_{ij}$. The total weight of the given pairing is the product of all the weights of the intersections.
\end{itemize}

\subsubsection{The $q\rightarrow 1$ limit}

To take the continuum limit $q\rightarrow 1$, it is convenient to define the matrix
\begin{equation}
    \tilde T \colonequals S \hat T S^{-1}\,,
\end{equation}
where $S$ is a diagonal matrix with entries $S_{ll} = (-1)^l$ \cite{Berkooz:2018qkz}. Notice that solving for the eigenvalues of the $\tilde T$ or $\hat T$ matrices resembles a scattering problem on the half-line, with the index $l$ of the vector measuring the distance from the origin. The asymptotic form of the $\tilde T$ matrix is made out of $(-1)$'s on the diagonal above the main one and on the diagonal below the main one
\begin{equation}
\tilde T  \sim  \frac{1}{\sqrt{1-q}} \begin{bmatrix}
\ddots & \ddots & \ddots & \vdots & \vdots & . &  .\\
\ddots & 0 & -1&  0  &  0 & 0 &  \dots   \\
\dots & -1 & 0 & -1 &  0 & 0 &  \dots  \\
\dots & 0 &  -1 & 0 & -1  & 0 & \dots \\
. & \vdots  &  \vdots & \ddots &  \ddots &  \ddots & \ddots \\
\end{bmatrix}\,.
\end{equation}
In the continuum limit at low energies, the above matrix becomes the second derivative operator. To make this more precise, define
\begin{equation}\label{eq:Phi to discrete index}
\phi \colonequals \ln(q) \ l\,.
\end{equation}
One can easily check that the $\tilde {T}$ operator in the continuum limit approaches
\begin{equation}
{\tilde T}\to \frac{1}{\sqrt{1-q}} \biggl( -2 - (\ln q)^2\partial_\phi^2 +  e^\phi \biggr) \,.
\label{eq:cont-T}
\end{equation}
Notice the potential term comes from the expansion $\sqrt{ 1-q^{l+1} \over 1-q } = {1 \over \sqrt{1-q} } (1 -  {q^{l+1} \over 2} +\dots )$ in $\eta_l $, as defined in \eqref{eq:def eta}, which is accurate since $l$ is large and $q\to 1$, from below. The eigenvalue problem $\tilde T \Psi = E \Psi$ then reduces to the quantum-mechanical eigenvalue problem (with $q = e^{-\lambda}$):
\begin{equation}\label{eq:Transfer Matrix Eigenvalue Eqn}
{\cal J} \left(-\lambda^{3 \over 2} \partial^2_\phi + {1 \over \sqrt {\lambda} } e^\phi\right) \Psi = (E-E_0)\Psi\,,
\end{equation}
where we have restored ${\cal J}$ and $E_0 \colonequals -{2 {\cal J}\over \sqrt \lambda}$.
As we will mention in the next subsection, this can be identified with the Liouville Hamiltonian form of the Schwarzian action.  

The Hilbert space we described here is related to the motion of a single boundary particle. A two-sided version extension of it can be found in \cite{Lin:2022rbf}.

\section{The transfer-matrix and $AdS_2$ spacetime}
\label{sec:CrdABK}

We have just seen that the transfer-matrix goes over to Liouville quantum mechanics. In this section we would like to extend this to a fuller map from the chord dynamics into spacetime $AdS_2$ dynamics. We will do this for the single-sided case, as it has a cleaner extension to the non-commutative case (but it is similar to the two-sided case \cite{Lin:2022rbf}).

In subsection \ref{subsec:Altland Bagrets} we recall the relation between the Schwarzian and Liouville quantum mechanics. In section \ref{sec:BootApp} we will recall the derivation of Liouville from a reduction of $AdS_3$ and the dynamics of the boundary particle on this reduction -- this will actually be our starting point for the non-commutative picture. In section \ref{subsec:Boundary trajectory to chords} we will discuss what this implies for the embedding of the chord dynamics into $AdS_2$.

\subsection{The Liouville form of the Schwarzian}\label{subsec:Altland Bagrets}

The low-energy degrees of freedom of the SYK model are reparametrization modes $\tau \rightarrow f(\tau)$ governed by a Schwarzian action \cite{Kitaev_talk,Maldacena:2016hyu}. In \cite{ABK1,ABK2}  the Schwarzian was brought to a more convenient description in terms of low-energy Liouville field $\phi(\tau)$ (related to the reparametrization modes as $f'(\tau) = e^{\phi(\tau)}$).

The action given there for the Liouville field is
\begin{equation}\label{eq:ABK Action}
    S[\phi]  \colonequals \int_{-{\beta \over 2}}^{\beta \over 2} d\tau \left(  {M \over 2}(\phi'^2 - 2 \phi'') + \gamma e^\phi \right) - \gamma t_H,\ \ \phi(\tau)=\phi(\tau+\beta)\,,
\end{equation}
where $\gamma$ is a Lagrange multiplier which imposes the constraint $\int_{-\beta \over 2}^{\beta \over 2} e^{\phi(\tau)} d\tau = t_H$, and $t_H$ is a regulator which will be taken to  $\infty$  eventually.\footnote{This is to allow for singular configurations, where $e^{\phi}$  diverges at $\tau = \pm {\beta \over 2}$.} Since it is a low-energy description, it applies -- for the $p=4$ SYK model which is the main focus of these papers -- at energies $E \le {64 \sqrt \pi J \over N \log N}$, and we need to take $ M = {N \log N \over 64 \sqrt \pi J}$.
The partition function is then given by
\begin{equation}
    Z_\beta \colonequals  \int_{\phi(-{\beta\over 2}) = \phi_0}^{\phi({\beta\over 2}) = \phi_0} [D\phi] e^{-S[\phi]}\,.
\end{equation}
The parameters of the model $\gamma, \phi_0$ are tuned such that we get finite results as $t_H \rightarrow \infty$ while maintaining the normalization condition  $\int_{-\beta \over 2}^{\beta \over 2} e^{\phi(\tau)} d\tau = t_H$. This will imply that one needs to take $\phi_0\rightarrow \infty$.

To match the double-scaled SYK, we can compare the Liouville Hamiltonian derived from action \eqref{eq:ABK Action}, that is $H = -{\partial_\phi^2 \over 2M} + \gamma e^\phi$, with the $q \rightarrow 1$ limit we obtained in equation (\ref{eq:Transfer Matrix Eigenvalue Eqn}). We obtain \cite{Berkooz:2018qkz}
\begin{equation}\label{eq: ABK M versus dSSYK lambda}
M = {1 \over 2 {\cal J} \lambda^{3 \over 2}}\,.
\end{equation}
We will soon verify this also by comparing partition functions in the two models.

Note that $\gamma$ can be absorbed in a shift of $\phi$. The term $\phi''$ is a total derivative, but it is required for finiteness of the answers. However, we can evaluate it explicitly and remove it from the action. The final result is therefore a Liouville action. Note that the boundary condition $\phi(\pm\beta/2)=\phi_0\rightarrow\infty$ means that the trajectories start and end under the Liouville wall.

Next, the partition function can be calculated via a saddle-point analysis. The equation of motion for the field $\phi'' = {\gamma \over M} e^{\phi}$ can be solved by
\begin{equation}\label{eq:Liouville Saddle Solution}
    \gamma e^{\bar \phi( \tau )} ={2 \pi^2 M (1-\epsilon)\over \beta^2  \cos^2{(\pi\tau(1-\epsilon)^2\over \beta}}\,,
\end{equation}
for arbitrary $\epsilon$.  Further imposing the constraint that $\int_{-\beta \over 2}^{\beta \over 2} e^{\phi(\tau)} d\tau = t_H$ by the function  $\phi(\tau) = \bar \phi(\tau)$ fixes $\epsilon$ to be
\begin{equation}
   \epsilon = {8 M \over \beta \gamma t_H}, \qquad \epsilon \ll 1\,,
\end{equation}
where we have assumed large $t_H$. One can also check that for this solution,
\begin{equation}
e^{\phi_0} = e^{\bar   \phi(\pm {\beta \over 2})}  = {\gamma t_H^2 \over 8 M}\,.
\end{equation}
We can now evaluate the action, equation (\ref{eq:ABK Action}), which we get as
\begin{equation}
    S[\bar \phi] = - {2 M \pi^2 \over \beta }\,.
\end{equation}
Note that we get finite results although $t_H \rightarrow \infty$. The Gaussian fluctuations around the above saddle solution were also computed in \cite{ABK2}, and they give rise to the correct the pre-exponential term in the final result which is:
\begin{equation}
    Z(\beta) \approx \left( {M \over \beta} \right)^{3 \over 2} e^{2 \pi^2 M \over \beta}\ \sim {1 \over (\beta {\cal J} )^{3 \over 2}} e^{ \pi^2 \over \beta {\cal J} \lambda^{3 \over 2}   }\,.
\end{equation}
In the second approximate equality we have used the identification \eqref{eq: ABK M versus dSSYK lambda} and dropped the overall $\beta$-independent terms. This exactly matches the partition function in the double-scaled SYK obtained in equation (4.13) of \cite{Berkooz:2018qkz} in the low-temperature limit $\lambda^{-{3 \over 2}} \gg  \beta \gg \lambda^{-{1 \over 2}}$ (after restoring $\cal J$ and dropping overall $\beta$-independent factors).

So the DS-SYK model gives rise, in the IR and $q\rightarrow 1$ limit, to the standard $AdS_2$ physics. But it gives more than that, since it solves the theory for any value of $q$ and energy. In particular, when $q$ goes down from the vicinity of 1 we expect to have new quantum gravity effects, and hence we would like to see what replaces $AdS_2$ for such general values of $q$.

\subsection{The Liouville action from $AdS_3$ and motion on $AdS_2$}\label{sec:BootApp}

One way of obtaining the Schwarzian action is by reduction of a Minkowskian $AdS_3$ \cite{Mertens:2017mtv}.
More precisely, we start with a Poincar{\'e} description of it, parameterized by $\phi, z, z^*$. For a Euclidean $AdS_3$, $z$ and $z^*$ are complex conjugate of each other, whereas for a Minkowskian one they are real and independent. The Lagrangian for a particle moving on this 3D space is:
\begin{equation}
    \mathcal{L}={1\over 2} {\dot\phi}^2 + \pi_z {\dot z} + \pi_{z^*}{{\dot z}^*}- \pi_z \pi_{z^*}e^{\phi}\ \ \ \text{ or }\ \
    \mathcal{L}={1\over 2} {\dot\phi}^2 + {\dot z}{{\dot z^*}}e^{-\phi}\,.
\end{equation}

In Minkowskian $AdS_3$, we integrate out $z^*$ (independently of $z$). Its conjugate momentum is a constant of motion, and we obtain
\begin{equation}\label{AdS2Lag}
    \mathcal{L}={1\over 2} {\dot\phi}^2 + \pi_z{\dot z} - \pi_z \pi_{z^*} e^{\phi}\,.
\end{equation}
Note that $\pi_{z^*}$ can be absorbed into a shift of $\phi$. We still have $\mathfrak{s}\mathfrak{l}(2,\mathbb{R}) \simeq \mathfrak{s}\mathfrak{u}(1,1)$ conserved charges given by\footnote{This is in a slightly different convention than \cite{Mertens:2017mtv}.}
\begin{equation}\label{eq:SL2Gen}
\begin{split}
        &B=i\pi_z=\partial_z\\
        &L_0=-i(z\pi_z+\pi_\phi)=-z\partial_z - \partial_\phi\\
        &C=-i(z^2\pi_z + 2z\pi_\phi + e^\phi)=-z^2\partial_z - 2z\partial_\phi - ie^\phi\\
        &[L_0,B]=B,\ \ [L_0,C]=-C,\ \ [B,C]=2L_0\,,
\end{split}
\end{equation}
and we introduced the notations $B$ and $C$ for the Chevalley generators which is common in the literature about quantum groups of rank one, and will be used later on in the paper.
The Casimir element (of the universal enveloping algebra) is
\begin{equation}\label{eq:CasC}
 \Omega=  (L_0)^2 + {1\over 2}\{B,C \}\,.
\end{equation}
In our conventions, it is $-H$, where the latter is the Hamiltonian on the space. This is an important point: all the Lagrangians that we discuss here are the Casimirs in the corresponding symmetry group representations that act on functions on the relevant spaces.

Next, we can integrate out $(\pi_z,z)$ (setting $\pi_z=\mu$) and bring the action to the Liouville form of the Schwarzian:\footnote{Actually, there is a slight discrepancy since \eqref{eq:CasC} behaves like $\partial_\phi^2+\partial_\phi$, whereas if we quantize \eqref{eq:1DLag}, naively we get $\partial_{\phi}^2$ without the linear term. The difference is a total derivative in the Lagrangian, or can be absorbed in a redefinition of the wave function.}
\begin{equation}\label{eq:1DLag}
    \mathcal{L}={1\over 2}{\dot\phi}^2 -\mu e^{\phi}, \ \ \pi_z=\mu\,.
\end{equation}
Finally, we can shift $\phi$ to eliminate $\mu$ as well. The Lagrangian that we obtained is the one that gives the Liouville quantum mechanics discussed in the previous section, up to some shifts and a rescaling of $\phi$. Thus, the Liouville action is just the Casimir acting on the (functions on the) space, properly reduced.

Several comments are in order:

1. Recall that the basic physics is that of the boundary particle moving on $AdS_2$. Hence in our reduction from $AdS_3$ to $AdS_2$, $\pi_{z^*}$ is not a dynamical degree of freedom that we need to integrate over, but rather specifies how we construct $AdS_2$ as a coset of $\mathbb{H}^3$ -- so it is a parameter in the 2D theory. Fortunately, it actually drops out. The status of $(\pi_z,z)$ is different. Since they are associated with a coordinate/momenta pair on $AdS_2$, we really need to integrate over them. This means that in the final expression we need to carry out a path integral over $\phi$ and a single integral over $\mu$.

Since $\phi$ can again be shifted to absorb $\mu$, this technically does not make a big difference at the level of the Liouville QM, which is the dynamics after we shift $\mu$ away. However, we are interested in lifting the transfer-matrix, in which the meaning of $\mu$ is also obscured, to an object associated with $AdS_2$. $\phi$ will be related to the chord number so most of the guesswork will have to do with how we reintroduce the non-commutative analogue of $\mu$ when we lift the transfer-matrix to the motion of the boundary particle on $AdS_2$.

2. We can work with the $(\phi,z)$ or $(\phi,\mu)$ coordinates -- either way they describe a particle moving on $AdS_2$. The easiest way for us to process this information into the SYK model partition function is to use the description given in \cite{Kitaev:2018wpr}.\footnote{It is the same as \cite{Maldacena:2016upp}, but the writing is slightly different.} The main formula that we will borrow from there is the relation between the SYK model's partition function and a propagator of the relevant particle on $AdS_2$, which is (see equation (46) there):
\begin{equation}\label{eq:Kitaev Suh}
Z_{\beta} \sim G_{AdS_2}(X,X;\beta)=\langle X | e^{-\beta H_{AdS_2}} | X\rangle,\ \ \ X\in AdS_2\,,
\end{equation}
where $|X\rangle$ can be any point in $AdS_2$. Using a fixed initial and starting point is the same as fixing the gauged $SL(2,\mathbb{R})$ group of isometries of the $AdS_2$. $H_{AdS_2}$ is the Casimir of the universal enveloping algebra of the Lie algebra of this group acting on $AdS_2$, in agreement with the fact that, as we saw earlier, the Liouville Hamitonian is the same Casimir.

The fact that the SYK partition function is a propagator of a particle on $AdS_2$, rather than a partition function there as well, is related to a statement that there is no one-sided Hilbert space in the effective bulk description. It is, however, a very mild way of 'not having a Hilbert space', since we still compute the propagator in some Hilbert space. This Hilbert space will be the chord Hilbert space even for the one-sided case.

3. Note that we need to take $\mu>0$ in order to match with the Liouville action that we had before, that is we consider only the subset of forward-moving trajectories in $z$ which is the Poincar{\'e} patch time coordinate. This is to be expected because, when quantizing $NAdS_2$ (following \cite{Maldacena:2016upp}), we impose the constraint $t'=z\epsilon$, where $\epsilon$ is the cut-off and $z=0$ is the boundary.

\subsection{The discrete trajectories of the boundary particle in the transfer-matrix approach}\label{subsec:Boundary trajectory to chords}

The previous two subsections were essentially a review, recast in a way useful for our purpose. In this subsection we will discuss how the chords fit into the picture above.

\subsubsection{The boundary particle trajectory in chord number}\label{subsec:Liouville to chords mapping}

We would like to provide another check of the the relation between $M$ and $\lambda$, and directly relate the motion of the boundary particle to the number of chords. This will allow us to identify the "bulk" of the chord diagram with the part of $AdS_2$ 'bounded' by the boundary particle.

The relation between the Liouville quantum mechanics and the transfer-matrix leads us to a relation of the form:
\begin{equation}\label{eq:Mapping of chord to Liouvlle}
    e^{\phi(\tau)} \propto q^{\hat n(\tau) }\,,
\end{equation}
where the constant of proportionality includes all the shifts  of $\phi$ that we have accumulated (which do not depend on time).

To test this statement we will match the saddle value of $e^{\phi(\tau)}$ on the Liouville side to the expectation value on the transfer-matrix side, say, in the thermal state characterized by $\beta$. Keeping the leading order in $t_H$ term (actually the correct dimensionless small parameter is $\epsilon = {8 M \over \beta \gamma t_H}$) in  the saddle-point solution equation (\ref{eq:Liouville Saddle Solution}),  we get the required mapping as
\begin{equation}
e^{\bar \phi(\tau)} =   {2 \pi^2 M \over \beta^2 \gamma \cos^2({\pi \tau \over \beta})}    \longleftrightarrow \langle 0 |  q^{\hat n(\tau)} | 0 \rangle_\beta =
\sum_n  \langle 0 |e^{-({\beta\over 2} - \tau ) \hat T } | n \rangle q^n \langle  n| e^{-({\beta\over 2} + \tau ) \hat T } | 0 \rangle\,.
\end{equation}

We now turn to computing the object on the right-hand side above. This is almost exactly the two-point function computed in \cite{Berkooz:2018qkz}. We review the calculation (along with needed redefinitions of variables) in appendix \ref{sec:q1toABK}.
In the low-temperature limit,
\begin{equation}
 {1 \over \sqrt \lambda }\ll  \beta \ll {1 \over \lambda^{3 \over 2} } ,   \qquad \tau \ll {\sqrt{\beta} \over \lambda^{3 \over 4} }\,,
\end{equation}
the result is
\begin{equation}
    \begin{split}
        \langle 0 |  q^{\hat n(\tau)} | 0 \rangle_\beta \propto
        {1 \over \cos^2({\pi \tau \over \beta})}\,,
    \end{split}
\end{equation}
as expected.

\subsubsection{The initial and final state, and the carving out of $AdS_2$}

\begin{figure}
    \centering
    \includegraphics[width=0.4\textwidth]{"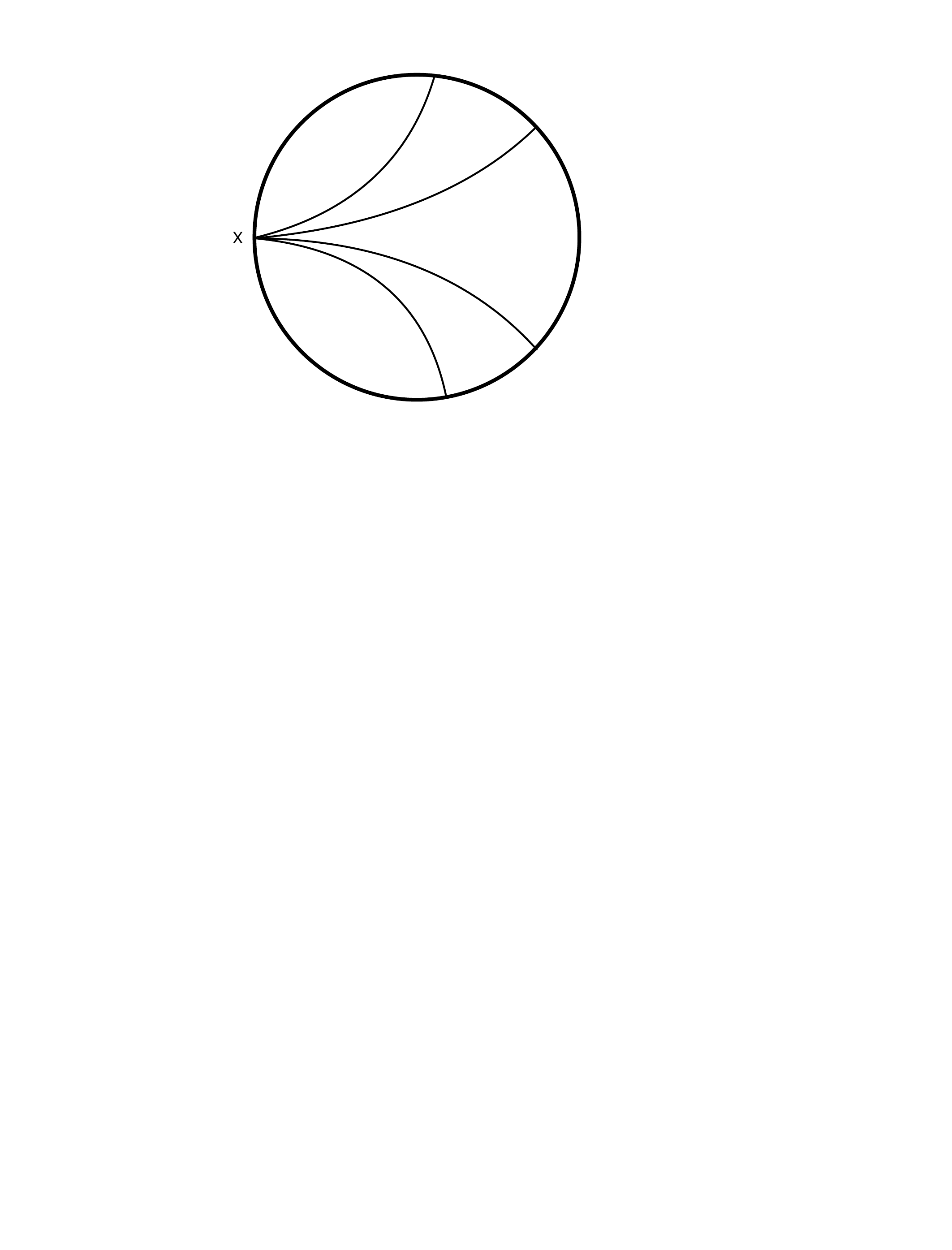"}
    \caption{Equal Poincar{\'e} time slicing of the disk.}
    \label{fig:PoinDis}
\end{figure}

The $AdS_2$ on which the boundary particle moves is parameterized by $(\phi,z)$. The slices of equal $z$ are depicted in figure \ref{fig:PoinDis}. Both in the Liouville description and in the transfer-matrix description, the initial and final states ($\langle 0|$ and $| 0\rangle$ for the transfer-matrix) correspond to values of $\phi$ as large as possible (recall equation \eqref{eq:Mapping of chord to Liouvlle}). We can take this also as the point $|X\rangle$ in equation \eqref{eq:Kitaev Suh}. The location of this point is therefore the special point on the boundary which maps to the Poincar{\'e} horizon, designated as X in the figure. The rest of the boundary of $AdS_2$ is at $\phi\rightarrow -\infty$, or $n\rightarrow\infty$.

\begin{figure}
    \centering
    \includegraphics[width=1\linewidth]{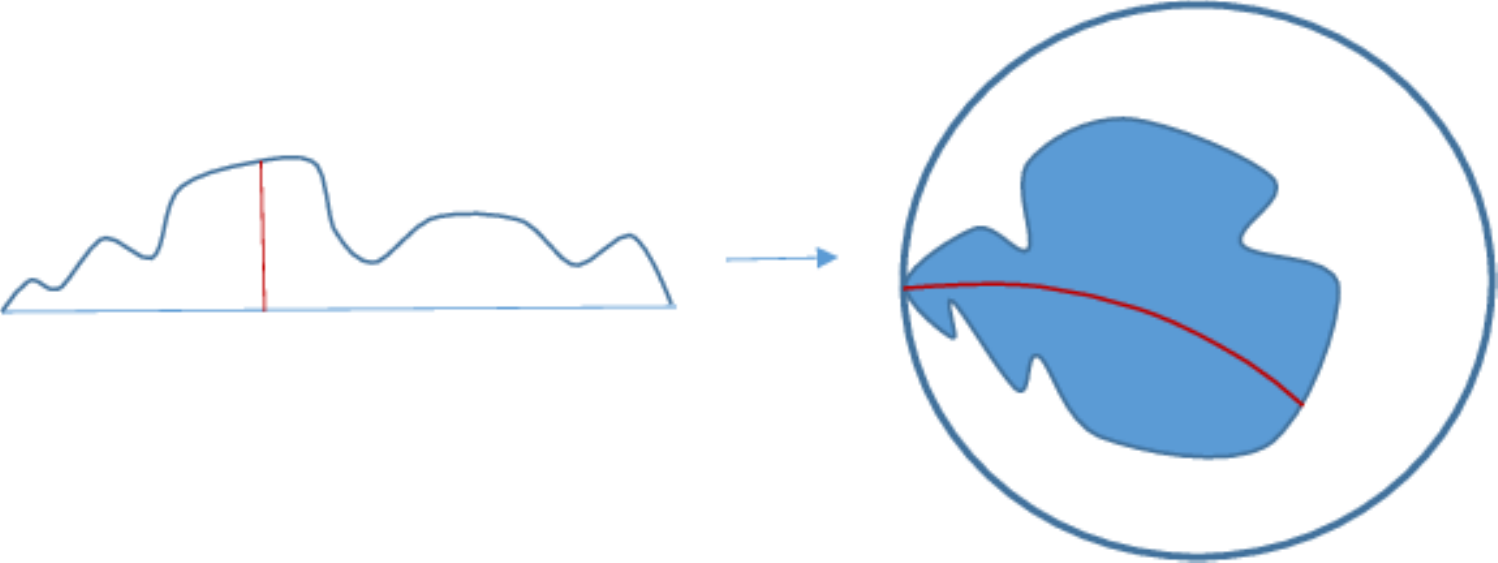}
    \caption{Chords to AdS: The left panel depicts the set of chord diagrams with a given envelope $n(t)$. The right panel is the area it carves out in $AdS_2$.}
    \label{fig:Poincare Disk}
\end{figure}

In figure \ref{fig:Poincare Disk} on the left we drew the envelope of the following chord process. Recall that before we discussed a chord that computes $\langle Tr(H^k)\rangle_J $, now it is more convenient to discuss a chord process which computes
\begin{equation}
\langle Tr(e^{-\beta H})\rangle_J
\rightarrow \langle 0| (1-{\epsilon \beta T})^{1/\epsilon}|0\rangle\,.
\end{equation}
The right-hand side is again described by chord opening and closing. One can, for example, use this formalism to compute fluctuations on lengths in the Hartle-Hawking state \cite{Okuyama:2022szh}.

Given a chord diagram CD (from the sum above), we can define its envelope as the following set of integers:
\begin{equation}
\begin{split}
&Env(CD) \colonequals \{n(t)| \ t=k\epsilon, k=0, \dots, \beta/\epsilon,\\
&n(0)=n(\beta/\epsilon)=0,\ n(t)>0,\ n(t+\epsilon)-n(t)=0,\pm 1 \}
\end{split}
\end{equation}
This is depicted in the left panel in figure \ref{fig:Poincare Disk}: the x-axis line is where we put the insertions of the transfer-matrix, and the vertical axis is how many chords we have open at that moment -- that is, $n(t)$. So, summing over all chord diagrams includes a sum over all chord envelopes as in the figure, and within each envelope we sum over all ways of opening and closing chords that give the same envelope.

Recall that we identify the $n(t)$ with the distance along a spatial slice (via the coordinate $\phi$) and that the bulk time ($z$) is increasing as we increase the time in the field theory. We therefore expect that we can map every time point in the chord diagram, characterized by $(t,n(t))$, into a specific time and associated radial distance in the Poincar{\'e} disk. This mapping is depicted in the right panel of the figure. The time axis in the chord description is mapped to the trajectory of the boundary particle (since this is where we insert operators), and the area under the curve in the chord process is mapped to the chunk of spacetime which is bounded by the boundary particle.

\section{A hands-on approach to non-commutative $AdS_2$}
\label{sec:hands-on}

When we quantize the boundary particle, then only after solving the constrained motion of the latter and  gauge-fixing we obtain the Liouville action. Alternatively, we can start with the 2D motion and diagonalize the momentum associated with one isometry, in order to reduce it to one dimension. In any case, one ends up with a 1D Liouville problem in which the original $AdS_2$ symmetries are obscured.

When we start with the statistical model and obtain the transfer-matrix, we are directly given the 1D dynamics. We would then like to see if there is some 2D lift which deforms $AdS_2$, just as the transfer-matrix is a deformation of the Liouville quantum mechanics.  

A hint for what we should be looking for comes from computing the crossed, or OTOC, 4-pt function in the DS-SYK model \cite{Berkooz:2018jqr}. In the usual $NAdS_2/SYK_1$ duality, the crossed 4-pt function is governed by a 6j-symbol of $\mathcal{U}(\mathfrak{s}\mathfrak{l}(2,\mathbb{R})) \simeq \mathcal{U}(\mathfrak{s}\mathfrak{u}(1,1))$  \cite{Mertens:2017mtv}, which is the symmetry of the space. In the double-scaled SYK model, the crossed 4-pt function is a 6j-symbol of $\mathcal{U}_{q^{1/2}}(\mathfrak{s}\mathfrak{u}(1,1))$. The natural guess, which we will implement here, is to construct a $q^{1/2}$-deformation of $AdS_2$ on which this '$q$-deformed group' acts.\footnote{A different quantum deformation of the JT gravity has been studied in \cite{Fan:2021bwt}. It is different from ours since they work with $U_q(sl(2,{\mathbb R}))$ where $q$ lies on the unit circle. This is a distinct, non-isomorphic real form of this rank one quantum group.}

In this section, we will do this in a hands-on way, by presenting some lattice realizations of such a space with properties that: 1) $\mathcal{U}_{q^{1/2}}(\mathfrak{s}\mathfrak{u}(1,1))$  acts on them, and 2) they go over to $AdS_2$ in the $q\rightarrow 1$ limit. These realizations were obtained by guesswork starting from simple plausible assumptions for the form of some of the generators, supplemented by imposing the algebra relations, and requiring that some reasonable transfer-matrix emerges from the Casimir element of the algebra.

Within these different realizations, one should ask which are equivalent to the others, where by equivalence we mean that there is a transformation taking the generators from one realization to another. We do not know the complete answer to this. The question is difficult for the following reasons:

1) One does not know exactly what is the space of functions on which these generators work. I.e., one needs to define the appropriate set of states, perhaps with appropriate inner product, and this turns out to be much more intricate than expected. In section \ref{sec:q-fourier} we discuss (a variant of) this problem.

2) When we have a geometric realization of some group of symmetries as vector fields on a manifold, then equivalence means equivalence under a smooth map of spaces, which takes points to points. In non-commutative geometry the very notion of a point is more subtle, and the set of allowed transformations is much larger (for example, if the coordinates do not commute, we can also mix momenta into the transformation, making it non-local on the lattice). To really show equivalence, one needs in principle to solve some spectral theory problem which is hard (related to point (1) above).

Thus, we will be left with realizations parameterized by several parameters (and perhaps more realizations exist). In the subsequent sections we will show that qualitatively similar realizations, albeit with fewer parameters, appear in some specific constructions from math literature, but the same issues apply there. So we will have to see what transfer-matrices they give and which of them might fit the required notion of a $AdS_{2,q^{1/2}}$, in the sense that it gives the DS-SYK transfer-matrix.

The outline of this section is as follows. In section \ref{sec:UqAlg} we define the  $\mathcal{U}_{q^{1/2}}(\mathfrak{s}\mathfrak{u}(1,1))$  algebra, and then in section \ref{sec:Hands_on_AdSq} we present the different lattice realizations of the algebra and compute the form of the Casimir. In section \ref{sec:q-fourier} we discuss some of the issues in carrying out the reduction from the Casimir to the transfer-matrix. The catch is that we need to diagonalize the $q$-analoque of the operator $\partial_z$ as in section \ref{sec:BootApp}, and what is known about $q$-Fourier transforms will allow us to solve this problem only partially. Finally, in section \ref{sec:NoBndry} we give some arguments why some $AdS_{2,q^{1/2}}$ reductions to transfer-matrices are more physically motivated than others.

\subsection{The $\mathcal{U}_{q^{1/2}}(\mathfrak{s}\mathfrak{u}(1,1))$  algebra}\label{sec:UqAlg}

Since the deformation parameter is actually $q^{1/2}$, and in order to conform with the standard notation on quantum groups, we are going to define
\begin{equation}
    \hq \colonequals q^{1\over 2}\,,
\end{equation}
and then use quantum groups notation with $\hq$.
The algebra $\mathcal{U}_{\hq}(\mathfrak{s}\mathfrak{u}(1,1))$ is generated by elements $A, B, C, D$ obeying the following relations:
\begin{equation}\label{SLq2}
    AD=DA=1, \qquad AB = \hq B A, \qquad A C  = \hq^{-1} C A,\qquad [B,C] = {A^2-D^2\over \hq -\hq^{-1}}\,.
\end{equation}
The Casimir element, generating the center of this algebra, is given by
\begin{equation}
        \Omega  \colonequals \left(  { \hq^{-1}A^2 + \hq D^2-2 \over (\hq^{-1} - \hq)^2} + B C \right)\,.
\end{equation}

To get a feeling on how it deforms the classical algebra, in the  $\hq \to 1$ limit, we start with the standard Pauli matrices and then define
\begin{equation}
    A_0=\hq^{{1\over 2}\sigma_3},\ \ B_0=\sigma_+={1\over 2}(\sigma_1+i\sigma_2),\ \ C_0=\sigma_-={1\over 2}(\sigma_1-i\sigma_2) \,,
\end{equation}
which satisfy the algebra \eqref{SLq2} to the leading order in $\hq-1$ when we take $\hq\rightarrow 1$. Of course, \eqref{SLq2} goes further into finite $\hq$, but the $\hq\rightarrow 1$ limit tells us which element of the quantum algebra behaves like an element of a group and which one behaves as an element of a Lie algebra. This will be important below, when we review the coproduct structure of (the Hopf algebra) $\mathcal{U}_{\hq}(\mathfrak{s}\mathfrak{u}(1,1))$.

Our goal is to define $AdS_{2,\hq}$, which will be a quantum version of $AdS_2$ in the sense that $\mathcal{U}_{\hq}(\mathfrak{s}\mathfrak{u}(1,1))$ will act on it, or more precisely, on functions on this space.
Unlike in the classical case, this does not seem to determine the space uniquely. Even what we mean by 'uniquely' is not clear, since to determine that we need to sort out the problem concerning equivalence of different realizations, as we discussed before.

In the following subsection we will motivate a large class of realizations of $AdS_{2,{\hat q}}$ candidates, obtained by trial and error, of the algebra \eqref{SLq2}, acting on functions supported on two-dimensional discrete (multiplicative) lattices. Some of these realizations will reduce to the transfer-matrix of the double-scaled SYK, and we will explain why they are distinguished.

\subsection{Lattice realizations of $AdS_{2,\hq}$: the $\hq$-deformed $AdS_2$ space }
\label{sec:Hands_on_AdSq}

Let us denote the basic operations, $T$ and $R$, of rescaling the arguments $\tilde z$ and $\tilde H$ of a two-variable function $F$
as follows:
\begin{equation}\begin{split}
    T \ F(\tilde H,\tilde z) &\colonequals  F(\hq^{1 \over 2} \tilde H,\tilde z) \\
    R \ F(\tilde H,\tilde z) &\colonequals  F(\tilde H,\hq \tilde z)\,. \\
    \end{split}
    \label{eq:def T and R}
\end{equation}
From now on we will use the symbol $T$ for this operator. The reason that we can do so is that we identified before the Liouville Hamiltonian as the Casimir element of $\mathcal{U}(\mathfrak{s}\mathfrak{u}(1,1)) \simeq \mathcal{U}(\mathfrak{s}\mathfrak{l}(2,\mathbb{R}))$ acting on $AdS_2$, so the correspondence is:
\begin{equation}
\text{Casimir of }\mathcal{U}_{\hq}(\mathfrak{s}\mathfrak{u}(1,1)) \rightarrow \text{DS-SYK transfer-matrix}\,.
\end{equation}
Thus, the object to track in the following sections is the Casimir element.

We will assume here that $({\tilde H},{\tilde z})$ live on a $\hq$-lattice
\begin{equation}
({\tilde H},{\tilde z})\in \left\{ (\hq^{m_1/2},\pm \hq^{m_2})\, | \, m_1,m_2\in \mathbb{Z} \right\} \equalscolon \mathbb{R}^2_{\hq} \subset \mathbb{R}^2 \,.
\end{equation}
Notice that we do not lose much generality considering this particular lattice:
if we would take a lattice of the form $(\hq^{\alpha_1+m_1/2},\pm \hq^{\alpha_2 + m_2})$, $\alpha_{1,2} \in\mathbb{R}$, the results would have been the same.
Compared to the section before, ${\tilde z}$ will be similar to $z$, and $\tilde H$ will be related to the radial direction $\phi$ by $\tilde H=e^{-\phi/2}$.

We  begin by considering the following realization of generators, which depend on the parameters $a$, ${\tilde a},\mu$:
\begin{equation}\label{eq:HandOn}
    \begin{split}
        A^{\HHsymb} & =  T R^{-1} \\
        B^{\HHsymb} & = {R^{a+2} - R^a \over \tilde  z (\hq^{a+2} - \hq^a)} \\
        C^{\HHsymb} & = {\tilde z \over \hq^{-1} - \hq} \left(  T^{-2} R^{-a} -   T^2 R^{-a-2}  \right) + i \mu \tilde H^{-2} T^{\tilde a}\\
        \Omega^\HHsymb    & \equiv \left(  { \hq^{-1}A^2 + \hq D^2-2 \over (\hq^{-1} - \hq)^2} + B^{\HHsymb}  C^{\HHsymb} \right)  \\
        &=  {\hq^{-1} T^{-2}  + \hq T^2-2 \over (\hq^{-1}-\hq)^2  } + {i \mu  \tilde H^{-2} } \ T^{\tilde a} \ B^\HHsymb \,,
    \end{split}
\end{equation}
and think about them as acting on the space of functions on our lattice
$\mathbb{R}^2_{\hq}$.
One can easily check that these generators satisfy the relations \eqref{SLq2},\footnote{In fact, one can, more generally, take a sum over different $\tilde a$'s in the definition of $C^{\HHsymb}$, and still satisfy the algebra relations.} so that we get a (highly reducible) representation of our algebra $\mathcal{U}_{\hq}\left(su(1,1)\right)$ on this space of functions. Obtaining these operators required some guesswork and hard digging, and hence we will label them with the superscript $\HHsymb$.  

In the remainder of this section, we will work with the realization \eqref{eq:HandOn} of our algebra. Let us note here that this realization is not unique: there are many more 'degrees of freedom'. For instance, since  $\tilde H^{-2} B^{\HHsymb}$, $T$ commute with $A^\HHsymb$ and $B^{\HHsymb}$, the transformation $C^{\HHsymb} \mapsto G C^{\HHsymb} G^{-1}$ for an arbitrary function $G = G(\tilde H^{-2} B^{\HHsymb}, T)$ still satisfies the algebra \eqref{SLq2}. There are two reasons that we do not consider such transformations here. The more essential one is that this changes the form of the Casimir in our realization \eqref{eq:HandOn},
and thus doesn't match the specific form of the SYK transfer-matrix that we need. A more technical reason is that, even if such a (non-trivial) trnasformation would leave the Casimir element invariant, an expression for it
generically involves infinitely many terms, so one first needs to go to an appropriate completion of this algebra to work with, which would open up many possibilities, thus taking us too far afield.

At this stage, the parameters $a$ and $\tilde a$ are arbitrary. Note that the Casimir in our representation \eqref{eq:HandOn} depends explicitly only on $\tilde a$.
The other parameter $a$ appears exclusively through $B^{\HHsymb}$, which we will shortly diagonalize, so we will really be sensitive only to the eigenvalues of $B^{\HHsymb}$, and not to $a$ directly (of course, $a$ determines these eigenvalues).

\smallskip

{\bf The $\hq\rightarrow 1$ limit:}  Note also that in the $\hq = e^{-\lambda / 2}\rightarrow 1$ limit, the generators given in equation (\eqref{eq:HandOn}) become:
\begin{equation}\begin{split}
A^{\HHsymb} -1  &= - \frac{\lambda}{2} \ \left({1 \over 2} \tilde H \partial_{\tilde H} - \tilde z \partial_{\tilde  z}  \right)  + {\cal O}(\lambda^2)\\
B^{\HHsymb}  &=\partial_{\tilde z} + {\cal O}(\lambda)\\
C^{\HHsymb} & = -{\tilde z}^2 \partial_{\tilde z} + \tilde H \tilde z \partial_{\tilde H} + i \mu {\tilde H}^{-2}  + {\cal O}(\lambda) \,,
\end{split} \end{equation}
which are the same as \eqref{eq:SL2Gen} with the identification $\tilde H = e^{-\phi / 2}$.

\smallskip

{\bf Reduction of the Casimir to the transfer-matrix:} Recall that the appropriate Casimir propagates the boundary particle in $AdS_2$. We would like to see how exactly it reduces to the DS-SYK transfer-matrix. To do it, let us look at the form of the Casimir in \eqref{eq:HandOn}. If we consider the values of the functions on the sublattice $\tilde H=\hq^{-n}$ as our vectors $v_n$, then the terms $T^2$ and $T^{-2}$ are 1 one diagonal above and one diagonal below the main one. If we take ${\tilde a}=\pm 2$ we will have $\tilde H^{-2}=\hq^{2n}$ in one of these diagonals as well.
Specifically, let us take $\tilde a=-2$. Then the Casimir looks like
\begin{equation}\label{eq:TrnsA}
\begin{split}
&q^{-1}(1-q)^{2}\Omega^{\HHsymb}=
 \begin{bmatrix}
\ddots & \ddots  & \vdots & \vdots & .\\
\ddots & -2 &  \hq^{-1}  &  0 & \dots   \\
\dots & \hq & -2 &  \hq^{-1} & \dots  \\
\dots & 0 &  \hq & -2  & \dots \\
. & \vdots  &  \vdots  &  \ddots &  \ddots  \\
\end{bmatrix} +
  i\mu(1-q)^2 \begin{bmatrix}
\ddots & \ddots  & \vdots & \vdots & .\\
\ddots & 0 &  q^{-1} B^{\HHsymb}  &  0 & \dots   \\
\dots & 0 & 0 &   q^0  B^{\HHsymb} & \dots  \\
\dots & 0 &  0 & 0  & \dots \\
. & \vdots  &  \vdots  &  \ddots &  \ddots  \\
\end{bmatrix}\,.
 \end{split}
\end{equation}

Suppose we take eigenfunctions of $B^\HHsymb$ such that
\begin{equation}
\mu B^{\HHsymb} = \frac{i \hat q^3}{(1-\hat q^2)^2}\, .
\end{equation}
In this case, we can restrict our vectors to $n \ge 0$ only, setting $v_{n<0}=0$ -- i.e, the transfer-matrix gets truncated; this is preserved under the Casimir action.

To match to the transfer-matrix from before,
we conjugate the Casimir by a diagonal matrix $\tilde S$ with elements
\begin{equation}
\tilde S_{nn} = q^{-n/2} \prod_{i=1}^n \sqrt{1-q^i} \quad \text{for } n \neq 0 \text{ and } \tilde S_{00}=1\,.
\end{equation}
Then we get that $\tilde S \left( q^{-1}(1-q)^{3/2}\Omega^{\HHsymb} + \frac{2}{\sqrt{1-q}}\right)\tilde S^{-1}$ is the same as our symmetric transfer-matrix $\hat T$ from \eqref{eq:The Sym Transfer Matrix}.

However, our prescription in the previous section called for summing over all $\partial_z$ momenta and then shifting $\phi$ appropriately. So we need to sum over all eigenvalues of $B^{\HHsymb}$, and still get the transfer-matrix for each of them. This can happen if the complete spectrum of $B^{\HHsymb}$ is of the form (or a subset of):
\begin{equation}
\mu B^{\HHsymb} = \frac{i \hat q^3}{(1-\hat q^2)^2} \cdot q^s, \ \ s\in \mathbb{Z}\,.
\end{equation}
In this case we can shift the index $n$ in the Casimir matrix and bring it to the form of the transfer-matrix. For all values of $s$ the matrix terminates when $n$ is a small enough integer and we simply set all values of $v_{n<s}=0$.

Since we can choose $\mu$ in the reduction, the criteria on the spetrum of $B^\HHsymb$ is that it is (a subset) of the form $ \alpha q^s, \ s\in \mathbb{Z}$ for some $\alpha$, or that the ratio of any two eigenvalues of $B^\HHsymb$ is in $q^{\mathbb{Z}}$.

\smallskip

Finally, one more comment is due which is why we refer to this an $AdS_{2,\hq}$. Consider for example the case of the non-commutative compact two-torus with a finite number of states. We can either write the $X$ and $Y$ coordinates as non-commuting operators (more precisely $U\sim e^{iX},\ V\sim e^{iY}$) acting on finite-dimensional vector space, or we can write it as a theory with a non-trivial *-product between functions of ordinary variables. In the latter case, since the space of functions is finite-dimensional, it is enough to take functions on a lattice. There is a lot of arbitrariness in the lattice which is compensated by the *-product. Here we are discussing such a lattice (and we don't need the *-product for our purposes). In section \ref{sec:q-groups-intro} and onwards we will incorporate the non-commuting coordinate approach.

\subsection{Eigenfunctions of $B^{\HHsymb}$ and the Fourier transform}\label{sec:q-fourier}

To complete the reduction to the transfer-matrix we need to specify the spectrum of $B^\HHsymb$. We would like to show that the spectrum is of the form $b=\hq^{2n}$. In this case for an appropriate value of $\mu$ (such that $\mu b=\hq^{2l}$ for some integer $l$) we obtain the terminating transfer-matrix from the Casimir.

This is a rather thorny issue. More precisely, we need to specify the Hilbert space on which $B^\HHsymb$ acts, identify the eigenfunctions, and show that they form a complete basis. The catch is that there are no satisfactory results for such an analysis, on the operator-algebraic level, for any value of $a$ (although $a=-1$ comes close). In the following we will present some of the existing results for certain kinds of generalized $\hq$-Fourier and $\hq$-Laplace transforms that give a spectrum which is $\hq^{2m}$. From the perspective of a general $a$, this is just supporting evidence that an appropriate q-Fourier theory exists for them.

We will discuss the cases $a=0$, which is preferred from the quantum group point of view (as we will see in equation (\ref{eq:Action of generators on L3 : Abstract})), and the value $a=-1$, for which there is some sort of $q$-Fourier theory.

\subsubsection{Eigenfunctions of the operator $B^{\HHsymb}$}

The easy part is to find the (formal) eigenvectors of the operator $B^{\HHsymb}$ (for any complex eigenvalue) defined via
\begin{equation}
B^{\HHsymb}_a  = { R^a- R^{a+2}  \over   z ( \hq^a-\hq^{a+2})}, \hspace{30mm} R\  F(z) \colonequals F(\hq z)\,.
\end{equation}
The hard part is to sort out the appropriate Hilbert space (i.e. specify the boundary conditions of the Hilbert space at the four infinities at $z=\pm \hq^n,\ n\rightarrow\pm\infty$), perform the spectral analysis of the operator on this space\footnote{E.g. in order to show rigorously that there is no absolutely continuous part in the spectral measure.} and see what eigenvalues are selected.

Let us define a family of generalizations of $\hq$-exponential via \footnote{One can check that as $\hq \rightarrow 1$, $e_a((1-\hq^2)z;\hq^2) \rightarrow e^z$ the usual exponential.}
\begin{equation}\label{eq:q-exponential with parameter a}
    e_a(z;\hq^2) \colonequals \sum_{n=0}^\infty { z^n \over \hq^{an(n-1) \over 2} (\hq^2 ;\hq^2)_n  }\,.
\end{equation}
For $0<\hq<1$, due to a Gaussian factor $\hq^{-an^2/2}$ this series comnverges absolutely for $a\in (-\infty, 0)$, defining an entire function of $z\in \mathbb{C}$. For $a=0$, the series converges only for $|z|<1$, but still can be analytically continued to an entire function on the whole complex plane.
One can easily verify the following fact:
\begin{equation}
    B_a^\HHsymb \ e_a(\mu z ; \hq^2) = {\mu \over 1-\hq^2} e_a(\mu z ; \hq^2) \,.
\end{equation}
Since $B^\HHsymb_b = R^{b-a} B^\HHsymb_a$,
it is also true that
\[
    B_a^\HHsymb \ e_b(\mu z ; \hq^2) = {\mu \over 1-\hq^2} R^{a-b} e_b(\mu z ; \hq^2)\,.
\]
For special cases, the function $e_a$ boils down to known $\hq$-exponentials:
\begin{itemize}
    \item $e_0(z;\hq^2) = (z;\hq^2)^{-1}$ is the standard, 'lowercase' $\hq$-exponential, defined e.g. in Gasper-Rahman book \cite{Gasper-Rahman-book}, equation (1.3.15).\footnote{ Also see equation (1.6) of \cite{2012arXiv1208.2521K}.}
    \item For the symmetric case $a=-1$, the function  $e_{-1}((1-\hq^2)z;\hq^2)$ is the same as $exp_{\hq}$ given in equation (1.3.26) of \cite{Gasper-Rahman-book}.  
    \item For completeness: what is actually known as the ('uppercase') $\hq$-exponential is $e_{-2}(-z;\hq^2) = (z;\hq^2)$, see (1.3.16) of \cite{Gasper-Rahman-book}.
\end{itemize}

Sometimes it might be useful to split the $q$-exponential function defined above into the analogues of cosine and sine functions via
\begin{equation}
    \begin{split}
        \cos_a(z;\hq^2) & \colonequals  \sum_{n=0,2,4,\dots}^\infty {(-1)^{n\over 2} z^n \over \hq^{an(n-1) \over 2} (\hq^2 ;\hq^2)_n  } \\
        \sin_a(z;\hq^2) & \colonequals \sum_{n=1,3,5,\dots}^\infty {(-1)^{n-1\over 2} z^n \over \hq^{an(n-1) \over 2} (\hq^2 ;\hq^2)_n  }\,,
    \end{split}
\end{equation}
such that $e_a(iz;q^2) = \cos_a(z;\hq^2) + i \sin_a(z;\hq^2)$. They satisfy
\begin{equation}
    \begin{split}
        B_a^\HHsymb \ \sin_b(\mu z;\hq^2) & = \  {\mu \over 1- \hq^2} \ \cos_b(\hq^{a-b}\mu z;\hq^2) \\
        B_a^\HHsymb \ \cos_b(\mu z;\hq^2) & =-{\mu \over 1- \hq^2}\ \sin_b(\hq^{a-b}
        \mu z;\hq^2)\,. \\
    \end{split}
\end{equation}

\subsubsection{$a=-1$}

The case $a=-1$ has a $q$-cos and a $q$-sin transform which are the closest to the standard cosine and sine transform, so we will start with it. In some sense, it is easiest to deal with because the operator is anti-self-adjoint under the inner product of functions on the lattice given by \cite{2008arXiv0801.0069B} which reads:
\begin{equation}
\langle f, g \rangle_q \colonequals
\sqrt{1-q}\sum_{n=-\infty}^{+\infty} q^n \left[ (f\overline{g})\left(\frac{q^n}{\sqrt{1-q}}\right) + (f\overline{g})\left(\frac{-q^n}{\sqrt{1-q}}\right) \right]\,.
\end{equation}
This $q$-integration goes over to the usual integration from $0$ to $\infty$ in the limit $q\rightarrow 1$. If we duplicate it to also go over the points at $-q^n,\ n\in\mathbb{Z}$, then we obtain an integration which goes over to the standard one in the range $-\infty$ to $\infty$.

First let us recall a usual Fourier cosine transform:
\begin{align}
    \hat{f}(\lambda) = \sqrt{\frac{2}{\pi}}\int_0^{\infty} f(x) \cos \lambda x dx, \quad
    f(x) = \sqrt{\frac{2}{\pi}}\int_0^{\infty} \hat{f}(\lambda) \cos \lambda x d\lambda\,.
\end{align}
It preserves the space of even $L^2$ functions on the real line and is a unitary operator between the $x$- and $\lambda$- $L^2$ spaces of even functions.
Similarly, for the sine transform.
The spectral decomposition of $-d^2/d^2x$ involves both sine and cosine. The eigenspaces in this case are two-dimensional. In the $q$-deformed case this degeneration gets lifted (because the eigenvalues shift).

For the case $a=-1$ there are a $q$-cos and $q$-sin transform, but, as it is usually stated, it does not diagonalize $B_{-1}^\HHsymb$, since the eigenvalues of the $q$-cos and $q$-sin functions are slightly shifted. It proceeds as follows (see \cite{2012arXiv1208.2521K}). We define:
\begin{equation}
    \begin{split}
        \cos(z;   \hq^4) & \colonequals \sum_{k=0}^\infty {(-1)^k \hq^{2k(k+1)} z^{2k} \over (\hq^2;\hq^2)_{2k} }  = \cos_{-1}(\hq^{{3 \over 2}}z;\hq^2)\\
        \sin(z;\hq^4) & \colonequals \sum_{k=0}^\infty {(-1)^k \hq^{2k(k+1)} z^{2k+1} \over (\hq^2;\hq^2)_{2k+1} } = \hq^{-{1\over 2}}  \sin_{-1}(\hq^{{1 \over 2}}z;\hq^2)\,.\\
    \end{split}
\end{equation}
One can check that as $q \rightarrow 1$, $\cos((1-\hq^2)z;q^4) \rightarrow \cos(z)$ and $\sin((1-\hq^2)z;q^4) \rightarrow \sin(z)$.
Then the relevant Fourier transform is (see below the equation (5.5) of \cite{2012arXiv1208.2521K}):
\begin{equation}
    \begin{split}
        g(\hq^{2n}) & = C_{\hq} \quad  (1-\hq^2) \sum_{k=-\infty}^\infty \hq^{2k} \cos( (1-\hq^2)\hq^{2(k+n)} ; \hq^4) f(\hq^{2k}) \\
        & = C_{\hq} \quad  (1-\hq^2) \sum_{k=-\infty}^\infty \hq^{2k} \cos_{-1}( (1-\hq^2)\hq^{2(k+n)+{3 \over 2}} ; \hq^2) f(\hq^{2k}) \\
        f(\hq^{2k}) & =C_{\hq}  \quad (1-\hq^2) \sum_{n=-\infty}^\infty \hq^{2n} \cos( (1-\hq^2)\hq^{2(k+n)} ; \hq^4) g(\hq^{2n}) \\
        & = C_{\hq} \quad (1-\hq^2) \sum_{n=-\infty}^\infty \hq^{2n} \cos_{-1}( (1-\hq^2)\hq^{2(k+n)+{3 \over 2}} ; \hq^2) g(\hq^{2n})\,,
    \end{split}
\end{equation}
where $C_{\hq} \colonequals {\sqrt{1+\hq^2} \over \Gamma_{\hq^4}({1 \over 2})  }$, and a similar one with $\sin$:
\begin{equation}\label{eq:SinTrans}
    \begin{split}
        g(\hq^{2n}) & = C_{\hq} \quad  (1-\hq^2) \sum_{k=-\infty}^\infty \hq^{2k} \sin( (1-\hq^2)\hq^{2(k+n)} ; \hq^4) f(\hq^{2k}) \\
        & = C_{\hq} \quad  (1-\hq^2) \sum_{k=-\infty}^\infty \hq^{2k-{1 \over 2}} \sin_{-1}( (1-\hq^2)\hq^{2(k+n)+{1\over 2}} ; \hq^2) f(\hq^{2k}) \\
        f(\hq^{2k}) & = C_{\hq} \quad (1-\hq^2) \sum_{n=-\infty}^\infty \hq^{2n} \sin( (1-\hq^2)\hq^{2(k+n)} ; \hq^4) g(\hq^{2n })\\
        & = C_{\hq} \quad (1-\hq^2) \sum_{n=-\infty}^\infty \hq^{2n-{1 \over 2}} \sin_{-1}( (1-\hq^2)\hq^{2(k+n)+{1\over 2}} ; \hq^2) g(\hq^{2n})\,.
    \end{split}
\end{equation}

Note however that the operator actions on these $q$-deformed trigonometric functions give $\tilde B_{-1 }\cos(z;\hq^4)  \propto \sin(q z;\hq^4)$ and $\tilde B_{-1 }\sin(z;\hq^4)  \propto \cos(q^{-1}  z;\hq^4)$. In other words, the expansion is in
\begin{equation}
\begin{split}
f(q^{2k})_{even} \in\ Span\{\cos_{-1} ( (1-{\hat q}^2) \hq^{2n+{3\over 2}} \hq^{2k}; {\hat q}^2)\ |\ n \in \mathbb{Z}  \}\\
f(q^{2k})_{odd} \in\ Span\{\sin_{-1} ( (1-{\hat q}^2) \hq^{2n+{1\over 2}} \hq^{2k}; {\hat q}^2)\ |\ n \in \mathbb{Z}  \}\,,
\end{split}
\end{equation}
and we cannot take linear combinations to form form the $q$-exponentials of $B^\HHsymb_{-1}$ since the lattices of "discrete momenta" are not the same.

Next, let us do the following adaptation of the above. Our operator takes us from an even lattice $\hq^{2k}$ to an odd lattice $\hq^{2k+1}$ and the reverse, and at the same time it switches even and odd functions (under $\hq^k\leftrightarrow -\hq^{k}$). So we will eliminate the redundancy by (granted a peculiar procedure of) choosing a subset of the functions to be even on the even lattice points and odd on the odd lattice points. In the $\hq\rightarrow 1$ limit we still have both odd and even functions on the entire line.

Thus, even functions sit on a lattice of points $\hq^{2k}$ and odd functions sit on a lattice $\hq^{2k+1}$. What does that do to the $q$-sine expansion? Consider an odd function ${\hat f}(z)$ defined on the lattice $z=\hq^{2k+1}$. Map it to $f(\hq^{2k})={\hat f}(\hq^{-1}z) = {\hat f}(q^{2k-1})$ and apply to $f$ the second expansion \eqref{eq:SinTrans}. We are still expanding $f$ in
$\sin_{-1}((1-{\hat q}^2)\hq^{2(k+n)+{1\over 2}};{\hat q}^2)$,
but in terms of $z=\hq^{2n+1}$ the function is
$\sin_{-1}((1-{\hat q}^2)\hq^{2n+{3\over 2}}z;{\hat q}^2)$. I.e., we now have an expansion:
\begin{equation}
\begin{split}
f(q^{2k})_{even} \in\ Span\{\cos_{-1} ( (1-{\hat q}^2) q^{2n+{3\over 2}} q^{2k}; {\hat q}^2)\ |\ n \in \mathbb{Z} \}\\
f(q^{2k+1})_{odd} \in\ Span\{\sin_{-1} ( (1-{\hat q}^2) q^{2n+{3\over 2}} q^{2k}; {\hat q}^2)\ |\ n \in \mathbb{Z}  \}\,.
\end{split}
\end{equation}
Next we have to turn to the algebra, in order to see if the algebra relations are compatible with this truncation. $B^\HHsymb_{-1}$ and $C^\HHsymb_{-1}$ are consistent (assuming $\tilde a$ is even as we will see later) with this truncation, since they take even functions on the even lattice to odd functions on the odd lattice. $A^\HHsymb$, on the other hand, is not compatible. Note however that for the algebra it is enough to discuss $A^{2}$ which is compatible with the truncation -- we can replace the relation (equation \ref{SLq2}) $A B={\hat q} BA$ by $({A}^2)B={\hat q}^2 B({A}^2)$ and otherwise only ${A}^2$ appears in the RHS of $[B,C]$. With $A^2=T^2R^{-2}$, it takes even functions on the even lattice to themselves, and odd functions on the odd lattice to themselves.

On this truncated set of functions we can diagonalize the Casimir with a spectrum which is ${\hat q}^{2n}$. This is exactly the spectrum that we need in order to have a well-defined transfer-matrix, as discussed in \ref{sec:Hands_on_AdSq}.

\subsubsection{$a=0$}

As we will see in the subsequent section, the case $a=0$ actually arises in variants of some constructions of Euclidean $AdS_3$ and their reductions. To go from $AdS_3$ to $AdS_2$ one makes some assumptions on the spectra of one '$q$-isometry', in order to proceed further from $AdS_2$ to the transfer-matrix one makes some assumptions about the spectra of another '$q$-isometry' -- the latter of course is the problematic one. Assumption about the spectra amounts to knowing a $q$-Fourier transform theory with respect to $B^{\HHsymb}_{0}$.  Unfortunately, for the case $a=0$ there is no explicitly known $q$-Fourier transform.

What is known for the $a=0$ case is:
\begin{itemize}
\item Certain combinatorial and bi-orthogonality relations between the $q$-exponential functions associated with $B^{\HHsymb}_{0}$ and $B^{\HHsymb}_{-1}$. We have not
succeeded in using
them to get an analytically satisfactory Fourier theory for $a=0$.
\item A very suggestive $q$-Laplace transform, which we detail next.
\end{itemize}

The known transform related to the operator $B_0^\HHsymb$, is a deformed Laplace transform of half a line: to do it, we need to expand the space of functions in $\tilde z$ in terms of the functions $e_0(z;\hq^2)$. In what follows, we will be using the results of \cite{FEINSILVER1989509}. Equation (3.6) of \cite{FEINSILVER1989509} (adapted to our notations) states that any function $f(v = \hq^{-2m}),\ m\ge 0$ can be expanded in terms of $q$-exponentials with coefficient functions $p(x= \hq^{2n}),\ n\ge 0$ as
\begin{equation}
\begin{split}
    f(v) &= \sum_{n=0}^\infty {e_0(x v;\hq^2) \over e_0(v;\hq^2) }  p(x) x,  \qquad v = \hq^{-2m}, x = \hq^{2n} \\
    p(x) &= \sum_{m=0}^\infty\ {(-1)^m \hq^{m(m-1)} \over (\hq^2,\hq^2)_m (\hq^2,\hq^2)_\infty }\  e_0(x v ;\hq^2) f(v),   \qquad v = \hq^{-2m}, x = \hq^{2n}\,. \\
\end{split}
\end{equation}

\subsection{The "no bound states" condition on the transfer-matrix}\label{sec:NoBndry}

From the discussion around \eqref{eq:TrnsA} it was important for us that the transfer-matrix truncates. This meant that the spectrum of $B^\HHsymb$ had to be of a specific nature, and we had to take $\tilde a=\pm 2$. But there is ample room for things to go wrong -- e.g. we could choose a value of $\mu$ such that the transfer-matrix does not truncate, or the spectrum of $B^\HHsymb$ could be misaligned, or we can try and work with another $\tilde a$. It doesn't give us the DS-SYK transfer-matrix, but are these reasonable models for a $q$-deformed $AdS_2$? In this section we will present an argument why physics dictates that it truncates, and what goes wrong if it does not. We will not do the full analysis, but show how this works if we take a value of $\mu$ such that the transfer-matrix does not truncate.

Basically, we will argue if the transfer-matrix does not truncate, then there are spurious modes on the Poincar{\'e} horizon. This is not acceptable physically because we are working in the thermal ensemble which is the quantum mechanics on an $S^1$, or in other words, the boundary of the Poincar{\'e} disk should form a translationally-invariant $S^1$ at the end of the day. The fact that there is an origin is only an artifact of gauge fixing (when going from the Schwarzian to the Liouville description) or by choosing a cutting point for the circular chord diagram.

As we showed in \cite{Berkooz:2018qkz}, \cite{Berkooz:2018jqr}, the symmetric version of our transfer-matrix, considered as operator on the 'one-sided'\footnote{By a one-sided Hilbert space here we mean the one for which the labels of an orthonormal basis are infinite only in one direction, e.g. take natural number values. This should not be confused with the one-sided (i.e. single-boundary) Hilbert space of section \ref{sec:transfer_matrix} vs. the two-boundary of \cite{Lin:2022rbf}.} Hilbert space of chords $\ell^2(\mathbb{Z}_{\geq 0})$, is a perfectly fine self-adjoint operator. It is diagonalized by the continuous $q$-Hermite polynomials and has only continuous spectrum $[-1;1]$. As soon as we extend it to the 'two-sided' Hilbert space $\mathcal{H}_{\mathbb{Z}} \colonequals  \ell^2\left(\mathbb{Z}\right)$, this introduces some subtleties: now the symmetric transfer-operator might have several self-adjoint extensions, or none at all, depending on the boundary conditions we choose at $n \to -\infty$, and each of these extensions will a priori have a different spectrum.

To fill in some details, we will now give a sketch of the spectral analysis
of our $\tilde a = \pm 2$ transfer-operators, perturbed with a parameter $\chi \in \mathbb{R}_{>0}$.
Thus, we consider two operators, acting on the Hilbert space $\mathcal{H}_{\mathbb{Z}}$, with an orthonormal basis $\{e_n\}_{n\in \mathbb{Z}}$, as:
\begin{align}
    & T^{(1)}_{\chi}\, e_n \colonequals e_{n-1} + (1-\chi q^{n+1})\,e_{n+1}, \quad n\in\mathbb{Z},\\
    & T^{(2)}_{\chi}\, e_n \colonequals e_{n+1} + (1-\chi q^{n})\,e_{n-1}, \quad n\in\mathbb{Z}\,.
\end{align}
Correspondingly, they act on the sequences $\{c_n\}_{n\in\mathbb{Z}}$, $\sum_n |c_n|^2 < \infty$ determining vectors $\sum_n c_n e_n \in \mathcal{H}_{\mathbb{Z}}$ as:
\begin{align}
    & T^{(1)}_{\chi}\, c_n = c_{n+1} + (1-\chi q^{n})\,c_{n-1}, \quad n\in\mathbb{Z},\\
    & T^{(2)}_{\chi}\, c_n = c_{n-1} + (1-\chi q^{n+1})\,c_{n+1}, \quad n\in\mathbb{Z}\,.
\end{align}

Let us first find the (formal) eigenvectors of these. Thus, we need to solve:
\begin{align}\label{eq:Tp-chi-eigen}
    T^{(p)}_{\chi} c_n^{(p)} = 2 x^{(p)} c_n^{(p)}, \quad \sum_{n\in\mathbb{Z}} \left|c_n^{(p)}\right|^2 < \infty, \quad p=1, 2\,.
\end{align}
To do this, we need to recall a bit of lore of the basic (that is, $q$-deformed) hypergeometric functions. One-variable such function can be defined by the series:
\begin{align}
   \pPq{r}{s}{a_1, \dots , a_s}{b_1, \dots , b_s}{q, z} \colonequals \sum_{k=0}^{\infty}\frac{\left(a_1, \dots , a_r ; q\right)_k}{\left(b_1, \dots , b_s, q ; q\right)_k} \left((-1)^k q^{\binom{k}{2}}\right)^{1+s-r} \, z^k\,,
\end{align}
and then appropriately analytically continued beyond the unit disc in $z$ if the parameter numbers satisfy $r=s+1$.\footnote{We do not consider the cases $r>s+1$ unless it is a polynomial.}
Now we define the following functions which are (Jackson's) $q$-analogues of modified Bessel functions:
\begin{align}
   &  I_{\nu}^{(1)}\left(z; q\right) \colonequals \frac{\left(q^{\nu + 1}; q\right)_{\infty}}{(q;q)_{\infty}}\left(\frac{z}{2}\right)^{\nu} \, \pPq{2}{1}{0,0}{q^{\nu+1}}{q, \frac{z^2}{4}} = \frac{\left(q^{\nu + 1}; q\right)_{\infty}}{(q;q)_{\infty}} \sum_{n=0}^{\infty} \frac{1}{\left(q^{\nu+1}, q ; q\right)_n} \left(\frac{z}{2}\right)^{\nu+2n}\\
   &  I_{\nu}^{(2)}\left(z; q\right) \colonequals \frac{\left(q^{\nu + 1}; q\right)_{\infty}}{(q;q)_{\infty}}\left(\frac{z}{2}\right)^{\nu} \, \pPq{0}{1}{-}{q^{\nu+1}}{q, \frac{z^2q^{\nu+1}}{4}} = \frac{\left(q^{\nu + 1}; q\right)_{\infty}}{(q;q)_{\infty}}\sum_{n=0}^{\infty}\frac{q^{n^2+\nu n}}{\left(q^{\nu+1}, q ; q\right)_n} \left(\frac{z}{2}\right)^{\nu+2n} \,.
\end{align}
The second series is absolutely convergent for $z\in \mathbb{C}^{\times}$, and the first absolutely converges for $0<|z|<2$.

Using appropriate contiguous relations for these basic hypergeometric functions, see e.g. propositions 3.5, 3.7 from \cite{OR-qBessel}, we get that each of the two eigenvalue equations in \eqref{eq:Tp-chi-eigen} has two formal solutions (that we label by $\pm$), expressed as follows via the modified $q$-Bessel functions:\footnote{Recall that $\lambda = - \ln q = 2p^2/N$.}
\begin{align}\label{eq:Bessel-solution}
    c_n^{(p), \pm} (x^{(p)}) \colonequals I^{(p)}_{\pm\frac{2i}{\lambda}\arccos{x^{(p)}}} \left(2 \chi^{1/2} q^{\frac{n+1}{2}} ; q\right)\, \quad p=1,2\,.
\end{align}
Notice that the modified $q$-Bessel functions $I^{(1)}$ and $I^{(2)}$ are related via:\cite{OR-qBessel}
\begin{align}\label{eq:I1-I2}
 I_{\nu}^{(1)}(z;q) = e_0\left(\frac{z^2}{4}; q\right) I_{\nu}^{(2)}(z;q)\,,
\end{align}
which, as can be seen from the formula \eqref{eq:Bessel-solution} and our discussion of the transfer-matrix in the subsection \ref{subsec:transfer-matrix}, corresponds (in the 'one-sided' case $\mathcal{H} = \ell(\mathbb{Z}_{\geq 0})$) to the similarity transform exchanging the transfer-operator having 1's above the diagonal with the transfer-operator having 1's below the diagonal. Because of this relation, essentially it is enough to work just with one of the operators $T^{(1)}_{\chi}$, $T^{(2)}_{\chi}$, say, the first.

To check linear independence of the two solutions $c_n^{(p), \pm}$, we use a $\hq$-analogue of the Wronskian:
\begin{align}
    W[f_1, f_2](z) \colonequals f_1(z) f_2(\hq z) - f_2(z) f_1(\hq z)\,.
\end{align}
Like its classical cousin, as long as it is non-zero, it guarantees linear independence of the two solutions. From \cite{OR-qBessel}, proposition 3.6, we see that
\begin{align}
   W\left( I_{\nu}^{(1)}(z;q), I_{-\nu}^{(1)}(z;q)\right) = \hq^{-\nu}\frac{\left(q^{\nu}, q^{1-\nu}; q\right)_{\infty}}{(q;q)^2_{\infty}}\, e_0\left(\frac{z^2}{4}; \hq^2\right)\,,
\end{align}
so that the only possible spectral points where the two ($\pm$) solutions \eqref{eq:Bessel-solution} fail to be independent are
\begin{align*}
    x^{(1)} = x^{(1)}_m \colonequals \pm \frac{\hq^m + \hq^{-m}}{2}, \quad m\in \mathbb{Z}\,.
\end{align*}

It turns out that, like in the classical case, there is only one combination of the modified Bessel functions that is actually bounded as $n\to -\infty$, this is an appropriate $q$-analogue of the Macdonald function. The only fact about it that we need here is that it is (for both $p=1,2$) proportional to a specific basic hypergeometric function $_2\Phi_1$ with the parameters that we give in \eqref{eq:2phi1-Macdonald}.

To see what are the actual spectra of the (self-adjoint extensions) of the operator, say, $T_{\chi}^{(1)}$, it's better to switch from the sequence space description of our recursion to the description as a difference operator acting on functions on a $\hq$-lattice. Take $H \colonequals \hq^{-n}$, $F(H) \colonequals c_n$. Notice that this coordinate $H$ is a natural one in the vicinity of the $n\to -\infty$ region (since the latter corresponds to $H \to 0$). Our ($p=1$) eigenvalue equation \eqref{eq:Tp-chi-eigen} will then read (after an overall shift by $\hq$):
\begin{align}
\left[\left(1-\frac{\chi}{\hq^2H^2}\right)\tau_{\hq}^2 - 2x^{(1)}\tau_{\hq}+1 \right]F(H) = 0\,,
\end{align}
where $\tau_{\hq}$ is the shift operator, acting on functions by multiplying the argument by $\hq$.
In the rescaled variable
\begin{align}
    t \colonequals \frac{H}{2\hq \chi^{1/2}} \equiv \frac{\hq^{-n-1}}{2\chi^{1/2}}\,,
\end{align}
the difference equation becomes
\begin{align}
\left[\left(1-\frac{1}{4\hq^4 t^2}\right)\tau_{\hq}^2 - 2x^{(1)}\tau_{\hq}+1 \right]F(t) = 0\,.
\end{align}
One can see that $t=0$ is an irregular singular point. Just as in the classical case of second-order differential equations, in this case it is natural to represent the solution as an appropriate exponential times another function having a power series expansion.

Namely, let us define the following functions:
\begin{align}
    \theta_{\hq}(t) \ \colonequals \sum_{n\in \mathbb{Z}} \hq^{\binom{n}{2}}t^n, \quad \mathcal{E}_{\xi}(t)\colonequals \frac{1}{\theta_{\hq}(-\xi t)}\,,
\end{align}
and take the parameter $\xi = \pm 2 \hq^{3/4}$ (both of these values work).
Factoring out as
\begin{align}
    F(t) \equalscolon \mathcal{E}_{\xi}(t)\, f(t)\,,
\end{align}
we get for $f(t)$ the difference equation:
\begin{align}
    \left[-(1-4\hq^4t^2)\tau_{\hq}^2 + 2x^{(1)}\xi t \tau_{\hq} + 1\right] f(t) = 0\,.
\end{align}
Then $f(\hq^{1/2}t)/(\alpha t ; \hq)_{\infty}$ satisfies:
\begin{align}\label{eq:hyperg-transfer-opr}
\left[-(1+\xi \hq t) \tau_{\hq}^2 + 2x^{(1)}\xi\hq^{1/2} t \tau_{\hq} + (1-\xi t)\right] \cdot \left[f(\hq^{1/2}t)/(\alpha t ; \hq)_{\infty}\right] = 0\,,
\end{align}
which is exactly the defining difference equation for the following $\hq$-hypergeometric function:
\begin{align}\label{eq:2phi1-Macdonald}
    \pPq{2}{1}{\hq^{1/2}\left(x^{(1)} + \sqrt{x^{(1)\,2}-1}\right), \hq^{1/2}\left(x^{(1)} - \sqrt{x^{(1)\,2}-1}\right)}{-\hq}{\hq ,\, \xi t}\,.
\end{align}
Spectral theory of this operator in the parameter range relevant for us was studied in \cite{Koelink2003}. The analysis there shows that it can be brought to a symmetric form and that the resulting operator has a continuous one-parameter family of self-adjoint extensions. Unless the parameters are such that the eigenfunction becomes polynomial, the spectrum always contains some bound states, see \cite{Koelink2003}, theorem 5.8 and remark 2.3.
Thus we get a family of self-adjoint extensions for our operators $T^{(p)}_{\chi}$, $p=1,2$. When the parameters are such that the eigenfunction becomes polynomial, this corresponds to the situation when the transfer-matrix recursion splits, and we have continuous $q$-Hermite polynomial eigenfunctions.\footnote{To be more precise, for generic $\chi$ the polynomial eigenfunctions are given by a $\chi$-deformation of the $q$-Hermite polynomials, called symmetric Al-Salam-Chihara polynomials. They go to the (appropriately shifted) continuous $q$-Hermite polynomials as $\chi \to q^m$, $m\in\mathbb{Z}$.}

A useful fact from functional analysis is Weyl's theorem (see e.g. \cite{Reed-Simon-book}) saying that essential spectrum
of the self-adjoint operator (that is, the spectrum unaffected by compact perturbations, such as additional eigenvectors corresponding to bound states) does not change when we go over the possible extensions. The essential spectrum in our case consists only of continuous spectrum on the interval $[-1;1]$. We interpret the bound states that appear on top of that as corresponding to operators sitting at the Poincar{\'e} horizon.

Since in our interpretation, the transfer-matrix describes only Hamiltonian chords and should reflect physics on a thermal circle, it is a physically reasonable choice to demand that no additional states are localized just on the Poincar{\'e} horizon. In this way, the spectral analysis outlined above singles out the boundary conditions corresponding to the terminating transfer-matrix.

\section{A short review of quantum groups}
\label{sec:q-groups-intro}

Effectively, in subsection \ref{sec:BootApp} we reviewed the derivation of the action by reducing a Minkowskian $AdS_3$ down to two dimensions, specifying the momentum along one of the light-cone directions. Here we are interested in a $\hq$-deformation of that construction. A closely related problem was actually discussed when starting with a Euclidean $AdS_3$\footnote{Recall that hyperbolic space, Lobachevsky space  and Euclidean AdS space are different names for the same object (in particular, we will always have in mind only one sheet of the corresponding two-sheeted hyperboloid). We occasionally use all of these names, depending on the context, but will prefer the notation $\mathbb{H}^3$ and $\mathbb{H}^2$ for the three- and two-dimensional Euclidean AdS spaces (and similarly for the corresponding $\hq$-deformed objects, see further).}
\cite{Olshanetsky:1993sw, Olshanetsky:2001}. We will review and follow these works, and only at the end carry out the appropriate analytic continuation to the Minkowski space. This will serve to explain the mathematical origin of some of the hands-on models that we discussed so far, and the ambiguities that it contains. In this section we will do a crash review of the quantum groups relevant for us
 and in the next section \ref{sec:Lobachevsky} we will carry out the actual reduction to the hyperbolic space and its $\hq$-analogues.

Recall that the group of Riemannian isometries of Euclidean $AdS_3$ is the Lie group\footnote{We denote a connected component of identity in a Lie group by subscript {\it e}. E.g. $SO_e(1,3)$ means the identity component of the Lie group $SO(1,3)$.} $SL(2,\mathbb{C})/\mathbb{Z}_2 \simeq SO_e(1,3)$, whereas for Minkowskian $AdS_3$ it is $\left(SL(2,\mathbb{R})\times SL(2,\mathbb{R})\right)/\mathbb{Z}_2 \simeq SO_e(2,2)$.
Since we will be starting with Euclidean $AdS_3$, we will need to $\hq$-deform the group algebra of $SL(2,\mathbb{C})$.

First, let us quickly recap how the isomorphism of Lie groups
\begin{align}\label{eq:iso-sl2c-so13}
SL(2, \mathbb{C})/\mathbb{Z}_2 \xlongrightarrow{\sim} SO_e(1,3)\,,
\end{align}
actually comes about. Consider a point $x \equiv (x_0, x_1,x_2, x_3)\in\mathbb{R}^{1,3}$ in the defining representation  of $SO(1,3)$ and put its coordinates into a Hermitian matrix:
\begin{align}
    h(x) = \begin{pmatrix}
    x_0 + x_1 & x_2 + ix_3\\
    x_2 -ix_3 & x_0 - x_1
    \end{pmatrix}\,.
\end{align}

We let an element $w \in SL(2, \mathbb{C})$ act on this matrix as
\begin{align}\label{eq:sl2c_adjoint_action}
    \pi(w):\,\,h(x) \mapsto w^\dagger h(x) w\,,
\end{align}
which clearly preserves the hermiticity of the matrix and its determinant $det \,\,h(x) = x_0^2 - x_1^2-x_2^2-x_3^3$. In other words, it gives an element of the Lie group $SO_e(1,3)$ acting in the $x$-space. This action has a $\mathbb{Z}_2$ redundancy, since $w=-{\mathbb I}_{2*2}$
stabilizes $h(x)$. We can definitely get any element of the group $SO_e(1,3)$ from some element of $SL(2,\mathbb{C})$ this way. Thus, after we mod out by the redundancy, we get an isomorphism of Lie groups.
Taking a differential of this isomorphism, we get it on the level of Lie algebras:
\begin{align}\label{eq:Lie_alg_iso_sl2C}
    \mathfrak{s}\mathfrak{o}(1,3) \xlongrightarrow{\sim} \mathfrak{s}\mathfrak{l}(2,\mathbb{C})\,,
\end{align}
with the corresponding action of $\mathfrak{s}\mathfrak{l}(2,\mathbb{C})$ on $h(x)$ given by:
\begin{align}\label{eq:sl2c_adjoint_action_Liealg}
    d\pi(Y):\,\,h(x) \mapsto  h(x) \cdot Y + Y^* \cdot h(x), \quad Y \in \mathfrak{s}\mathfrak{l}(2,\mathbb{C})\,.
\end{align}
We will use these isomorphisms
in this section and in section \ref{sec:Lobachevsky}, where the Lobachevsky space and its $\hq$-analogues, as well as a $\hq$-deformed version of the coset reduction, are discussed.

The main objects we will need to describe in the next two sections are:
\begin{itemize}
    \item  The generalization of the Lie algebra $\mathfrak{s}\mathfrak{l}(2, \mathbb{C})$ and its $\hq$-deformations. More precisely, since we would like to be able to multiply elements in the algebra, rather then just compute Lie brackets, we first need to go to the universal enveloping algebra ${\cal U}(\mathfrak{s}\mathfrak{l}(2, \mathbb{C}))$, and then $\hq$-deform that into a more interesting Hopf algebra ${\cal U}_{\hq}(\mathfrak{s}\mathfrak{l}(2, \mathbb{C}))$ (and its multiparametric cousin ${\cal U}_{\hq, r, s}(\mathfrak{s}\mathfrak{l}(2, \mathbb{C}))$).
    \item The group $SL(2,\mathbb{C})$, the algebra of functions on it, and subsequently (the different versions of) the $\hq$-deformation ${\cal A}_{\hq}(SL(2, \mathbb{C}))$ of the latter.\footnote{We are going to ignore the subtleties of the non-compact quantum groups in our discussion. In particular, typically, the non-compact case requires to refine the Hopf-algebraic description that works well for compact quantum group case, see \cite{Kustermans-Vaes} for the reference. E.g. the antipode is usually not defined on the entire non-compact quantum group, but only on a *-strongly-dense core. While a finer analysis is very interesting in order to fully appreciate the operator-algebraic content of our model, we largely gloss over these distinctions in the present paper. The subsequent discussion will be purely algebraic.} This by itself will not be necessary for the reduction, but will provide an avenue into the the next object which is  
    \item A $\hq$-deformed Euclidean $EAdS_3 \equiv \mathbb{H}^3$. This will also come in two versions, labelled $\mathbb{H}^3_{\hq}$ and $\mathbb{H}^3_{\hq, \kappa}$. We note here that these versions are the ones explicitly discussed in the literature. Since there is no easy classification of such spaces, there might be additional variants.
\end{itemize}

Before we go on to consider our quantum groups, let us clarify and introduce some notations.

\subsection{Notation}

In this subsection we will briefly summarize our mathematical notation needed in the subsequent parts of the paper.

Recall that $\hq = q^{1/2} = e^{-\lambda/2}$. All the formulas will be given only for the relevant range of parameter $0< \hq <1$.

We remind the '$\hq$-number' definitions commonly used to compactify the notation:
\begin{equation}
\begin{split}
    \mbox{$\hq$-number  }  \hspace{12mm} [l]_{\hq^2}  & \colonequals {{\hq}^l -\hq^{-l} \over \hq -\hq^{-1}}, \quad l \in \mathbb{Z}_{\geq 0} \\
    \mbox{$\hq$-factorial  } \hspace{10mm} [l]!_{\hq^2}   & \colonequals [l]_{\hq^2} [l-1]_{\hq^2} \dots [1]_{\hq^2}, \qquad [0]_{\hq^2} \colonequals 1 \\
    \mbox{$\hq$-binomial } \hspace{5mm} \binom{l}{k}_{\hq^2}   & \colonequals { [l]!_{\hq^2} \over [k]!_{\hq^2} [l-k]!_{\hq^2} }, \quad k, l \in \mathbb{Z}_{\geq 0}, \quad k \leq l\,. \\
\end{split}    
\end{equation}

We will discuss various mathematical objects ranging from very general to specific. The following are our conventions:

\begin{itemize}

\item A letter without upper or lower scripts refers to a general object of the type under discussion at that point.
   
\item Fraktur script is reserved for the Lie algebras: both abstract $\mathfrak{g} = Lie\, G$ (that is, $\mathfrak{g}$ is a Lie algebra\footnote{When we make examples involving a general Lie algebra $\mathfrak{g}$, we always assume it to be semisimple, i.e. not containing any $\mathfrak{u}(1)$ factors.} of Lie group $G$) and concrete, such as $\mathfrak{s}\mathfrak{l}(2, \mathbb{C})$, $\mathfrak{s}\mathfrak{u}(1,1)$. The universal enveloping algebra of a Lie algebra $\mathfrak{g}$ is denoted as $\mathcal{U}\mathfrak{g}$.

\item Curly capital letters are reserved for the spaces $\mathcal{X}$ carrying an action of some Lie group $G$,\footnote{All the classical spaces we consider here are smooth, so this automatically implies that $\mathcal{X}$ carries an action of the Lie algebra $Lie\, G$, too.} or for an algebra of functions $\mathcal{B}$ on such space.

\item In most cases we explicitly label the $\hq$-deformations of objects by a subscript $\hq$, at least when keeping track of the deformation parameter is important to us. For example, a $\hq$-deformed version of the algebra of functions $\mathcal{B}$ on the $G$-space $\mathcal{X}$ is denoted as $\mathcal{B}_{\hq}$. While there are other alternatives (see \cite{Koornwinder-Dijkhuizen} for an overview of possible approaches in the context of $q$-homogeneous spaces, on the algebraic level), we choose to describe such deformed $G$-spaces as $\mathcal{U}_{\hq}$-module algebras for the corresponding quantum group $\mathcal{U}_{\hq}$ of symmetries, see section \ref{sec:Modules of U} for details.

\item Hopf algebras are a framework to describe deformations of structures related to the classical Lie groups. The constituents of a Hopf algebra, say, $\mathcal{A} = (A, \Delta, \epsilon, S)$ are labeled in a traditional way: $A$ is an (associative unital) algebra, $\Delta$ is a coproduct map $\Delta:\,\, A \to A\otimes A$, $\epsilon:\,\, A \to \mathbb{C}$ is a counit and $S:\,\, A \to A$ is the antipode map. We prefer not to use notations for the algebra multiplication and unit maps explicitly. We will also often allow ourselves to abuse the notation in calling an algebra part $A$ of the Hopf algebra structure of $\mathcal{A}$ by the same curly letter $\mathcal{A}$: we hope it will not lead to a confusion, since it should be clear in each case which structure we mean. Explicit definitions of these maps and the particular Hopf algebras relevant for us are introduced in subsections \ref{subsec:U_hopf_alg}, \ref{subsec:A_hopf_alg}.

When there is a choice in how to label a general Hopf algebra, we prefer using letters $\mathcal{A}$ if we have in mind Hopf algebras that are related to deformations of an algebra of functions on a Lie group, and letters $\mathcal{U}$ if we rather think of Hopf algebras that are related to deformations of a universal enveloping algebra of some Lie algebra.

    \item
    On the other hand, to label concrete objects of the types we just described, we will add to the notations for generic such objects two superscripts: $\mathbb{C}$ and $\mathbb{R}$.
    In particular, a letter with superscript $\mathbb{C}$ will label a concrete object related to Lie group $G^{\mathbb{C}} \colonequals SL(2,\mathbb{C})$ or its deformations, for example:
    \begin{itemize}
    \item Lie algebra $\mathfrak{g}^{\mathbb{C}} \colonequals \mathfrak{s}\mathfrak{l}(2, \mathbb{C})$,
    \item Lobachevsky 3-space (that is, Euclidean $AdS_3$) $\mathcal{X}^{\mathbb{C}} \colonequals \mathbb{H}^3 \simeq SU(2)\backslash SL(2, \mathbb{C})$,
    \item the universal enveloping algebra of $\mathfrak{s}\mathfrak{l}(2,\mathbb{C})$, $\mathcal{U}^{\mathbb{C}} \colonequals \mathcal{U}(\mathfrak{s}\mathfrak{l}(2,\mathbb{C}))$
    \item quantum group $\mathcal{U}_{\hq}^{\mathbb{C}} \colonequals \mathcal{U}_{\hq}(\mathfrak{s}\mathfrak{l}(2,\mathbb{C}))$ which is one of possible $\hq$-deformations of $\mathcal{U}^{\mathbb{C}}$.
\end{itemize}

    \item A letter with superscript $\mathbb{R}$ will label objects that are pertinent to the real form $G^{\mathbb{R}} \colonequals SU(1,1)$ of the Lie group $SL(2,\mathbb{C})$, or its deformations\footnote{When the deformation parameter $\hq \neq 1$, the real form $\mathcal{U}_{\hq}\,\mathfrak{s}\mathfrak{u}(1,1)$ is not to be confused with the other real form of this quantum group, denoted $\mathcal{U}_{\hq}\,\mathfrak{s}\mathfrak{l}(2,\mathbb{R})$ in the literature. This latter form exists only for $\hq$ being a pure phase and will not concern us in this paper.}, for example:
    \begin{itemize}
    \item a Lie algebra $\mathfrak{g}^{\mathbb{R}} \colonequals \mathfrak{s}\mathfrak{u}(1,1)$,
    \item its universal enveloping algebra $\mathcal{U}^{\mathbb{R}} \colonequals \mathcal{U} (\mathfrak{s}\mathfrak{u}(1,1))$
    \item Euclidean $AdS_2$ space $\mathcal{X}^{\mathbb{R}} \colonequals \mathbb{H}^2 \simeq U(1)\backslash SU(1,1) $, and
    \item $\hq$-deformation of the Hopf algebra of polynomial functions on the group $SU(1,1)$: $\mathcal{A}^{\mathbb{R}}_{\hq} \colonequals \mathcal A_{\hq}\,(SU(1,1))$.
    \end{itemize}
   
\item Another aspect of our notation is worth reiterating: $\hq$-deformed $AdS_3$ and $AdS_2$ will either be described using non-commuting variables, or by using discrete lattices. The coordinates of the former will be denoted by letters without a tilde ($z,H,z^*$), and of the latter by letters carrying tildes (${\tilde z},{\tilde H}, {\tilde z}^*$).
   
\end{itemize}

\subsection{$\hq$-deformations of ${\cal U}(\mathfrak{s}\mathfrak{l}(2),\mathbb{C})$}
\label{subsec:U_hopf_alg}

In section \ref{sec:hands-on} we discussed the algebra\footnote{The full Hopf algebra structure was irrelevant to us in section \ref{sec:hands-on}.} $\mathcal{U}_{\hq} \left(\mathfrak{s}\mathfrak{u}(1,1)\right) \equiv \mathcal{U}_{\hq}^{\mathbb{R}}$.
It's now time to introduce its complexification
$\mathcal{U}_{\hq}^{\mathbb{C}}$.
Just as in the classical case, we first need to 'double' the number of generators of the algebra. In other words, let us take the generators $A,B,C,D$ without any restrictions of reality
and append the list of generators by $A^*,B^*,C^*,D^*$. The algebra that they satisfy is \eqref{SLq2},
and the relations for $A^*$, $B^*$, $C^*$ and $D^*$ are obtained from it by using the relation $(uv)^*=v^*u^*$, $u,v \in \mathcal{U}_{\hq}^{\mathbb{C}}$ for the conjugation anti-automorphism.
Looking back at the classical isomorphism \eqref{eq:Lie_alg_iso_sl2C},
we can see that, since in the 'geometric' action \eqref{eq:sl2c_adjoint_action_Liealg} on the matrix $h(x)$ a particular generator and its conjugate counterpart act from different sides, they commute.
We impose that they still commute in our $\hq$-deformation.

The basic objects that go into the definition of a
Hopf algebra, a mathematical object naturally describing both usual and $\hq$-deformed symmetries (in the latter case e.g. such as the ones combining into quantum groups), are the coproduct, counit and the antipode. We will first state the formulas for those in the case of specific quantum groups $\mathcal{U}_{\hq}^{\mathbb{C}}$ and $\mathcal{U}_{\hq, r,s}^{\mathbb{C}}$,\footnote{As well as for the arbitrary 'classical' Hopf algebra $\mathcal{U}\mathfrak{g}$, such as $\mathcal{U}^{\mathbb{C}}$, to build some intuition for the $\hq$-deformed cases.}  
and then explain a little what they are and how they are used.

A coproduct representing a 'minimal' deformation of the coproduct for the classical universal enveloping algebra $\mathcal{U}^{\mathbb{C}}$ reads \cite{Woronowicz1994}:
\begin{align}\label{eq:U_coproduct}
    \Delta(A) = A\otimes A, \quad \Delta(B) = A \otimes B + B \otimes D, \quad \Delta (C) = A\otimes C + C \otimes D\,.
\end{align}
We will denote the Hopf algebra
corresponding to this coproduct as $\mathcal{U}_{\hq}^{\mathbb{C}}$.
Notice that the action of this coproduct on generators without stars doesn't involve any generators with stars, i.e.  the starred and unstarred 'sectors' do not mix. In this way this coproduct can also be
restricted to the real forms of this Hopf algebra, including $\mathcal{U}_{\hq}(\mathfrak{s}\mathfrak{u}(1,1))$ that we discussed in section \ref{sec:hands-on}.

The counit and antipode maps of $\mathcal{U}_{\hq}^{\mathbb{C}}$ are
\begin{equation}
    \begin{split}\label{eq:Counit for U1sl2c}
        \epsilon (A) = 1, \quad \epsilon (D) = 1,\quad \epsilon (B) = 0 ,\quad \epsilon (C) = 0\,,
    \end{split}
\end{equation}
\begin{equation}\label{eq:Antipode for U1sl2c}
    S(A) = D,\quad S(D) = A,\quad S(B) = -\hq^{-1} B,\quad S(C) = -\hq C\,.
\end{equation}

However, instead of \eqref{eq:U_coproduct}, we can also allow for a more general coproduct that mixes the two 'sectors' of generators (the ones carrying and not carrying stars). This then defines a different coalgebra (and, correspondingly, Hopf algebra) structure, and is related to the previous coproduct by {\it twisting}\footnote{More precisely, this corresponds to choosing a specific invertible element in $\mathcal{U}^{\otimes 2}$ called (left) Drinfeld twist \cite{Drinfeld1990}. In order to render the new coproduct coassociative and counital, Drinfeld twist should satisfy some further compatibility conditions \cite{Reshetikhin1990}.
}
\cite{Reshetikhin1990}:
\begin{equation}\label{eq:U_coproduct_twisted}
\begin{split}
    \Delta(A) &= A\otimes A, \quad \Delta(B) = \left(A^*\right)^{-r}A \otimes B + B \otimes D\left(A^*\right)^s,\\ \quad \Delta (C) &= \left(A^*\right)^{r}A\otimes C + C \otimes D\left(A^*\right)^{-s}\,\ \ \ r,s \in \mathbb{Z}.
\end{split}
\end{equation}
We will call the corresponding Hopf algebra by $\mathcal{U}_{\hq,r,s}^{\mathbb{C}}$. Compared to $\mathcal{U}^{\mathbb{C}}_{\hq}$, the counit stays the same and the antipode changes into
\begin{equation}
    \label{eq:Twisted Antipode}
S(A) = D,\quad S(D) = A,\quad S(B) = -\hq^{-1} {A^*}^{r-s} B,\quad S(C) = -\hq {A^*}^{s-r} C\,.
\end{equation}

Lastly, we should add here that all these structures are extended to the generators labeled with a star by demanding the usual compatibility conditions between the conjugation $*$ and the maps $\Delta$, $\epsilon$:\footnote{The underlying algebra is then also assumed to be a $^*$-algebra, i.e. an algebra equipped with an anti-involution preserving the algebra structure.}
\begin{align}\label{eq:star-properties}
    (*\otimes *) \circ \Delta = \Delta \circ * \quad \text{ on } \mathcal{U}, \quad \epsilon(u^*) = \overline{\epsilon(u)}, \quad u\in \mathcal{U}\,,
\end{align}
where $\circ$ denotes composition of different maps (from right to left) and, for instance, $\mathcal{U} = \mathcal{U}^{\mathbb{C}}, \mathcal{U}_{\hq}^{\mathbb{C}}, \,\mathcal{U}_{\hq,r,s}^{\mathbb{C}}$.\footnote{Using these and Hopf algebra compatibility conditions discussed below, it is then quick to check that
\begin{align*}
    S \circ * \circ S \circ * = \text{1}\,.
\end{align*}
}

\medskip

The coproduct, counit and antipode are used in the following way.

\smallskip

{\bf Coproduct:} The {\it coproduct} is used to define how the algebra acts on a tensor product of its modules.\footnote{We make no distinction between a representation and a module for an algebra, and e.g. often use the word 'representation' even
when we only care about the module structure.
} It is a linear map $\Delta:\  \mathcal{U} \to \mathcal{U} \otimes \mathcal{U}$ which basically tells us
what an element $u\in \mathcal{U}$ does when we apply it
on a tensor product of two representations $V_1$, $V_2$.
We will sometimes use Sweedler's notation for the coproduct of Hopf algebras:
\begin{align}\label{eq:Sweedler}
        \Delta(u) = \sum_{(u)} u_{(1)} \otimes u_{(2)}\,, \quad u, u_1, u_2 \in \mathcal{U}\,.
\end{align}
Thus, for instance,
\begin{equation}
\Delta(u)(v_1\otimes v_2) = \sum_{(u)}\
\bigl( u_{(1)} v_1  \bigr) \otimes  
\bigl( u_{(2)} v_2 \bigr)\,,
\end{equation}
where all $u$ and $v$ are again elements of our Hopf algebra in question.

For example for a classical Lie algebra $\mathfrak{g} \subset \mathcal{U}\mathfrak{g}$, the coproduct always reads\footnote{
Thus, $X_{(1)} = \{X, 1\}$ and $X_{(2)} = \{1, X\}$.
}
\begin{align}\label{eq:coproduct_Lie_alg}
    \Delta(X) = X \otimes 1 + 1 \otimes X, \quad X \in \mathfrak{g}\,,
\end{align}
so that in a tensor product of two $\mathfrak{g}$-representations $V_1, V_2$, $X$ will act as
\begin{align}
    \Delta(X) (v_1 \otimes v_2) = (X \otimes 1 + 1 \otimes X)(v_1 \otimes v_2) = X(v_1) \otimes v_2 + v_1 \otimes X(v_2), \quad v_i \in V_i, i = 1,2,
\end{align}
which just reproduces the usual addition rule of, say, angular momenta. Of course, this then extends to the action of the Hopf algebra $\mathcal{U} = \mathcal{U}\mathfrak{g}$, the universal enveloping algebra of $\mathfrak{g}$,
on $V_i$. In the limit $\hq\rightarrow 1$, since $A,A^*\sim \hq^{L_0},\hq^{L_0^*}\rightarrow 1$, the coproducts \eqref{eq:U_coproduct}, \eqref{eq:U_coproduct_twisted} contract to the non $\hq$-deformed coproduct of the form \eqref{eq:coproduct_Lie_alg}.

As part of its definition, the co-product also needs to satisfy the relation
\begin{align}\label{eq:coassoc}
(\Delta\otimes id) \circ \Delta = (id\otimes\Delta) \circ \Delta \quad \text{ on } \mathcal{U}\,.
\end{align}
This property ('coassociativity') makes sure that it doesn't matter in which order we do tensor product of three (and more) representations.
Besides, the coproduct should be a map of algebras, which just means
\begin{align}
    \Delta(u_1 u_2) = \Delta (u_1) \Delta(u_2),\, \quad u_1, u_2 \in \mathcal{U}, \quad \quad  \quad \Delta(\text{1}) = \text{1}\otimes \text{1}\,,
\end{align}
where 1 denotes a unit element in the algebra.
Thus, in the classical example of the Hopf algebra $\mathcal{U} = \mathcal{U}\mathfrak{g}$, the second formula defines the coproduct on constants (by linearity, $\Delta(p) = p \cdot \text{1}\otimes \text{1}$ for a complex number $p$), and then the first formula allows us to extend the coproduct from just 'affine linear functions' $pX + q$, $p,q \in \mathbb{C}$ of Lie algebra elements $X \in \mathfrak{g} \subset \mathcal{U}\mathfrak{g}$ further to arbitrary polynomials in these elements (which span the whole algebra $\mathcal{U}\mathfrak{g}$).

We will need the coproduct of the $\hq$-deformed $\mathcal{U}^{\mathbb{C}}$ as follows. The $\hq$-deformed $\mathbb{H}^3\equiv EAdS_3$ will be obtained by a certain manipulation with the Hopf algebras
$\widetilde{{\cal A}}_{\hq, s}^{\mathbb{C}}$, ${\cal A}_{\hq, t}^{\mathbb{C}}$ (deformations of $\mathcal{A}^{\mathbb{C}}$, both defined in the next subsection \ref{subsec:A_hopf_alg}), which mimics how the classical coset reduction produces $\mathbb{H}^3$ from $SL(2,\mathbb{C})$ (see subsection \ref{subsec:L3}). This will go in two stages:

1.  First,\footnote{We will exemplify this first stage only for {\it one} particular deformation of $\mathcal{A}$, called $\widetilde{\mathcal{A}}_{\hq, s}^{\mathbb{C}}$, in subsection \ref{sec:OR1} and appendix \ref{app:details_qLobachevsky}.} we will need to figure out how elements of
the $\hq$-deformed $\mathcal{U}^{\mathbb{C}}$
act on
the $\hq$-deformation of $\mathcal{A}^{\mathbb{C}}$.
The action is built from defining the action of $A,B,C,D$ on simple elements (generators) of the latter, and then taking products (basically, building polynomials from the single letters). In order to see how $A,B,C,D$ act on the more complicated monomials, we use the coproduct.

2.  Going through
the $\hq$-deformed $\mathcal{A}^{\mathbb{C}}$
is in fact just an intermediate stage, since we are actually interested in how $A,B,C,D$ act on functions on the $\hq$-deformed $EAdS_3$. We again start by how they act on simple functions, and then obtain their action on more complicated monomials. To figure out the latter, we again need to use the coproduct.

\smallskip

{\bf Counit:} While for quantum groups that we will discuss in the next subsection, the counit has a natural interpretation as an 'evaluation map at the identity', for (universal enveloping algebras of) Lie algebras and their $\hq$-deformations, the intuition behind the counit is different, and might seem more artificial. Namely, in the case of a classical Lie algebra $\mathfrak{g}$, given a polynomial in the Lie algebra elements (that is, an element of $\mathcal{U}\mathfrak{g}$), the {\it counit} $\epsilon:\,\, \mathcal{U} \to \mathbb{C}$ is an operation taking out a constant term of that polynomial. In other words, we have
\begin{align}
    \epsilon(\mathbf{1}) = 1, \quad \epsilon(X) = 0, \quad X \in\mathfrak{g}\,,
\end{align}
and then we extend this to the entire $\mathcal{U}\mathfrak{g}$ using its universal property and that $\epsilon$ should be an algebra map, $\epsilon(X_1X_2) = \epsilon(X_1)\epsilon(X_2)$. If we are to compare this definition with the counit \eqref{eq:Counit for U1sl2c} of the deformed Hopf algebra $\mathcal{U}_{\hq}^{\mathbb{C}}$, notice that $A\sim\hq^{L_0}$ for $\hq\sim 1$, so $A$ looks more like a group element close to the identity, rather than an element of a Lie algebra: in the Hopf algebra jargon, such elements $A\in\mathcal{U}_{\hq}^{\mathbb{C}}$ are called group-like. Thus, it is only natural that the counit value of $A$ is $1$.

Besides, in this context one usually demands\footnote{This is completely analogous to the situation when, having specified a unit element 1 in the algebra, to make it into a {\it unital algebra} we moreover demand multiplication to be compatible with the unit, i.e. $a \times \text{1} = a$ for any algebra element $a$.} that counit is compatible with the coproduct, which gives an additional condition:
\begin{align}\label{eq:counital}
    (\epsilon \otimes id) \circ \Delta = (id \otimes \epsilon) \circ \Delta = id \quad \text{ on } \mathcal{U}\,.
\end{align}
For $\mathcal{U} =  \mathcal{U}\mathfrak{g}$ this
essentially just means that tensoring with identity doesn't change the constant term.

{\bf Antipode:} Finally, the {\it antipode} is a linear\footnote{Linearity is enough here. It turns out that the axioms we already imposed constrain it further to be an anti-algebra map, i.e. a homomorphism of algebras reversing multiplication.} map $S:\,\, \mathcal{U} \to \mathcal{U}$. It acts as identity on numbers $\mathbb{C} \subset \mathcal{U}\mathfrak{g}$, whereas on the Lie algebra $\mathfrak{g}$ (when we view it as a subspace within $\mathcal{U}\mathfrak{g}$) the antipode acts by taking the elements to $(-1)*$themselves:
\begin{align}\label{eq:antipode_classical}
   S(\text{1}) = \text{1}, \quad S(X) = -X, \quad X \in \mathfrak{g}\,,
\end{align}
which extends to the full $\mathcal{U}\mathfrak{g}$ as a map negating the odd-degree pieces of a corresponding polynomial in $\mathcal{U}\mathfrak{g}$.
If we think of \eqref{eq:Antipode for U1sl2c} as a deformation of a classical Lie algebra, we see that $B$ and $C$ still behave a lot like Lie algebra elements, as in the limit $\hq\rightarrow 1$ they indeed pick up a minus sign under the antipode action. On the other hand, as we just said, $A$ is group-like ($\sim\hq^{L_0}$),
so the action of the antipode \eqref{eq:Antipode for U1sl2c} on it by inversion becomes in the $\hq\rightarrow 1$ limit precisely the antipode action by negation on $L_0$.
 
One can check that the antipode map \eqref{eq:antipode_classical} on $\mathcal{U} = \mathcal{U}\mathfrak{g}$ (trivially) satisfies the following condition:
\begin{align}\label{eq:antipode_condition}
    m \circ (id \otimes S) \circ \Delta (u) = m \circ (S \otimes id) \circ \Delta (u) = \epsilon(u) \cdot \text{1}\,, \quad u\in \mathcal{U}\,.
\end{align}
The linear map $S:\, \,\mathcal{U} \to \mathcal{U}$ satisfying this same condition
is the definition of the antipode for the general
Hopf algebra case. In particular, for such more general Hopf algebras (e.g. for our quantum groups $\mathcal{U}_{\hq}^{\mathbb{C}}$, $\mathcal{U}_{\hq,r,s}^{\mathbb{C}}$), \eqref{eq:antipode_condition} becomes a non-trivial constraint.

{\bf All together:}
The whole bunch of structures and compatibility conditions that we just listed
turn $\mathcal{U}\mathfrak{g}$, $\mathcal{U}_{\hq}^{\mathbb{C}}$, $\mathcal{U}_{\hq,r,s}^{\mathbb{C}}$ into {\it Hopf algebras} (actually, {\it Hopf *-algebras} if we also take into account the properties of our conjugation *).
For more details, we refer to one of the classical textbooks, e.g. \cite{Kl-Schm}.
Shortly, a Hopf algebra is a
formalization of
what one could expect from, say, a set of polynomials in the Lie algebra generators
(or a reasonable set of functions on a group as we will see shortly in the next section) in a way that is flexible enough to allow interesting generalizations.\footnote{See also our footnote at the beginning of this section regarding further subtleties in a non-compact group case.}

\subsection{$\hq$-deformations of ${\cal A}(SL(2, \mathbb{C}))$}
\label{subsec:A_hopf_alg}

Now it's time to introduce a second quantum group (or rather a family of them) that we need in our story, the one which consists of the '$\hq$-deformed functions' on the classical group $G^{\mathbb{C}}$. Let us first
refer to
the classical case. Here we look at a matrix group $SL(2,\mathbb{C}) \equiv G^{\mathbb{C}}$, whose elements are parametrized as\footnote{Notice that $\mathfrak g$ is completely different from $g$.}
\begin{align}
    g = \begin{pmatrix}\alpha & \beta\\
    \gamma & \delta
    \end{pmatrix}, \quad \alpha, \beta, \gamma, \delta \in \mathbb{C}, \quad \alpha \delta - \gamma \beta = 1\,.
\end{align}
Clearly, $\alpha, \beta, \gamma$ and $\delta$ can be considered as coordinates on the group, or more precisely, values of the corresponding coordinate functions $G^{\mathbb{C}} \to \mathbb{C}$ on a group element $g\in G^{\mathbb{C}}$. If we form all possible polynomials in these four basic variables, and use the determinant one condition (when needed) to simplify those, we get an algebra of polynomial functions on our group, ${\cal A}(SL(2, \mathbb{C})) \equiv \mathcal{A}^{\mathbb{C}}$.

We need a non-commutative version of the $SL(2,{\mathbb C})$,
i.e. something that might be appropriately called $SL_{\hq}(2,{\mathbb C})$.
To tackle it, following the usual formalism, we will rather deal with
the corresponding $\hq$-deformation(s) of the algebra of functions $\mathcal{A}^{\mathbb{C}}$.
Let us now quickly summarize some relevant facts about such deformations and highlight the one(s) we need.

Possible $\hq$-deformations of the Lorentz group algebra $\mathcal{A}^{\mathbb{C}}$, with the constraint that they keep the finite-dimensional representation theory of the group unchanged, were classified in \cite{Woronowicz1994}.
Each variant of such a $\hq$-deformation is generated by
eight generators
which we denote by the letters $\alpha,\beta,\gamma,\delta$ and
$\alpha^*,\beta^*,\gamma^*,\delta^*$.
The latter four are the adjoints of the former.

All of the different deformations listed in \cite{Woronowicz1994}
share the following set of relations
\begin{equation}\label{eq:Algebra A_qs}
\begin{split}
 &\alpha \beta = \hq \beta \alpha, \qquad \alpha \gamma = \hq \gamma \alpha, \qquad \alpha \delta - \hq \beta \gamma = 1,\qquad \beta\gamma = \gamma \beta, \qquad \beta \delta = \hq \delta \beta \\
        &\beta\gamma=\gamma\beta,\qquad \gamma \delta = \hq \delta \gamma,\qquad \delta \alpha -{\hq}^{-1} \beta \gamma = 1\,,
\end{split}
\end{equation}
and their $*$-conjugates.
The variants come in how the generators and their conjugates behave with respect to multiplication.\footnote{As mentioned before, we will allow ourselves an ambiguity in notation here, by calling an algebra and its corresponding Hopf algebra (i.e. an algebra supplemented by appropriate compatible coproduct, counit and antipode) by the same letter $\mathcal{A}$, with sub/super scripts as needed. Since we only assign a unique coproduct, counit and antipode to each case of the $\mathcal{A}$ algebra, we hope this will not cause confusion, since it is clear from the context which one we mean.} We will focus on the following variant that we call $\widetilde{\mathcal{A}}_{\hq,s}^{\mathbb{C}}$:
\begin{equation}\label{eq:Alg313}
\begin{split}
&\widetilde{\mathcal{A}}_{\hq,s}^{\mathbb{C}}:\\
     &\alpha\alpha^*+s\gamma\gamma^*=\alpha^*\alpha,\ \alpha\beta^*+s\gamma\delta^*=\hq^{-1}\beta^*\alpha,\ \beta\beta^*+s\delta\delta^*=\beta^*\beta+s\alpha^*\alpha,\ \alpha\gamma^*=\hq\gamma^*\alpha,\\ &\beta\gamma^*=\gamma^*\beta,\ \gamma\gamma^*=\gamma^*\gamma,\ \alpha\delta^*=\delta^*\alpha,\hq^{-1}\beta\delta^*=\delta^*\beta+s\gamma^*\alpha\\
    &\delta\delta^*=\delta^*\delta+s\gamma^*\gamma\,.
\end{split}
\end{equation}
The parameter $s$ in the second case can actually be rescaled
 away,\footnote{Namely, if one rescales the generators of $\widetilde{\mathcal{A}}_{\hq,s}^{\mathbb{C}}$ as:
\begin{equation*}
    \alpha \equalscolon {\hat\alpha},\ \beta \equalscolon |s|^{1/2}{\hat\beta}, \ \gamma \equalscolon |s|^{-1/2}{\hat \gamma},\ \delta \equalscolon {\hat\delta}\,,
\end{equation*}
the new generators (we distinguish them by hats) will generate the algebra $\widetilde{\mathcal{A}}_{\hq,\text{sgn}\,s}^{\mathbb{C}}$.
This reduces the freedom we have here just to two possible values $s=\pm 1$.
} which results in just two distinct algebras $\widetilde{\mathcal{A}}_{\hq,s = \pm 1}^{\mathbb{C}}$. However, for our use we need to keep $s$ away from $\pm 1$: in order to obtain an appropriate $\hq\rightarrow 1$ limit, $s$ should be tuned differently.
We will choose $s=-(1-\hq^2)$ and drop the subscript $s$ from $\widetilde{\mathcal{A}}^{\mathbb{C}}_{\hq, s}$ in what follows:
\begin{align}
  \widetilde{\mathcal{A}}^{\mathbb{C}}_{\hq} \colonequals \widetilde{\mathcal{A}}^{\mathbb{C}}_{\hq, s = -(1-\hq^2)}\,.  
\end{align}

Another continuous family of possibilities, $\mathcal{A}_{q,t}^{\mathbb{C}}$, is:
\begin{equation}\label{eq:Alg311}
\begin{split}
\mathcal{A}_{q,t}^{\mathbb{C}}:\ \
    &\alpha\alpha^*=\alpha^*\alpha,\ \alpha\beta^*=t\beta^*\alpha,\ \beta\beta^*=\beta^*\beta,\ \alpha\gamma^*=t^{-1}\gamma^*\alpha,\\ &\beta\gamma^*=\gamma^*\beta,\ \gamma\gamma^*=\gamma^*\gamma,\ \alpha\delta^*=\delta^*\alpha,\ \beta\delta^*=t^{-1}\delta^*\beta\\
    &\delta\delta^*=\delta^*\delta\,.
\end{split}
\end{equation}
There exist also a few sporadic, discrete-parameter cases.

Next we will now list all the ingredients of the Hopf algebra structure pertinent to the algebras we just described.

\smallskip

{\bf Product and unit:}  For completeness, let us first just recall how an algebra structure comes about when considering functions on some compact manifold (not necessarily having a group structure). Take, for example, an ordinary compact Lie group $G$. Consider the set of all (polynomial) functions on it, $\mathcal{A}(G)$.
Clearly, we can add the functions together, multiply them with complex numbers and multiply the functions among themselves, this endows the set $\mathcal{A}(G)$ with the structure of an (associative $\mathbb{C}$-)algebra. Among these functions, there is definitely a unit function (i.e. identically one on the whole $G$), which gives a special element in our algebra, unit element $1_{\mathcal{A}}$.

{\bf Coproduct:} The coproduct on the polynomial group algebra $\mathcal{A}(G)$ encodes the product on the Lie group $G$ in the following way.  A product on $G$ is given by a map
\begin{align*}
\mu:\, G\times G\rightarrow G, \quad  \mu(g_1,g_2) \colonequals g_1g_2 \quad g_1, g_2 \in G \,.
\end{align*}
On the algebra of functions $\mathcal{A}(G)$ ('reversing the arrows'), we can then define the following map using $\mu$. One goes from functions on $G$, $\mathcal{A}(G)$, to functions on $G \times G$ by precomposition with $\mu$:
\begin{align*}
    \Delta:\,\,\mathcal{A}(G)\rightarrow \mathcal{A}(G \times G), \quad \Delta \colonequals \underline{\phantom{a}} \circ \mu\,,
\end{align*}
that is, simply,
\begin{align*}
    \Delta(F)(g_1,g_2) \colonequals F(g_1g_2), \quad F\in\mathcal{A}(G), \quad g_1, g_2 \in G\,.
\end{align*}
Notice that, for a compact Lie group $G$, Peter-Weyl theorem implies that $\mathcal{A}(G\times G) \simeq \mathcal{A}(G) \otimes \mathcal{A}(G)$.\footnote{Here we need it only for polynomial functions. A bit more generally, Peter-Weyl theorem says that (square-integrable) functions on the group are (in the completion of) the linear span of the matrix coefficients of finite-dimensional representations.
The isomorphism in the text follows from the rules of matrix multiplication.} Thus, we have obtained a map of (unital) algebras $\Delta:\,\, \mathcal{A}(G) \to \mathcal{A}(G) \otimes \mathcal{A}(G)$, which is our coproduct for $\mathcal{A}(G)$.  The coproduct
for $\mathcal{A}^{\mathbb{C}}_{\hq}$, $\mathcal{A}_{\hq,t}^{\mathbb{C}}$ and $\widetilde{\mathcal{A}}_{\hq,s}^{\mathbb{C}}$ that we are about to describe
is just an analogue of this construction for the $\hq$-deformed groups.\footnote{For non-compact Lie groups, this simple description gets modified in two ways. For one, polynomials on a non-compact Lie group are not good enough set of functions to inform us of the properties of that group (e.g. their integrals with respect to the Haar measure would typically diverge). To fix this, one needs to look at some better-controlled classes of functions, e.g. those that have compact support or those that are square-integrable. Secondly, one needs to be more careful with completion in various places. E.g. formulas like $L^2(G\times G) \simeq L^2(G) \otimes L^2(G)$ would only hold after suitable completion of the tensor product on the right-hand side (turning it into a Hilbert space). Similar subtleties will also make their way to the quantum group level. Finally, not all the constituents of the Hopf algebra structure might be well-defined in the quantum group case, e.g. antipode is typically only defined on a *-strongly-dense subspace of the algebra. Making sense of these facts requires an appropriate generalization of the notion of the Hopf algebra structure:
depending on operators one is interested in, such generalizations come in a variety of flavours, most interesting of them being (the two variants of) the C*-algebraic and the von Neumann-algebraic quantum groups \cite{Kustermans-Vaes2003}.}.

In the above compact Lie group case, the elements of the matrix actually represented a 'good' set of functions, whose products span
the entire algebra $\mathcal{A}(G)$
(again by Peter-Weyl theorem). Practically, this means that it is enough to define $\Delta$ only on the matrix entries of the group element $g\in G$. Thus, it is no surprise that, say, for the Hopf algebra $\mathcal{A}^{\mathbb{C}}$, the coproduct rule from above is
equivalent to
\begin{equation}\label{eq:coproduct_A}
        \Delta \begin{pmatrix}
            \alpha  & \beta \\
            \gamma & \delta
        \end{pmatrix} \colonequals \begin{pmatrix}
            \alpha  & \beta \\
            \gamma & \delta
        \end{pmatrix} \otimes \begin{pmatrix}
            \alpha  & \beta \\
            \gamma & \delta
        \end{pmatrix}\,,
    \end{equation}
and its *-conjugate.\footnote{This equality should be read entrywise, i.e. as four equalities of the matrix entries.}
For example, $\alpha(g_1g_2)=\alpha(g_1)\alpha(g_2)+\beta(g_1)\gamma(g_2)$, when we view the matrix entries as functions on the group. The coassociativity \eqref{eq:coassoc}
of the coproduct in a Hopf algebra that we postulated in subsection \ref{subsec:U_hopf_alg}, here is just a statement about the product of elements in the underlying group $G^{\mathbb{C}}$, which is associative.

Our coproduct \eqref{eq:coproduct_A} is, in fact, already good enough, so that we can extend the same rule (and the *-conjugate of it) unchanged to the quantum group case, for each of the Hopf algebras $\mathcal{A}^{\mathbb{C}}_{\hq}$, $\mathcal{A}_{\hq,t}^{\mathbb{C}}$ and $\widetilde{\mathcal{A}}_{\hq,s}^{\mathbb{C}}$.

{\bf Counit:} The most natural way to think of a counit in the case of algebra of functions on the group is as evaluation map. I.e. if we think about the elements $\alpha,\beta,\gamma,\delta$ as functions on the group $G^{\mathbb{C}}$, we can ask what is their value at the identity. This is what is measured by the counit, an algebra map $\epsilon:\, \mathcal{A} \to \mathbb{C}$. In the case of the algebra of functions $\mathcal{A}^{\mathbb{C}}$
it reads as
\begin{equation}
    \epsilon \begin{pmatrix}
            \alpha  & \beta \\
            \gamma & \delta
        \end{pmatrix} \colonequals \begin{pmatrix}
          1  & 0 \\
          0  & 1
        \end{pmatrix}\,.
        \end{equation}

\smallskip
Similarly to the coproduct, it passes unchanged to the quantum group case, for each of the Hopf algebras $\mathcal{A}^{\mathbb{C}}_{\hq}$, $\mathcal{A}_{\hq,t}^{\mathbb{C}}$ and $\widetilde{\mathcal{A}}_{\hq,s}^{\mathbb{C}}$.
Properties of counit \eqref{eq:counital}
here just ensure that it satisifes what we expect from evaluation of a function on the Lie group $G$ at the identity element. In particular, it should 'remember' that $e\times g = g = g \times e$ for any group element $g\in G$.

{\bf Antipode:} As before, antipode is the last bit of structure that we need for specifying our Hopf algebras. In the case of algebra of functions on the group, antipode should be thought of as 'remembering' the inverse operation on the group. That is, in the classical example of a compact Lie group $G$, $S(F)(g) \colonequals F(g^{-1})$, for $F\in \mathcal{A}(G)$, $g\in G$. In the $\hq$-deformed case, antipode can be obtained as an inversion of a two-by-two matrix of the quantum group generators.\footnote{I.e. a matrix of the defining corepresentation.} Explicitly, it reads:
\begin{equation}
    S \begin{pmatrix}
            \alpha  & \beta \\
            \gamma & \delta
        \end{pmatrix} = \begin{pmatrix}
          \delta  & -\hq^{-1}\beta \\
          -\hq\gamma  & \alpha
        \end{pmatrix}\,,
        \end{equation}
which, in combination with the properties of the conjugation stated around the equation \eqref{eq:star-properties}, define it for all our quantum groups $\mathcal{A}^{\mathbb{C}}_{\hq}$, $\mathcal{A}_{\hq,t}^{\mathbb{C}}$, $\widetilde{\mathcal{A}}_{\hq,s}^{\mathbb{C}}$.

{\bf All together:} As before, in the case of $\mathcal{U}$ Hopf algebras in subsection \eqref{subsec:U_hopf_alg}, all these structures, satisfying their compatibility conditions, form the corresponding Hopf algebras $\mathcal{A}(G)$, $\mathcal{A}^{\mathbb{C}}_{\hq}$, $\mathcal{A}_{\hq,t}^{\mathbb{C}}$, $\widetilde{\mathcal{A}}_{\hq,s}^{\mathbb{C}}$. Arguably, working with these objects in full detail is a living hell, but in any case fortunately we only need a small set of additional information about the relation between the $\hq$-deformed universal enveloping algebra and the $\hq$-deformed algebra of functions on the group. We'll explain these now.

\subsection{Pairing of the Hopf algebras}
\label{subsec:pairing}

Consider the Lie algebra as vector fields on the Lie group where they act as first-order differential operators in the directions determined by the elements of the Lie algebra. We can consider situations when the Lie algebra acts on the group from the left or from the right, which will give rise to right-invariant and left-invariant vector fields on the group. The non-degenerate bilinear pairing of Hopf algebras provides the generalization of these
interrelations to the $\hq$-deformed case.

In the classical case the usual formulas for the left- or right-invariant vector fields on the Lie group $G$ are\footnote{Recall that right shifts give a left-invariant vector field, and left shifts -- a right-invariant one.}
\begin{equation}\label{eq:Lie_grp_shifts}
\begin{split}
    &\text{Right shifts:}\ (X.f)(g) \colonequals \partial_t f(ge^{t X})|_{t=0},\ \  \\
     &\text{Left shifts:}\ (f.X)(g) \colonequals \partial_t f(e^{t X}g)|_{t=0}\,,
\end{split}
\end{equation}
where $X \in \mathfrak{g}$ denotes the corresponding element of the Lie algebra.
Somewhat confusingly, in group theory one calls the first action above {\it left action} and the second as {\it right action}. This is due to the way how two such actions compose, namely we have:
\begin{align}
    X_1.(X_2.f) = (X_1X_2).f, \quad (f.X_1).X_2 = f.(X_1X_2)\,.
\end{align}

 Let us introduce the notation $\langle X,f\rangle$ for a {\it non-degenerate bilinear pairing}\footnote{Non-degeneracy of a bilinear form means that the only element able to annihilate all possible elements put into the other entry of the form, is $0$.} of functions on $G$ and elements of its Lie algebra $\mathfrak{g}$:
\begin{align}
    \langle , \rangle\  :\  \mathfrak{g} \times \mathcal{A}(G) \to \mathbb{C}\,,
\end{align}
by which here we just mean evaluation of the corresponding vector field action at the identity of the group:
\begin{align}\label{eq:pairing-eval}
    \langle X, f \rangle \colonequals (X.f)(e) = \partial_t f(e^{t X})|_{t=0}\,.
\end{align}
Notice that now we can suggestively rewrite the classical equations \eqref{eq:Lie_grp_shifts} as
\begin{equation}\label{eq:Lie_grp_pairing}
\begin{split}
    &X.f = (id \otimes X) \circ \Delta(f),\ \  \\
    &f.X = (X \otimes id) \circ \Delta(f)\,,
\end{split}
\end{equation}
where the pairing is used implicitly, to contract element $X$ with one of the legs of the coproduct. This is the form we will generalize to the quantum group case in a moment.

Notice that in the classical case, in particular, for $\mathfrak{g} = \mathfrak{g}^{\mathbb{C}} \equiv \mathfrak{s}\mathfrak{l}(2, \mathbb{C})$, the definition \eqref{eq:pairing-eval} prescribes that,
considering
the left-invariant actions generated by
\begin{equation}
    L_0=
    \frac{1}{2}\begin{pmatrix}
            1  & 0 \\
            0 & -1
        \end{pmatrix},\
    B =
    \begin{pmatrix}
            0 & 1 \\
            0 & 0
        \end{pmatrix},\
    C =
    \begin{pmatrix}
            0 & 0 \\
            1 & 0
        \end{pmatrix},\
\end{equation}
the corresponding bilinear pairing is
   \begin{equation}\begin{split}\label{eq:sl2C-gens-matrix}
        \langle L_0 , \begin{pmatrix}
            \alpha  & \beta \\
             \gamma & \delta
        \end{pmatrix}  \rangle = \frac{1}{2}\begin{pmatrix}
            1  &  0 \\
           0 &  -1
        \end{pmatrix} ,\ \ \langle B , \begin{pmatrix}
            \alpha  & \beta \\
            \gamma & \delta
        \end{pmatrix}  \rangle = \begin{pmatrix}
             0 &  1 \\
           0 &  0
        \end{pmatrix}, \ \
        \langle C , \begin{pmatrix}
            \alpha  & \beta \\
            \gamma & \delta
        \end{pmatrix}  \rangle = \begin{pmatrix}
             0 & 0 \\
           1 &  0
         \end{pmatrix}\,,
    \end{split}\end{equation}
where the
matrix equalities
should be read entrywise.    

Recall that the algebra of polynomials on our Lie group $G^{\mathbb{C}}$, $\mathcal{A}^{\mathbb{C}}$, is spanned by linear combinations of products
of generators $\alpha,\beta,\gamma,\delta$ (viewed as functions on the group). Thus, one can start suspecting that it might be enough to define the pairing between the generators and just these basic functions. In fact, it {\it is} enough: the pairing extends to the full polynomial algebra,
once one imposes certain natural properties.
We will discuss them shortly.

To cook up something similar to \eqref{eq:sl2C-gens-matrix} in the case of a $\hq$-deformed universal enveloping algebra $\mathcal{U}_{\hq}^{\mathbb{C}}$ is not hard: recall that, to leading order in $1-\hq$, $A\sim{\hq}^{L_0}$. One can then define the non-degenerate bilinear form $\langle , \rangle:\,\, \mathcal{U}_{\hq}^{\mathbb{C}} \times \mathcal{A}_{\hq}^{\mathbb{C}} \to \mathbb{C}$
that, on generators, reads:
   \begin{equation}\label{eq:pairing-Uq-Aq}
   \begin{split}
        \langle A , \begin{pmatrix}
            \alpha  & \beta \\
            \gamma & \delta
        \end{pmatrix}  \rangle = \begin{pmatrix}
           {\hq}^{1 \over 2}  &  0 \\
           0 & {\hq}^{-{1 \over 2}}
        \end{pmatrix} ,\qquad \langle B , \begin{pmatrix}
            \alpha  & \beta \\
            \gamma & \delta
        \end{pmatrix}  \rangle = \begin{pmatrix}
             0 &  1 \\
           0 &  0
        \end{pmatrix} \\
        \langle C , \begin{pmatrix}
            \alpha  & \beta \\
            \gamma & \delta
        \end{pmatrix}  \rangle = \begin{pmatrix}
             0 & 0 \\
           1 &  0
        \end{pmatrix},\qquad \langle D , \begin{pmatrix}
            \alpha  & \beta \\
            \gamma & \delta
        \end{pmatrix}  \rangle = \begin{pmatrix}
           {\hq}^{-{1 \over 2}}  &  0 \\
           0 & {\hq}^{{1 \over 2}}
        \end{pmatrix}\,,
    \end{split}\end{equation}
and check that it satisfies all the needed properties, detailed in equation \eqref{eq:Consistency of Pairing}.
To reiterate, these formulas are to be understood as the $\hq$-analogue of taking a function on the (compact Lie) group manifold (in this case, one of the matrix element functions generating $\mathcal{A}(G)$) and a generator of the algebra $\mathcal{U}\mathfrak{g}$, viewed as a vector field $X$ acting on the group,
and then evaluating $X\cdot f|_e$.    
   
Now let us see what is actually required from the non-degenerate bilinear pairing to extend it to the full polynomial algebra.
In our classical example above, we have described an action of the Lie algebra (of left-invariant vector fields) $\mathfrak{g}$ only at the origin (where we can identify its elements with vectors in the tangent space), and only on a specific set of generators of the group $\alpha$, $\beta$, $\gamma$, $\delta$. Classically, what we think about next is to extend this action in two 'directions': the first is to extend it to any 'polynomial function' on the quantum group:
\begin{align*}
  f \mapsto \partial_t f(e^{t T})|_{t=0}, \quad f\in \mathcal{A}(G) \,,
\end{align*}
and to any number of derivatives in different directions.
The second is to extend it so that we can evaluate the expression at any point on the quantum group:
\begin{align*}
f \mapsto \partial_t f(g \times e^{t T})|_{t=0}, \quad f\in \mathcal{A}(G),\quad g\in G\,.
\end{align*}

Analysis shows that the first step is performed by requiring the properties\footnote{As is common in the literature, we are using the same notation for coproduct, unit etc in both ${\cal U}_{\hq}$ and ${\cal A}_{\hq}$.}
\begin{equation}
    \begin{split}\label{eq:Consistency of Pairing}
        \langle \Delta (u), a \otimes b \rangle = \langle u , a  b \rangle,\quad \langle u \otimes v , \Delta(a)  \rangle = \langle u v , a \rangle \\
        \langle 1, a \rangle = \epsilon(a),\quad \langle u, 1 \rangle = \epsilon(u), \quad \langle S(u) , a \rangle = \langle u, S(a) \rangle
    \end{split}\,.
\end{equation}
Here, the first property in the first line classically is just the Leibniz rule of differentiation, and the second property in the first line amounts to the usual rule of extending action of Lie algebra elements (understood as, say, left-invariant vector field) on a manifold to the action of polynomials in the vector fields (the latter generate the full universal enveloping algebra $\mathfrak{g}$). One can check that constructing a pairing with these properties is a 'minimal' way to ensure such an interpretation.

In order to extend to an 'action of Lie vector fields on functions on the quantum group', we define the left\footnote{We note here that Olshanetsky and Rogov in \cite{Olshanetsky:2001} call it a right action, although their action actually satisfies the composition rule for a left action. To avoid confusion, in our exposition we always only use left actions for the unstarred generators.} action\footnote{As in the classical case \eqref{eq:Lie_grp_pairing}, following the usual notational convention we leave implicit the contraction of elements of $u\in\cal U$ and $a\in\cal A$ via bilinear pairing $\langle,\rangle$. I.e. any product of elements from these two different algebras encountered in the formulas should be understood as a number obtained by evaluating $\langle u,a\rangle$.}
    \begin{equation}\label{eq:def left action quantum group}
        u \cdot a  = (id\otimes u)\circ \Delta(a)
    \end{equation}
    where $a \in {\cal A}_{\hq}, u \in {\cal U}_{\hq}$.
    From the properties of pairing, \eqref{eq:Consistency of Pairing}, one can see that this satisfies\footnote{We remind that Sweedler's notation used in these formulas was defined in \eqref{eq:Sweedler}. }
    \begin{equation}
    \begin{split}
       \mbox{Leibniz rule : } &\qquad u \cdot (ab) = \sum_{(u)} ( u_{(1)} \cdot a)  (u_{(2)} \cdot b) \\
       \mbox{Product rule : } &\qquad (uv) \cdot a = u \cdot (v \cdot a)\,.\\
    \end{split}
    \end{equation}
    For our example one can work out
    \begin{equation}\label{eq:AlgActGrp}
    \begin{split}
        &A \cdot \begin{pmatrix}
            \alpha  & \beta \\
            \gamma & \delta
        \end{pmatrix}  = \begin{pmatrix}
         {\hq}^{1 \over 2}  \alpha  &{\hq}^{-{1 \over 2}} \beta \\
         {\hq}^{1 \over 2}  \gamma &{\hq}^{-{1 \over 2}}  \delta
        \end{pmatrix} \qquad    B \cdot   \begin{pmatrix}
            \alpha  & \beta \\
            \gamma & \delta
        \end{pmatrix} = \begin{pmatrix}
          0 &  \alpha  \\
          0 & \gamma
        \end{pmatrix} \\
        & C \cdot \begin{pmatrix}
            \alpha  & \beta \\
            \gamma & \delta
        \end{pmatrix}   = \begin{pmatrix}
         \beta &  0 \\
         \delta & 0
        \end{pmatrix} \qquad  \qquad  D \cdot    \begin{pmatrix}
            \alpha  & \beta \\
            \gamma & \delta
        \end{pmatrix}   = \begin{pmatrix}
         {\hq}^{-{1 \over 2}}  \alpha  &{\hq}^{{1 \over 2}} \beta \\
         {\hq}^{-{1 \over 2}}  \gamma &{\hq}^{{1 \over 2}}  \delta
        \end{pmatrix}\,.
    \end{split}
    \end{equation}
    Action of generators on the monomials of $\alpha, \beta, \gamma, \delta$ can be derived via Leibniz rule. In the case of $\widetilde{\mathcal{A}}_{\hq,s=-(1-\hq^2)}^{\mathbb{C}} \equiv \widetilde{\mathcal{A}}_{\hq}^{\mathbb{C}}$ paired up with $\mathcal{U}_{\hq}^{\mathbb{C}}$, it is also required that the generators $A,B,C,D$ have no action on the starred generators $\alpha^*,\beta^*$, $\gamma^*$ and $\delta^*$.
   
\subsection{Modules over the Hopf algebras}
\label{sec:Modules of U}

Classically a Lie algebra of symmetries $\mathfrak{g}$ can act by flows on various spaces (say, smooth manifolds) $\mathcal{X}$ carrying the corresponding Lie group action. For our purposes, even before the quantum deformation, we  formulate this action by specifying how the algebra, realized via vector fields, acts on functions on the space. By building polynomials in these vector fields, we can then always extend this to an $\mathcal{U}\mathfrak{g}$-action.\footnote{Because of the universal property of $\mathcal{U}\mathfrak{g}$.}

When we $\hq$-deform this picture, we simultaneously $\hq$-deform:
\begin{enumerate}
    \item the algebra $\mathcal{U}$  (which is classically equal to $\mathcal{U}\mathfrak{g}$) that acts on the space
    \item the space $\mathcal{B}$ of "functions" on which it acts (which is classically equal to an appropriate algebra of functions on the space $\mathcal{X}$)
    \item the action of (1) on (2).
\end{enumerate}
So there is quite a large measure of arbitrariness in each of these steps, and in particular in (2) and (3).\footnote{If we put a restriction that the deformation we make is a formal power series in the deformation parameter $\lambda$, $\hq=e^{-\lambda/2}$, and if the Lie algebra $\mathfrak{g}$ we start with is semisimple (as it is in our example $\mathfrak{g} = \mathfrak{s}\mathfrak{l}(2, \mathbb{C})$), then cohomological considerations show that all such deformations of $\mathcal{U}\mathfrak{g}$ are isomorphic as algebras, i.e. a large fraction of the freedom in the first item is fixed \cite{Drinfeld1990}, \cite{Gerstenhaber1964}, \cite{Drinfeld1990b}. }

There are several approaches to describe the $\hq$-deformation of a space $\mathcal{X}$ carrying Lie group action. We will take the one where the $\hq$-deformed space is defined by its algebra of coordinate functions\footnote{In other words, a vector space of functions on $\mathcal{X}$, with a law of multiplication.} $\mathcal{B}_{\hq}$
carrying a (left) action of the Hopf algebra $\mathcal{U}_{\hq}$\footnote{Other closely related natural approach is to take the same algebra $\mathcal{B}_{\hq}$, but now carrying a {\it coaction} of the deformed algebra of the coordinate functions (i.e. $\mathcal{A}_{\hq}$ with various sub/superscripts, in our notation), satisfying certain constraints making it to resemble a classical homogeneous space of a Lie group.} (or of any of its cousins with other sub/superscripts that we discussed in section \ref{subsec:U_hopf_alg}\footnote{All of these being $\hq$-deformations of the corresponding universal enveloping algebras $\mathcal{U}$ or $\mathcal{U}^{\mathbb{C}}$ of the classical case.}).
This action of $\mathcal{U}_{\hq}$ gives the algebra $\mathcal{B}_{\hq}$ a structure of a $\mathcal{U}_{\hq}${\it-module}, i.e. a vector space with an action of $\mathcal{U}_{\hq}$.

The way that the $\hq$-deformed algebra $\mathcal{U}_{\hq}$ acts on the '$\hq$-deformed space $\mathcal{X}_{\hq}$' is encoded
in the laws $b_\alpha.u_i$, describing how the algebra generators $u_i\in {\cal U}_{\hq}$ act on the algebra generators $b_\alpha\in \mathcal{B}_{\hq}$. This action needs to satisfy the axioms of associativity and unit
\begin{align}
    (uv).b = u.(v.b), \quad \text{1}.b = b, \quad u,v\in {\cal U}_{\hq}, b\in \mathcal{B}_{\hq}\,.
\end{align}
Moreover, we want the action to respect multiplication in $\mathcal{B}_{\hq}$. We can abstractly write this condition as:
\begin{align}\label{eq:Leibniz-q-homog}
    u.m(b_1\otimes b_2) = m\left(\Delta(u).(b_1\otimes b_2)\right), \quad b_1, b_2\in \mathcal{B}_{\hq}, u \in \mathcal{U}_{\hq}\,,
\end{align}
where $m:\,\, \mathcal{B}_{\hq} \otimes \mathcal{B}_{\hq} \to \mathcal{B}_{\hq}$, $m(b_1\otimes b_2) = b_1b_2$ is multiplication map of the space $\mathcal{B}_{\hq}$. This indeed looks as a covariance condition prescribing how to interchange operation $m$ and action by an element $u\in \mathcal{U}_{\hq}$. For the classical case, this is just the Leibniz rule. The resulting mix, i.e. an algebra $\mathcal{B}_{\hq}$ carrying such an action of the Hopf algebra $\mathcal{U}_{\hq}$, is called $\mathcal{U}_{\hq}${\it-module algebra}, or {\it quantum space}, in the mathematical literature.\cite{Kl-Schm}

Let us again illustrate how this reduces to a classical example in the case when
$\mathcal{U}$ is a universal enveloping algebra $\mathcal{U}\mathfrak{g}$ of the Lie algebra of a compact Lie group $G$ and $\mathcal{B}$ an algebra of (polynomial) functions on the homogeneous space $\mathcal{X}$ for $G$. Let us define:
\begin{align}
    (Y.f)(x) \colonequals \frac{d}{dt}f(e^{-tY}x)|_{t=0}\,, \quad Y\in\mathfrak{g}\subset\mathcal{U}, \quad f\in \mathcal{B}, \quad x\in \mathcal{X}\,.
\end{align}
Now it is easy to see that \eqref{eq:Leibniz-q-homog} indeed reduces to the Leibniz rule, after we use the coproduct \eqref{eq:coproduct_Lie_alg}.

\section{It's Lobachevsky spaces all over the place}
\label{sec:Lobachevsky}

Different variants of non-commutative $AdS_3$, in the setting of $\hq$-homogeneous spaces, were constructed in \cite{Olshanetsky:1993sw}, \cite{Olshanetsky:2001}.\footnote{A complete classification of the possibilities is not known to us, but presumably can be obtained by analysis of the possible coproduct twists in the corresponding quantum groups, and $\hq$-coset reduction thereof.}
All of them go over to the usual geometric $AdS_3$ in the limit $\hq\rightarrow 1$. This section is an expanded version of the constructions in these papers (with some correction). Importantly, the space that they construct is not exactly what we need since these works actually describe deformations of a Euclidean $AdS_3$, i.e. the Lobachevsky space $\mathbb{H}^3$, whereas we want to reduce a Minkowskian $AdS_3$. Hence we will need to carry out certain analytic continuations (encoded by the spectral properties of corresponding operators) and deviate from it at some point. We will highlight this junction when we arrive at it.

\subsection{The classical Lobachevsky space}
\label{subsec:L3}

The construction begins with recalling that the three-dimensional Lobachevsky space $\mathbb{H}^3$ can be described as a coset
$\mathcal{X}^{\mathbb{C}} \colonequals SO(3)\backslash SO_e(1,3) \simeq SU(2)\backslash SL(2,{\mathbb C})$. That is, denoting by $g$ an element of $SL(2,{\mathbb C})$ and by $h$ an element of its $SU(2)$ subgroup, our space is
\begin{equation}
    \mathcal{X}^{\mathbb{C}} = \{ g\in SL(2,\mathbb{C}) \}/ \{g\sim h g,\ h\in SU(2)\} \simeq {\mathbb{H}^3}\,.
\end{equation}

A convenient way of parametrizing this coset is to consider the combination of $SL(2,\mathbb{C})$ group elements, invariant under the action of $SU(2)$:
\begin{equation}
    x \colonequals g^\dagger g\,.
\end{equation}
Then we can use the matrix entries of $x$ (or rather their combinations) as coordinates $(H, z, z^*)$ on the
space $\mathcal{X}^{\mathbb{C}}$:
\begin{equation}\label{eq:LubPlan1}
    x=\begin{pmatrix}
    x_{11} & x_{12} \\
    x_{21} & x_{22}
    \end{pmatrix} =
    \begin{pmatrix}
    \alpha^* \alpha + \gamma^* \gamma & \alpha^* \beta + \gamma^* \delta \\
    \beta^* \alpha + \delta^* \gamma & \beta^* \beta + \delta^* \delta
    \end{pmatrix} \equalscolon \begin{pmatrix}
    H & Hz \\
    z^* H & z^* H z + H^{-1}
    \end{pmatrix}\,.
\end{equation}

 If we also recall the reality conditions for the group elements $\overline{\alpha} = \alpha^*$ etc., then equation \eqref{eq:LubPlan1} can be considered as a definition\footnote{To be precise, the fact that the lower right corner elements of the last two matrices in equation \eqref{eq:LubPlan1} are equal is not a definition, but rather consistency condition coming from the determinant one condition in $SL(2, \mathbb{C})$. Thus, $H$, $z$ and $z^*$ are well defined by equating the remaining three matrix entries.}
 of $H, z, z^*$,
 that obey the following reality conditions:\footnote{We denote the complex conjugation by $\overline{\phantom{z} }\,$.}
 \begin{align}
     \overline{H} = H, \quad \overline{z} = z^*\,.
 \end{align}
 In other words,
 \eqref{eq:LubPlan1} tells us
 that the diagonal elements are real (and non-negative), that the off-diagonal elements are complex conjugate of each other, and that $x_{11}x_{22}=x_{12}x_{21}+1$ (because $\alpha\delta-\beta\gamma=1$). If we define
 \begin{align}
 x_{11} \equalscolon x_0+x_1, \quad x_{22} \equalscolon x_0-x_1, \quad x_{12} \equalscolon x_2+ix_3\,,
 \end{align}
 we obtain the familiar hypersurface $x_0^2-x_1^2-x_2^2-x_3^2=1$, $x_0>0$ (i.e. upper sheet of the two-sheeted hyperboloid) in $\mathbb{R}^4$.

We can also think about this space more geometrically. The standard hyperbolic metric on $\mathbb{H}^3$ is  
\begin{align}
ds^2_{\mathbb{H}^3} = {dH^2 \over H^2} + H^2 dz dz^* \,.
\end{align}
So these are the coordinates on the Euclidean disk used before.

Upon Wick rotation, this will also describe the Poincar{\'e} patch of Minkowskian $AdS_3$. Namely, if, as we just did, we take $z$ to be complex and $z^* = \bar z$, then this is Euclidean $AdS_3$ ($\equiv\mathbb{H}^3$), whereas taking $z,z^*$ to be real and independent gives Lorentzian  $AdS_3$.\footnote{Here we consider both Euclidean and Lorentzian manifolds as real slices of the ambient complexified hyperbolic space. The latter can be realized either as a coset manifold $\left(SL(2, \mathbb{C}) \times SL(2, \mathbb{C})\right)/SL(2, \mathbb{C})$,
or by an embedding into $\mathbb{C}^4$.}

The action of $G^{\mathbb{C}} = SL(2,{\mathbb C})$ on $(H,z,z^*)$  is obtained from the action \eqref{eq:sl2c_adjoint_action} of $w\in G^{\mathbb{C}}$
\begin{equation}
    \pi(w):\,\,  x \mapsto w^\dagger x w, \quad x \in \mathcal{X}^{\mathbb{C}}, \quad w\in G^{\mathbb{C}}\,.
\end{equation}
As we saw there, on the level of the Lie algebra $\mathfrak{g}^{\mathbb{C}} = Lie \left( G^{\mathbb{C}}\right) = \mathfrak{s}\mathfrak{l}(2,\mathbb{C})$ this gives an action
\begin{equation}\label{eq:sl2C-action}
    d\pi(Y):\,\,  x \mapsto x\cdot Y + Y^*\cdot x, \quad x \in \mathcal{X}^{\mathbb{C}}, \quad Y \in \mathfrak{g}^{\mathbb{C}}\,.
\end{equation}
From here,
we get to an action of $\mathfrak{g}^{\mathbb{C}}$ (and its universal enveloping algebra) on the functions on our hyperbolic space $\mathbb{H}^3$. Recalling that right shift of the group element leads to the left action, we obtain:
\begin{equation}\label{eq:sl2C-geom-action}
    d\pi^{\bullet}(Y):\,\,  f \mapsto Y \cdot f + f \cdot Y^* \equalscolon Y \vartriangleright f, \quad f \in \mathcal{B}^{\mathbb{C}}, \quad Y \in \mathfrak{g}^{\mathbb{C}}\,.
\end{equation}
Of course, by building polynomials in $Y$, we can further uniquely extend this to an action of the universal enveloping algebra $\mathcal{U}\mathfrak{g}^{\mathbb{C}}$ on the space $\mathcal{B}^{\mathbb{C}}$. We denote the geometric action (which combines 'left' action of $X$ and 'right' action of $X^*$) as $\vartriangleright$.
The last definition might seem like nit-picking, but it is the one generalized in non-commutative geometry.

Note that for computational convenience we will still split the full 'geometric' action $\vartriangleright$ into two pieces as in \eqref{eq:sl2C-geom-action}, formally considering independent action of $Y$ (e.g. of the 'unstarred' generators $L_0, B,C$) on $f\in \mathcal{B}^{\mathbb{C}}$ on the left, and the action of $Y^*$ (in this case, of $L_0^*, B^*, C^*$) on $f$ on the right.  This makes the computation simpler because we need to deal with only one coproduct at a time.
But in the end, in order to get an actual, 'geometric' action of the $\mathfrak{s}\mathfrak{l}(2,\mathbb{C})$ generators on $\mathbb{H}^3$,  we will always add up the action of these generators from the left with the action of their corresponding conjugate counterparts from the right.

It is straightforward to work out the explicit action of generators on the coordinate functions on $\mathbb{H}^3$. We give them and the corresponding vector fields below:
\begin{equation}\label{eq:Classical action og generators on L3}
\begin{split}
    L_0 \cdot  H  = {H \over 2} ,\ L_0 \cdot  z =-z,\ L_0 \cdot z^*=0 \qquad \implies & \quad  L_0={1 \over 2 } H\partial_H-z\partial_z\\
    B \cdot H = 0,\ B \cdot  z = 1,\ B \cdot  z^* = 0\hspace{10mm} \implies & \quad B = \partial_z \\
    C \cdot H = H z,\ C \cdot  z  = -z^2,\ C \cdot  z^*  = H^{-2} \hspace{6mm} \implies & \quad C = H z \partial_H - z^2 \partial_z+ H^{-2} \partial_{z^*} \\
         H \cdot L_0^* = {H \over 2},\  z \cdot L_0^* = 0, \ z^* \cdot L_0^* = - z^* \hspace{6mm} \implies & \quad L_0^* ={1 \over 2 } H \partial_H- z^* \partial_{z^*} \\
    H \cdot B^* = 0,\  z\cdot B^* = 0,\ z^* \cdot B^* =  z^* \hspace{6mm} \implies & \quad  B^* = \partial_{ z^*}  \\
    H \cdot C^* = H z^*, \  z\cdot C^* = H^{-2},\  z^* \cdot C^* =  -{z^*}^2 \implies &  \quad   C^* = H z^*\partial_H + H^{-2} \partial_z - {z^*}^2  \partial_{z^*}
\end{split}
\end{equation}

In the case of Euclidean $AdS_3$ the reality condition reads $\overline{z} = z^*$, hence the symmetry generators are $c L_0+c^* L_0^*$, and so on, for some complex
phase $c$: for instance, for $c= i$ the generator $i (L_0 - L_0^*)$  corresponds to rotations in the angular directions in the complex plane $z$. In the case of Lorentzian $AdS_3$, we take $z$ and $z^*$ to be independent and real, hence the generators act with independent real coefficients. If one wants to compare to the generators in equation \eqref{eq:SL2Gen} (which are formulas for $AdS_2$), one needs to take $z$ and $z^*$ to be independent, and then match the generators that commute with $\partial_{z^*}$. This singles out $L_0$, $B$ and $C$ which reduce to \eqref{eq:SL2Gen} (with $\partial_{z^*}=i\pi_{z^*}$ as it should be).

The center of the algebra $\mathcal{U}\mathfrak{g}^{\mathbb{C}}$ is spanned by two Casimir elements $\Omega$, $\Omega^*$. One of them has the form
\begin{equation}
\Omega =
{1 \over 4} + L_0^2 + {BC + CB \over 2 }
\end{equation}
(which is here shifted by a constant with respect to \eqref{eq:CasC}). The other Casimir element $\Omega^*$ has an identical expression, but in terms of the conjugated counterparts of the Lie algebra generators $L_0, B, C$; we will not need it in the subsequent discussion. The two Casimirs act identically on functions on $AdS_3$.

As an element of the center of $\mathcal{U}\mathfrak{g}$, $\Omega$ gives rise to the second-order $G^{\mathbb{C}}$-invariant differential operator on $\mathbb{H}^3$, the Laplacian. The wavefunction $\Phi_\Lambda$ of a particle moving on this space\footnote{Using the standard (quasi-invariant) measure on the homogeneous space $\mathbb{H}^3$, we can define a $G^{\mathbb{C}}$-invariant scalar product of functions on this space, and then use it to define the Hilbert space of wavefunctions $L^2(\mathbb{H}^3)$.} satisfies
\begin{equation}
    \Omega \cdot  \Phi_\Lambda(H,z,\bar z) = - \Lambda^2 \Phi_\Lambda (H,z,\bar z)\,.
\end{equation}
We can reduce the problem by considering fixed eigenvalue solution under the generators $B$ and $B^*$
\begin{equation}
    \Phi_\Lambda (H,z,\bar z)  = e^{i \mu z + i \bar \mu \bar z }\, \Phi_{\Lambda,\mu,\bar \mu}(H)
\end{equation}
where $\mu$ and $\bar\mu$ are complex conjugate of each other in the Euclidean case and real, independent in the Minkowski case. If we further define  
\begin{align}
    \Phi_{\Lambda,\mu,\bar \mu}(H) \equalscolon H^{-1} \Psi_{\Lambda,\mu,\bar \mu}(\phi),\qquad H \equalscolon e^{-{\phi \over 2}}\,,
\end{align}
the Casimir equation  becomes
\begin{equation}
    \left( -  \partial_{\phi}^2 +  \mu \bar \mu e^{\phi}\right) \Psi_{\Lambda,\mu,\bar \mu}(\phi)= \Lambda^2 \Psi_{\Lambda,\mu,\bar \mu}(\phi)\,,
\end{equation}
which is the Liouville equation.

\subsection{The quantum Lobachevsky spaces ${\mathbb{H}}_{\hq}^3$, ${\mathbb{H}}_{\hq, \kappa}^3$}
\label{subsec:Particle moving on quantum group}

Next we proceed to construct a couple of variants of quantum Lobachevsky space, which we will call ${\mathbb{H}}_{\hq}^3$ (discussed in subsection \ref{sec:OR1}) and ${\mathbb{H}}_{\hq,\kappa}^3$ (discussed in subsection \ref{sec:OR2}). In both cases we define these non-commutative spaces as module algebras of the Hopf algebras $\mathcal{U}_{\hq}^{\mathbb{C}}, \mathcal{U}_{\hq,r,s}^{\mathbb{C}}$. First, we will define the module structure on the algebra generators, i.e. on coordinate functions on ${\mathbb{H}}_{\hq}^3$, ${\mathbb{H}}_{\hq,\kappa}^3$, and then will derive how they act on functions on these spaces.

\subsubsection{The quantum Lobachevsky space ${\mathbb{H}}_{\hq}^3$}\label{sec:OR1}

To get this variant of the $\hq$-Lobachevsky space, we can start with the quantum group $\widetilde{\mathcal{A}}^{\mathbb{C}}_{\hq}$ (defined in section \ref{subsec:A_hopf_alg}) and perform a $\hq$-analogue of coset reduction that we reviewed in subsection \ref{subsec:L3}. Our discussion here follows \cite{Olshanetsky:1993sw}, with some additional details and correction collected in appendix \ref{app:details_qLobachevsky}.

Our $\hq$-Lobachevsky space ${\mathbb{H}}_{\hq}^3$ will still be generated (as algebra) by the generators $H,z,z^*$ which we still define by the equation \eqref{eq:LubPlan1}, except that
on both sides we will now have non-commuting objects: the matrix in the center will contain the elements of the Hopf algebra $\widetilde{\mathcal{A}}^{\mathbb{C}}_{\hq}$  and the rightmost matrix will involve elements of our ${\mathbb{H}}_{\hq}^3$.  
These generators can be shown to satisfy the relations\footnote{The corresponding algebra of functions on this $\hq$-Lobachevsky plane is denoted $\mathcal{B}_{\hq}^{\mathbb{C}}$, according to our conventions.}
\begin{equation}\begin{split}
    H=H^*,\ Hz = \hq^2 z H  ,\quad z^* H = \hq^2 H z^* \\
    z z^* = \hq^2 z^* z + (\hq^2-1) (1-H^{-2})\,.
\end{split}    \label{eq:OR1 Lobachevsky commutations}
\end{equation}

Just as in classical case, discussed in subsection \ref{sec:Modules of U}, this algebra is moreover a module over $\mathcal{U}_{\hq}^{\mathbb{C}}$. To obtain the action of $A,B,C,D$ on the coordinate functions\footnote{The appropriate version of Poincar{\'e}-Birkhoff-Witt (PBW) theorem for the algebra $\mathcal{B}_{\hq}^{\mathbb{C}}$ says that linear combinations of ordered products of $z^*, H, z$ form a basis for the algebra, which we should interpret as algebra of polynomial functions on our $\hq$-deformed space.} $z^*, H, z$ , we need to compute objects of the form $A \cdot H=A \cdot (\alpha^*\alpha+\gamma^*\gamma)$ and check that the Leibniz rule is satisfied. We do so using the definition \eqref{eq:def left action quantum group} for the dot-action.
This gives the following action of generators on coordinates \cite{Olshanetsky:1993sw} \footnote{The details of this derivation are explained in appendix \ref{app:details_qLobachevsky}.}
\begin{equation}\begin{split}
     A \cdot H = \hq^{1 \over 2} H ,& \quad   A \cdot z = \hq^{-1} z, \quad  A \cdot z^*  = z^*  \\
     B \cdot H = 0 ,& \quad  B \cdot z =\hq^{-{1 \over 2}}  , \quad  B \cdot z^*  = 0  \\
     C \cdot H =  H z  ,& \quad  C \cdot z = -\hq^{1 \over 2} z^2 , \quad  C \cdot z^*  = \hq^{-{1 \over 2}} H^{-2} \,,
\end{split}\end{equation}
and the corresponding right action of starred generators on the coordinates.\footnote{Since the $*$-structures of $\mathcal{U}_{\hq}^{\mathbb{C}}$ and $\mathcal{B}_{\hq}^{\mathbb{C}}$ agree, we have that $f\cdot Y^* = (Y \cdot f)^*$ for $f\in\mathcal{B}_{\hq}^{\mathbb{C}}$ and $Y\in \mathcal{U}^{\mathbb{C}}_{\hq}$, e.g. for $Y = A, B, C, D$.} It is easy to see that in the $\hq \rightarrow 1$ limit this reproduces equation (\ref{eq:Classical action og generators on L3}).

To find the action on products of coordinates, we can define the arbitrary functions on ${\mathbb{H}}_{\hq}^3$ as a span of the ordered monomials
\begin{equation}\label{eq:H-monom-1}
    v^{m,r,n} \colonequals   (\hq^{1\over 2}z^*)^{m} H^r(\hq^{1\over 2}z)^{n}\,.
\end{equation}
With these definitions, one can verify that
the action of generators is (for more details see Appendix \ref{app:details_qLobachevsky}):
\begin{equation}\label{eq:OSAdS2}
    \begin{split}
        A \cdot v^{m,r,n} & = \hq^{{r \over 2}-n} v^{m,r,n}  \\
        D \cdot v^{m,r,n} & = \hq^{-{r \over 2}+n} v^{m,r,n}  \\
        B \cdot v^{m,r,n} & =  [n]_{\hq^2}{\hq^{r\over 2}}  v^{m,r,n-1} \\
        C \cdot v^{m,r,n} & =  \hq^{m+n+{r-2 \over 2}} [m]_{\hq^2} v^{m-1,r-2,n}  - \hq^{-{r\over 2}} [n-r]_{\hq^2}  v^{m,r,n+1} \,,
    \end{split}
\end{equation}
whereas the Casimir generator becomes
\begin{equation}\begin{split}
    \Omega \cdot v^{m,r,n}& \equiv\left(  { \hq^{-1}A^2 + \hq D^2-2 \over (\hq^{-1} - \hq)^2} + B C \right) \cdot v^{m,r,n } \\
    & = [m]_{\hq^2}[n]_{\hq^2}\hq^{m+r+n-2} v^{m-1, r-2, n-1} + [{r+1 \over 2}]^2_{\hq^2} v^{m,r,n}\,.
\end{split}
\label{eq:Casimir Quantum Group on Monomial}
\end{equation}
Once we have an action on monomials we can define it on a general function $f(z^*,H,z)$. At this stage this is still a function of non-commuting variables and we can think of the ordering $z^*-H-z$ as a normal ordering convention.

In order to compare with the results of section \ref{sec:hands-on}, it is more convenient
to use
the following automorphism of the algebra $\mathcal{U}_{\hq}^{\mathbb{C}}$:
\begin{equation}
A' \colonequals A, \ \ D' \colonequals D, \ \  B' \colonequals \sqrt{\hq} D B , \ \  C'\colonequals \sqrt{\hq} A C \,.
\end{equation}
Thus,
\begin{equation}
\label{eq:Action of Conjugated generators}\begin{split}
    A' \cdot v^{m,r,n} & = \hq^{{r \over 2}-n} v^{m,r,n}  \\
        D'  \cdot v^{m,r,n} & = \hq^{-{r \over 2}+n} v^{m,r,n}  \\
        B' \cdot v^{m,r,n} & = \hq^{n-{1 \over 2}} [n]_{\hq^2} v^{m,r,n-1} \\
        C'  \cdot v^{m,r,n}
        & = [m]_{\hq^2} \hq^{m+ r-{3 \over 2}} v^{m-1,r-2,n} - \hq^{-n - {1 \over 2}} [n-r]_{\hq^2}  v^{m,r,n+1}\,,\\
\end{split} \end{equation}
and similarly for the starred generators.
One can readily check the automorphism property (i.e. that the algebra stays the same):
\begin{equation}
    A'  B' = \hq  B'  A',\quad  A'  C' = \hq^{-1}  C'  A',\quad [ B' ,  C' ] = { (A')^2 - (D')^2\over \hq - \hq^{-1} }\,.
\end{equation}
Therefore the Casimir will still be as in equation (\ref{eq:Casimir Quantum Group on Monomial}), but in terms of generators $A', B',  C', D'$. However, the algebra automorphism we applied is not Hopf algebra automorphism, in particular the coproduct and antipode will change. Since we will not need them for our discussion, we do not write the actual new expressions.

We now define the action of generators on functions of the coordinates on $\mathbb{H}_{\hq}^3$. While we could continue working with non-commuting coordinates, it will be easier to shift to commuting ones, although this might seem like a sleight of hand. Basically, since we have the functions of the non-commuting variables normal-ordered, we can simply think about them as ordinary functions of commuting variables with a more complicated product (the $*$-product), by virtue of the so called symbol map of deformation quantization. In other words, suppose we will use $\tilde z,\tilde z^*,\tilde H$ for the commuting coordinates. Then a monomial built out of these new coordinates $ ( \hq^{1\over 2} \tilde z^*)^{m} \tilde H^r( \hq^{1\over 2} \tilde  z)^{n}$ is simply mapped to the monomial of the non-commuting variables $ ( \hq^{1\over 2}  z^*)^{m}  H^r( \hq^{1\over 2}   z)^{n}$, and all operations on the latter are pulled back to the former.
In particular, the $q$-deformed algebra is defined to have the same action as in \eqref{eq:OSAdS2} on the commuting variables.

We will now switch to functions $f(\tilde z^*,\tilde H,\tilde z)$ of these commuting coordinates and give the action of generators.  In terms of basic scaling operations that we defined in section \ref{sec:hands-on},
\begin{equation}\label{eq:Simple Operators on  H and z}
\begin{split}
& T \cdot f(\tilde z^*,\tilde H,\tilde z) \colonequals  f(\tilde z^*,\hq^{1 \over 2}\tilde H,\tilde z) \\
R \cdot f(\tilde z^*,\tilde H,\tilde z) \colonequals & f(\tilde z^*,\tilde H,\hq \tilde z),\quad  R^* \cdot f(\tilde z^*,\tilde H,\tilde z) \colonequals  f(\hq\tilde z^*,\tilde H,\tilde  z) \,,
\end{split}
\end{equation}
the generators are given by
\begin{equation}\label{eq:Action of generators on L3 : Abstract}
    \begin{split}
        \tilde A' & = T \cdot R^{-1} \\
        \tilde D' & = T^{-1} \cdot R \\
        \tilde B' & = { 1 - R^2 \over (1-\hq^2)\tilde z },\quad \tilde B^{*\prime} = { 1 - {R^*}^2 \over (1-\hq^2)\tilde z^*}  \\
        \tilde C' & = \tilde z {(T^{-2} - T^2 R^{-2})\over \hq^{-1} - \hq} + { \tilde H^{-2} \over \hq }\ \tilde B^{*\prime}T^2 \\
        \tilde \Omega'  & = {\hq^{-1} T^{-2} + \hq T^2 -2 \over  (\hq^{-1} -\hq)^2 } + \hq^{-1} \tilde H^{-2} T^2 \tilde B \tilde B^{*\prime}\,.
    \end{split}
\end{equation}
We note here that if we do a reduction on $\tilde z^*$ by diagonalizing the operator $\tilde B^{*\prime}$, this matches to the $a = 0,\, \tilde a =2$ choice in the hands-on model that we considered in section \ref{sec:hands-on}.

\subsubsection{The quantum Lobachevsky space ${\mathbb{H}}_{\hq, \kappa}^3$}\label{sec:OR2}

In this subsection we will follow \cite{Olshanetsky:2001}. The parameter $\kappa$ that we use here is real, and $0<\hq <1$ as before.  Just like $\mathbb{H}_{\hq}^3$ in previous subsection, the other version of $\hq$-Lobachevsky space, ${\mathbb{H}}_{\hq, \kappa}^3$,\footnote{The corresponding algebra of functions on this $\hq$-Lobachevsky plane is denoted $\mathcal{B}_{\hq, \kappa}^{\mathbb{C}}$, according to our conventions.} is also generated by three coordinate functions $z^*,H, z$, but now satisfying the different relations:
\begin{equation}\label{eq:Commutation of coords L_kappa}
\begin{split}
&zH=\kappa^{-1}Hz,\\
&zz^*=az^*z-b H^{-2},\ a={\kappa^2\over \hq^2},\ b= {\kappa\over \hq^2} (1-{\hq^2})\,.
\end{split}
\end{equation}
One can check that the following action of generators of $\mathcal{U}_{\hq, 0, s}^{\mathbb{C}}$  on the coordinates
makes this algebra into a $\mathcal{U}_{\hq, 0, s}^{\mathbb{C}}$-module:
\begin{equation}\label{eq:Action on chords L_kappa}
    \begin{split}
        A \cdot H = \hq^{1 \over 2} H ,& \quad A \cdot z = \hq^{-1} z , \quad A \cdot z^* =  z^* \\
        D \cdot H = \hq^{-{1 \over 2}} H ,& \quad D \cdot z = \hq z , \quad\ \  D \cdot z^* =  z^* \\
        B \cdot H = 0 ,& \quad B \cdot z = \hq^{-{1 \over 2}} , \quad B \cdot z^* = 0 \\
        C \cdot H = H z,& \quad C \cdot z = -\hq^{1 \over 2} z^2 , \quad C \cdot z^* = \hq^{3 \over 2} \kappa^{-1} H^{-2}\\
        A^* \cdot H = \left({\hq \over \kappa }\right)^{1 \over s} H ,& \quad A^* \cdot z =  z , \quad A^* \cdot z^* = \left({\kappa \over \hq}\right)^{2 \over s} z^* \\
         D^* \cdot H = \left({\kappa \over \hq }\right)^{1 \over s} H ,& \quad D^* \cdot z =  z , \quad D^* \cdot z^* = \left({\hq\over \kappa}\right)^{2 \over s} z^* \\
        B^* \cdot b = 0& = C^* \cdot b \mbox{ for any } b \in \mathcal{B}^{\mathbb{C}}_{\hq, \kappa}\,.
    \end{split}
\end{equation}
This action\footnote{The right action of generators branded with a $*$ is related by conjugation to the left action, since $f\cdot Y^* = (Y \cdot f)^*$ for $f\in\mathcal{B}_{\hq, \kappa}^{\mathbb{C}}$ and $Y\in \mathcal{U}^{\mathbb{C}}_{\hq, 0, s}$, e.g. for $Y = A, B, C, D$.} is compatible with the commutation relations (\ref{eq:Commutation of coords L_kappa}) and the twisted coproduct \eqref{eq:U_coproduct_twisted}.

We can now write the action of generators on the monomials of coordinates, just as we did in case of $\mathbb{H}^3_{\hq}$. Namely, defining\footnote{Note that the monomials we use here are shifted by powers of $\hq$ in comparison to \cite{Olshanetsky:2001}.}
\begin{equation}\label{eq:H-monom-2}
    v^{m,r,n} \colonequals (\hq^{1 \over 2}z^*)^m H^r (\hq^{1 \over 2}z)^n \,,
\end{equation}
we have that (recall $[n]_{\hq^2} \colonequals { \hq^n-\hq^{-n} \over \hq-\hq^{-1} }$):
\begin{equation}\begin{split}
    A \cdot v^{m,r,n} & = \hq^{-n + {r \over 2}}\  v^{m,r,n}    \\
    D \cdot v^{m,r,n} & = \hq^{n - {r \over 2}}\  v^{m,r,n}    \\
    B \cdot v^{m,r,n} & = \hq^{r \over 2} [n]_{\hq^2} \  v^{m,r,n-1}\\
    C \cdot v^{m,r,n} &= \hq^{n+m -{3r \over 2} +1} \kappa^{r-1} [m]_{\hq^2} \ v^{m-1,r-2,n} -\hq^{-{r \over 2}} [n-r]_{\hq^2}\  v^{m,r,n+1}   \\
    v^{m,r,n} \cdot A^* & = \hq^{-m+{r \over 2}}\  v^{m,r,n} \\
    v^{m,r,n} \cdot D^* & = \hq^{m-{r \over 2}}\  v^{m,r,n} \\
    v^{m,r,n} \cdot B^* & = \hq^{r \over 2} [m]_{\hq^2}\   v^{m-1,r,n} \\
    v^{m,r,n} \cdot C^* & = \hq^{m +n -{3 r \over 2} + 1} \kappa^{r-1}  [n]_{\hq^2}  v^{m,r-2,n-1} - \hq^{-{r \over 2}} [m-r]_{\hq^2}\  v^{m+1,r,n}\,,
 \end{split}\end{equation}
and the action of the Casimir element $\Omega$ becomes
\begin{equation}\begin{split}
    \Omega \cdot v^{m,r,n}& \equiv\left(  { \hq^{-1}A^2 + \hq D^2 -2\over (\hq^{-1} - \hq)^2} + B C \right) \cdot v^{m,r,n } \\
    & = [m]_{\hq^2}[n]_{\hq^2} \kappa^{r-1} \hq^{m-r+n} v^{m-1, r-2, n-1} + [{r+1 \over 2}]^2_{\hq^2} v^{m,r,n}\,.
\end{split}
\label{eq:Casimir Quantum Group on Monomial OR2}
\end{equation}
Again, we will find it useful to use an automorphism of the algebra $\mathcal{U}_{\hq}^{\mathbb{C}}$ in order to compare with the results of section \ref{sec:hands-on}.  
Namely, as before, let us pass to the new generators
\begin{align}
    A' \colonequals A, \quad  D' \colonequals D,  \quad  B' \colonequals \sqrt{\hq} D B, \quad  C' \colonequals \sqrt{\hq} A C \,,
\end{align}
obtaining
\begin{equation}\begin{split}\label{eq:OR2ConjOps}
    A' \cdot v^{m,r,n} & = \hq^{{r \over 2}-n} v^{m,r,n}  \\
        D'  \cdot v^{m,r,n} & = \hq^{-{r \over 2}+n} v^{m,r,n}  \\
        B' \cdot v^{m,r,n} & = \hq^{n-{1 \over 2}} [n]_{\hq^2} v^{m,r,n-1} \\
        C'  \cdot v^{m,r,n}
        & = [m]_{\hq^2}\kappa^{r-1} \hq^{m- r+{1 \over 2}} v^{m-1,r-2,n} - \hq^{-n - {1 \over 2}} [n-r]_{\hq^2}  v^{m,r,n+1},\\
\end{split} \end{equation}
and similarly for the action of conjugate generators.
As before, the algebra stays the same, but the Hopf algebra structure is not preserved by the algebra automorphism. One can push the Hopf algebra structure through the automorphism and find that the counit stays the same (on new generators), but the coproduct and antipode change.

In terms of operators we defined in (\ref{eq:Simple Operators on  H and z}) for the corresponding commuting coordinates $\tilde z^*,\tilde H,\tilde z$, we have
\begin{equation}\label{eq:Action of generators on L3 : Abstract OR2}
    \begin{split}
        {\tilde A}' & = T \cdot R^{-1} \\
        {\tilde D}' & = T^{-1} \cdot R \\
        {\tilde B}' & = { 1 - R^2 \over (1-\hq^2)\tilde z } ,\quad \tilde B^{*\prime} = { 1 - {R^*}^2 \over (1-\hq^2)\tilde z^*} \\
        {\tilde C}' & = \tilde z {(T^{-2} - T^2 R^{-2})\over \hq^{-1} - \hq} + \hq^{-{\tilde a \over 2}}\tilde H^{-2}  \ \tilde B^{*\prime} T^{\tilde a } \\
        \tilde \Omega'  & = {\hq^{-1} T^{-2} + \hq T^2 -2 \over  (\hq^{-1} -\hq)^2 } +  \hq^{-{\tilde a \over 2}} \tilde H^{-2} T^{\tilde a} {\tilde B}' {\tilde B}^{*\prime}\,,
    \end{split}
\end{equation}
where we have defined $\kappa \colonequals \hq^{1 + {\tilde a \over 2} }$. Note that for $\tilde a = 2$, we get identical results as ${\mathbb{H}}_{\hq}^3$.   We note here that if we do a reduction on $\tilde z^*$ by diagonalizing the operator $\tilde B^{*\prime}$, we obtain the same generators as in section \ref{sec:hands-on} -- in fact the parameter $\tilde a$ is the same as what we encountered in \eqref{eq:HandOn} (and still $a=0$).

\section{From $AdS_3$ to $AdS_2$ to the transfer-matrix}
\label{sec:down-to-AdS2}

In section \ref{sec:hands-on} we constructed some discretized version of $AdS_{2,\hq}$. The approach in the previous section relied on non-commutative variables, and we actually focused on a non-commutative $AdS_{3,\hq}$. In this section we will:
\begin{itemize}
\item Carry out a reduction of non-commutative $AdS_{3, \hq}$ to a non-commutative $AdS_{2,\hq}$, still written in terms of non-commutative coordinates. The reduction is a NCG version of \ref{sec:BootApp}. This will rely on \cite{Olshanetsky:1993sw,Olshanetsky:2001} which carries out a reduction from $AdS_{3,\hq}$ to the transfer-matrix. We will add the intermediate steps in the reduction, in order to obtain $AdS_{2,\hq}$.
\item We will then transform these models into models on lattices of commuting variables. The relation between the commuting
lattice variables
and the non-commuting variables was briefly discussed in section \ref{sec:hands-on}, see also section \ref{sec:Lobachevsky}. We will show how it works in a simple way here. The result will be a subset of the models in section \ref{sec:hands-on}.
\item We will then proceed to reduce them to the transfer-matrix (this part of the discussion overlaps with the one in section \ref{sec:hands-on}).
\end{itemize}

As before, our conventions are that the tilded objects, such as ${\tilde H}, {\tilde z}$, designate commuting variables (on an appropriate lattice) whereas untilded objects, such as $H,z$, denote non-commuting variables.

Another important point is that at this stage we are going to switch from the Euclidean quantum Lobachevsky plane of the previous section to the Minkowskian one. This amount to take $z$ and $z^*$ to be two independent real non-commuting variables. This will allow us to take two independent reductions as in section \ref{sec:CrdABK}.

\smallskip

\paragraph{Reducing to $AdS_2$ lattice}
First, consider a family of functions of the form
\begin{equation}
F_{\mu}(z^*,H,z) \colonequals  e_0(i \mu (1-\hq^2)\hq^{1 \over 2 } z^*;{\hq^2}) F_\mu(H,z)= e_0(i \mu (1-\hq^2)\hq^{1 \over 2 } z^*;{\hq^2}) \sum_{r,n}  a_{r+1,n} H^r (\hq^{1 \over 2}z)^n \,,
\label{eq:Non Commuting AdS3 to AdS2}
\end{equation}
where $a_{r+1,n}$ are some constants and the $q$-exponential $e_0(\mu ; \hq^2)$ was defined in \eqref{eq:q-exponential with parameter a}. It can be verified explicitly that the Ansatz \eqref{eq:Non Commuting AdS3 to AdS2} diagonalizes ${\tilde B}^{*\prime}$. \eqref{eq:Action of generators on L3 : Abstract OR2} is written in the language of the commuting variables, and, with a slight care of ordering, it also applies to the non-commuting variables.

We find that the action of the (conjugated) generators on these functions is given by (from using equations \eqref{eq:Action of Conjugated generators} and \eqref{eq:OR2ConjOps} on the Ansatz above)
\begin{equation}
\begin{split}
 A' \cdot F_{\mu}(H,z)   & = F_{\mu}(\hq^{1 \over 2} H,\hq^{-1} z)  \\
D' \cdot F_{\mu}(H,z)   & = F_{\mu}(\hq^{-{1 \over 2}} H,\hq z)  \\
B' \cdot F_{\mu}(H,z)   & ={  F_{\mu}(H, z) - F_{\mu}( H,\hq^2 z)  \over  (1-\hq^2)  }  \ z^{-1} \\
        C' \cdot F_{\mu}(H,z)  &=  {1  \over \hq^{-1} - \hq }  \left(   F_{\mu}(\hq^{-2} H,z) - F_{\mu}(\hq^{2} H,\hq^{-2} z) \right) z \\\
        & \hspace{20mm} + i \mu \hq^{{1 -\tilde a \over 2}} F_\mu(\hq^{\tilde a \over 2} H,\hq^{-2-\tilde a} z) H^{-2} \\
         \Omega'  \cdot F_\mu( H, z)  & = {1 \over (\hq^{-1}-\hq)^2 } \left(\hq^{-1} F_\mu( \hq^{-1} H , z) + \hq F_\mu(\hq H, z) -2 F_\mu(H, z)  \right) \\
 & \hspace{20mm} + { i \mu \hq^{1 -\tilde a \over 2}  \over (1- \hq^2 ) } \left( F_\mu(\hq^{\tilde a \over 2} H, \hq^{-2-\tilde a}z) - F_\mu( \hq^{\tilde a \over 2} H, \hq^{-\tilde a }  z)  \right) H^{-2}   z^{-1}\,.
    \end{split}
\end{equation}
Recall that $\tilde a =0$ for ${\mathbb{H}}_{\hq}^3$, whereas $\tilde a$ is an arbitrary parameter for ${\mathbb{H}}_{\hq, \kappa}^3$ (with $\kappa = q^{1 + {\tilde a \over 2}}$). We emphasize that, since these are non-commuting coordinates, the positioning of the coordinates is important above.

It is easy to verify the following action of generators on the functions of commuting coordinates (with $F_\mu(\tilde z^*,\tilde H,\tilde z)$ being defined analogously to \eqref{eq:Non Commuting AdS3 to AdS2}).  We can begin by the actions just described or we can start with \eqref{eq:Action of generators on L3 : Abstract} and \eqref{eq:Action of generators on L3 : Abstract OR2} and note that the Ansatz \eqref{eq:Non Commuting AdS3 to AdS2} diagonalizes ${\tilde B}^{*\prime}$:
\begin{equation}\label{eq:ORCommAlg}
\begin{split}
\tilde A' \cdot F_{\mu}(\tilde H,\tilde z)   & = F_{\mu}(\hq^{1 \over 2} \tilde H,\hq^{-1} \tilde z)  \\
\tilde D' \cdot F_{\mu}(\tilde H,\tilde z)   & = F_{\mu}(\hq^{-{1 \over 2}} \tilde H,\hq \tilde z)  \\
\tilde B' \cdot F_{\mu}(\tilde H,\tilde z)   & ={  F_{\mu}(\tilde  H,  \tilde z) - F_{\mu}(  \tilde H,\hq^2 \tilde z)  \over  (1-\hq^2)  }   \tilde z^{-1} \\
       \tilde  C' \cdot F_{\mu}(\tilde H,\tilde z)  &=  {1  \over \hq^{-1} - \hq }  \left(   F_{\mu}(\hq^{-2} \tilde H,\tilde z) - F_{\mu}(\hq^{2} \tilde H,\hq^{-2} \tilde z) \right) \tilde z \\\
        & \hspace{20mm} + i \mu \hq^{1 -\tilde a \over 2} F_\mu(\hq^{\tilde a \over 2} \tilde H, \tilde z) \tilde H^{-2} \\
         \tilde \Omega'  \cdot F_\mu( \tilde H, \tilde z)  & = {1 \over (\hq^{-1}-\hq)^2 } \left(\hq^{-1} F_\mu(\hq^{-1} \tilde H , \tilde z) + \hq F_\mu(\hq \tilde H, \tilde z) -2 F_\mu( \tilde  H, \tilde z)  \right) \\
 & \hspace{20mm} + { i \mu  \hq^{1 - \tilde a \over 2} \over (1 - \hq^2 ) } \left( F_\mu( \hq^{\tilde a \over 2}  \tilde H,  \tilde z) - F_\mu(\hq^{\tilde a \over 2} \tilde H, \hq^{2} \tilde z)  \right) \tilde H^{-2}   \tilde z^{-1}\,.
    \end{split}
\end{equation}
From now on, we will work with only commuting coordinates.

We can now compare \eqref{eq:ORCommAlg} with the hands-on models in \eqref{eq:HandOn}. The dependence on ${\tilde z}^*$ is fixed by \eqref{eq:Non Commuting AdS3 to AdS2} so these functions really depend on ${\tilde z}$ and $\tilde H$. We see that out of the hands-on models we obtain those with $a=0$ --- this can be seen by comparing the $B$ generators in two cases (recall, however, that this is a value which does not have a satisfactory $q$-Fourier transform in our exposition). The parameter $\tilde a$ is mapped to the parameter $\tilde a$ in
 \eqref{eq:HandOn}. Thus, we see that a subset of the hands-on models can be realized concretely as modules of the quantum group. Of course, here as there, the requirement of obtaining the transfer-matrix sets ${\tilde a}$ to a specific value, but that is not enough to fully constrain the model.

 An interesting question is: what is the status of the additional models in \eqref{eq:HandOn}. It may very well be that they also have realizations in the NCG language of quantum groups and $\hq$-homogeneous spaces used in section \ref{sec:q-groups-intro}. The possible $\hq$-deformations of $\mathfrak{s}\mathfrak{l}(2,\mathbb{R})$ and $\mathfrak{s}\mathfrak{l}(2,\mathbb{C})$ (as well as of the algebras of functions on the corresponding Lie groups) are fully classified. The classification of possibilities for $\hq$-deformed coset spaces and of the full set of relevant module algebras requires additional tools (e.g. deformation theory of algebras) and is not easily accessible from the existing literature.
 
 Even in the approach in these sections, it seems that just requiring $\mathcal{U}_{\hq}\left(\mathfrak{s}\mathfrak{l} \,2\right)$ symmetry is not quite enough to fix the relevant $AdS_{2,\hq}$ uniquely. One still needs to appeal to an argument of the type we used in section \ref{sec:NoBndry}.

\smallskip

\paragraph{Reducing to $H$ lattice}
For completeness, we provide the details of the reduction to the $H$ lattice. We will be reducing by choosing a specific Ansatz for $z$-dependence as well. Consider $G_{\mu,\Lambda,\nu}(\tilde z^* ,\tilde  H ,\tilde z )$ which is defined via
\begin{equation}\label{eq:FUllReduc}
    G_{\mu ,\Lambda,\nu} = e_0(i \mu (1-\hq^2)\hq^{1 \over 2 } z^*;{\hq^2}) \  F_{\Lambda}(H) \  e_0(i \nu (1-\hq^2)\hq^{1 \over 2 } z;{\hq^2}) \,.
\end{equation}
Since we are reducing to one dimension, there is no notion of non-commuting coordinates anymore, and the procedure works in the same way for both commuting and non-commuting variables.

We are interested in eigenvectors of Casimir. Take the Ansatz above
with $F_\Lambda(\tilde H) \colonequals \sum_r a_{r+1} \tilde H^r$ a polynomial function. To make this into an actual (say, $L^2$) basis of functions, in principle, we may need to appropriately complete, but here we will just solve for it. Note that
\begin{equation}\begin{split}
    &\Omega \cdot G_{\mu ,\Lambda,\nu}   = e_0(i \mu (1-\hq^2)\hq^{1 \over 2 } z^*;{\hq^2})  \times \\
    &\qquad \left({ \hq F_{\Lambda}(\hq \tilde H) + \hq^{-1} F_{\Lambda}(\hq^{-1} \tilde H) -2  F_{\Lambda}( \tilde H) \over (\hq-\hq^{-1})^2}  - \mu \nu \hq^{1 - {\tilde a \over 2}} \tilde  H^{-2} F_{\Lambda}(\hq^{\tilde a \over 2} \tilde H)\right)  e_0(i \nu (1-\hq^2)\hq^{1 \over 2 } z;{\hq^2}) \,.
\end{split} \end{equation}
Therefore demanding that $G_{\mu ,\Lambda,\nu}$ be an eigenfunction of $\Omega$ with the eigenvalue $-\Lambda^2$ gives the following recursion relation:
\begin{equation}
{\hq F_{\Lambda}(\hq \tilde H) + \hq^{-1} F_{\Lambda}(\hq^{-1} \tilde H) -2  F_{\Lambda}( \tilde H) \over (\hq-\hq^{-1})^2}  - \mu \nu \hq^{1 - {\tilde a \over 2}} \tilde H^{-2} F_{\Lambda}(\hq^{\tilde a \over 2} \tilde H) = - \Lambda^2   F_{\Lambda}( \tilde H)  \,.
\end{equation}

For the case of ${\mathbb{H}}_{\hq}^3$ with $\tilde a = 2$ and also for the special case of ${\mathbb{H}}_{\hq,\kappa}^3$ with $\tilde a = - 2$ (i.e $\kappa=1$), we can relate this equation to the transfer-matrix of the double-scaled SYK given in section \ref{sec:transfer_matrix}. In what follows, we will only take $\tilde a = \pm 2$.

One more piece of assumption we need to do is
\begin{equation}
    \mu \nu   =\hq^{\tilde a -1} (\hq- \hq^{-1})^{-2}\,,
\end{equation}
after which the recursion relation becomes
\begin{equation}\label{eq:H lattice eqn}
(1-\tilde H^{-2}) \hq^\pm F_\Lambda(\hq^\pm \tilde H) + \hq^{\mp} F_\Lambda(\hq^{\mp} \tilde H) =  \left( 2- \Lambda^2(\hq-\hq^{-1})^2 \right) F_\Lambda( \tilde H) \,,
\end{equation}
where '+' sign is for ${\mathbb{H}}_{\hq}^3$ and '-' sign is for ${\mathbb{H}}_{\hq,\kappa=1}^3$.

For the case of ${\mathbb{H}}_{\hq}^3$,  we can define new variables
\begin{equation}
\hq^{-n} \colonequals \tilde  H, \qquad     c_n (x) \colonequals  {\hq^{-n} \over (\hq^2 ; \hq^2)_n}  F_\Lambda(\hq^{-n}),\qquad  2x \colonequals   2- \Lambda^2(\hq-\hq^{-1})^2\,,
\end{equation}
and the recursion relation \eqref{eq:H lattice eqn} becomes
\begin{equation}
    c_{n-1}(x)+ (1-\hq^{2n+2})c_{n+1}(x)  = 2 x c_n(x)\,.
\end{equation}
This is the same as solving for the eigenvalues of the matrix
\begin{equation}
\begin{bmatrix}
0 & 1-\hq^2 &  0  &  0 & 0 &  \dots   \\
1 & 0 &   1-\hq^4  &  0 & 0 &  \dots  \\
0 &  1 & 0 & 1-\hq^6  & 0 & \dots \\
\vdots  &  \ddots & \ddots & \ddots &  \ddots & \ddots \\
\end{bmatrix}\,.
\end{equation}
This matrix is almost a conjugation of the transfer-matrix $T$ of \eqref{eq:T Matrix form}. More precisely, if we define a diagonal matrix $(S_1)_{ll} = {1\over (1-\hq^2)^{l \over 2} }$, then the above matrix is just $\sqrt{1-\hq^2} S_1 T S_1^{-1}$.

For the case of ${\mathbb{H}}_{\hq, \kappa}^3$,  we can define new variables
\begin{equation}
\hq^{-n} \colonequals \tilde  H, \qquad     c_n (x) \colonequals \hq^{-n}\  (\hq^2 ; \hq^2)_{n-1} \ F_\lambda(\hq^{-n}),\qquad  2x \colonequals   2- \lambda^2(\hq-\hq^{-1})^2\,,
\end{equation}
and the recursion relation \eqref{eq:H lattice eqn} becomes
\begin{equation}
     (1-\hq^{2n}) c_{n-1}(x) + c_{n+1}(x)  = 2 x c_n(x)\,.
\end{equation}
As in the first case, this is the same as solving for the eigenvalues of the matrix
\begin{equation}
\begin{bmatrix}
0 & 1 &  0  &  0 & 0 &  \dots   \\
1 -\hq^2& 0 &   1  &  0 & 0 &  \dots  \\
0 &  1-\hq^4 & 0 & 1  & 0 & \dots \\
\vdots  &  \ddots & \ddots & \ddots &  \ddots & \ddots \\
\end{bmatrix}\,.
\end{equation}
This matrix is also almost a conjugation of the transfer-matrix $T$ of \eqref{eq:T Matrix form}. Namely, if we define a diagonal matrix $(S_2)_{ll} = { (\hq^2 ;\hq^2)_l \over (1-\hq^2)^{l \over 2} }$, then the above matrix is $\sqrt{1-\hq^2} S_2 T S_2^{-1}$.

\section{Conclusion}
\label{sec:conclusion}

In this paper we discussed the relation between the transfer-matrix solution of the double-scaled SYK model and the boundary particle moving on the Poincar{\'e} disk (in a Euclidean setting). We explained why we need to replace the standard $\mathfrak{s}\mathfrak{l}(2,\mathbb{R}) \simeq \mathfrak{s}\mathfrak{u}(1,1)$ symmetry structure by its specific $\hq$-deformed analogue $\mathcal{U}_{\hq}(\mathfrak{s}\mathfrak{u}(1,1))$. We then discussed some hands-on lattice versions of $AdS_2$ which realize the new quantum group symmetry and discussed under what circumstances they give rise to the DS-SYK transfer-matrix. These lattice realizations should be thought of as a non-commutative $AdS_{2,\hq}$.

Next we turned to the discussion of constructing a non-commutative $AdS_{2,\hq}$ as a $\hq$-homogeneous space, using tools from the theory of quantum groups. Our starting point included some known constructions of a $\hq$-deformed $AdS_3$ and how to reduce them to transfer-matrices with a specific Ansatz. We showed where $AdS_{2,\hq}$ appears and we suggested a commuting lattice realization of these spaces. The models we obtain are a subset of the hands-on models brought before.

For the usual NAdS${}_2$/SYK${}_1$ duality, we can start with the motion of the boundary particle on AdS$_2$ and obtain the Schwarzian 1D action. Here we are in the opposite situation: we are given the 1D Hamiltonian (the transfer-matrix) and are looking for the the 2D generalization. In the situation where there are several options for such a lift one needs to go back to the microscopics of the SYK model and see how $\mathcal{U}_{\hq}(\mathfrak{s}\mathfrak{u}(1,1))$ emerges there. This is a difficult task which we leave to future work.

\section*{Acknowledgements}
We would like to thank A.~Almheiri, E.~Koelink, H.~Lin, J.~Maldacena, S.F.~Ross, M.~Rozali, J.~Stokman, W.~van Suijlekom, H.~Verlinde, and J.~Yoon for interesting discussions related to the project. The work of MB is supported by an ISF center for excellence grant (grant number 2289/18), by the German Research Foundation through a DIP grant ”Holography and the Swampland”, and by the ISF grant no. 2159/22.
PN would like to acknowledge SERB grant MTR/2021/000145.

\appendix

\section{Appendix: The chord path of the boundary particle}
\label{sec:q1toABK}

We would like to evaluate $\langle q^{n(t)} \rangle$ as in  
subsection \ref{subsec:Liouville to chords mapping}.
Inserting identity as a sum of projectors onto the complete set of states $\sum_{n \geq 0} | n \rangle \langle n | $, where $n$ denotes the number of open chords, we get:
\begin{equation}
\langle 0 |  q^{\hat n(\tau)} | 0 \rangle_\beta = \sum_n \langle 0 | e^{-({\beta \over 2}-\tau) \hat T }  | n \rangle \  q^n \ \langle n |  e^{-({\beta \over 2}+\tau)  \hat T } | 0 \rangle \,.
\end{equation}
An expectation value of this form is exactly what was computed in \cite{Berkooz:2018qkz}, in the context of a two-point function,\footnote{See equation (6.21) in that paper, with the substitutions $m \mapsto 1$ and $i t \mapsto \tau$.} from which we get:
\begin{equation}\begin{split}
\langle 0 |  q^{\hat n(\tau)} | 0 \rangle_\beta = 16 (q^{2};q)_\infty  {(q;q)_\infty^2 \over (2\pi)^2}   \  \int _0^\pi d\theta_1 \int_0^\pi d\theta_2 e^{-{2\cos(\theta_1)({\beta\over 2}-\tau)\over\sqrt{(1-q})}} e^{-{2\cos(\theta_2) ({\beta\over 2}+ \tau)\over\sqrt{(1-q})}} \ \sin \theta_1 \sin \theta_2
\\
\times \sin({\theta_1 + \theta_2 \over 2}) \sin({\theta_1 - \theta_2 \over 2})     \ \  {\vartheta_1({\theta_1 \over \pi } | {i \lambda \over 2\pi}) \vartheta_1({\theta_2 \over \pi } | {i \lambda \over 2\pi})  \over  \vartheta_1({\theta_1 + \theta_2 \over 2\pi } | {i \lambda \over 2\pi}) \vartheta_1({\theta_1-\theta_2 \over 2\pi } | {i \lambda \over 2\pi}) }\,,
\end{split} \end{equation}
where $\vartheta_1$ is the Jacobi theta function.

We are interested in the classical limit $q \rightarrow 1$ where we can compare with the Liouville case. Using modular properties of $\vartheta_1$ functions, one can obtain the $q  \rightarrow 1$ limit (or, since $q= e^{-\lambda}$, equivalently $\lambda \rightarrow 0$ limit) as:
\begin{equation}
\langle 0 |  q^{\hat n(\tau)} | 0 \rangle_\beta \approx {32} (q^{2};q)_\infty  {(q;q)_\infty^2 \over (2\pi)^2}   \   I(\beta,\tau,q)\,,
\end{equation}
where we have defined $I(\beta,\tau,q)$ as
\begin{equation}
\begin{split}
I(\beta,\tau,q) \colonequals  & \int _0^\pi d\theta_1 \int_0^\pi d\theta_2 e^{-{2\cos(\theta_1)({\beta \over 2}-\tau)\over\sqrt{\lambda}}} e^{-{2\cos(\theta_2)({\beta \over 2} +\tau)\over\sqrt{\lambda} }}
\\
& \times \sin({\theta_1 + \theta_2 \over 2}) \sin({\theta_1 - \theta_2 \over 2})    \sin \theta_1 \sin \theta_2\ \  e^{-{3\pi^2 \over {2} \lambda}  -{1\over \lambda} {(  \theta_1  -{ \pi \over 2} )^2} -  {1\over \lambda}  {( \theta_2 -{ \pi \over 2} )^2}  }\\
& \times  { \sinh({2 \pi \theta_1\over \lambda}) \sinh({2\pi(\pi-\theta_1)\over  \lambda})  \over \sinh({ 2\pi ({\theta_1+\theta_2\over 2})\over \lambda}) \sinh({ 2\pi (\pi - {\theta_1+\theta_2\over 2})\over \lambda})  } \ { \sinh({2 \pi \theta_2\over \lambda})  \sinh({2 \pi (\pi-\theta_2) \over \lambda})  \over \sinh({ 2\pi ({\theta_1-\theta_2\over 2})\over \lambda}) (1-2\cosh({2\pi(\theta_1-\theta_2)\over \lambda })e^{-4\pi^2/\lambda}  )  }\,.   \end{split}
\label{eq:exact2pt}
\end{equation}
To proceed, we have to specify the range of $\beta$ and $\tau$, and how they scale with $\lambda$ in the $\lambda \rightarrow 0$ limit. Recall that the Schwarzian dynamics (or, equivalently, Liouville dynamics) emerges at low temperatures, along with the $\lambda \rightarrow 0$ limit. In the regime
\begin{equation}
 {1 \over \sqrt \lambda }\ll  \beta \ll {1 \over \lambda^{3 \over 2} } ,   \qquad \tau \ll {\sqrt{\beta} \over \lambda^{3 \over 4} }\,,
\end{equation}
the above integral was evaluated exactly in equation (6.30) of \cite{Berkooz:2018qkz} (replace $it \mapsto \tau$), to obtain:
\begin{equation}\label{eq:q1Expn}
I(\beta,\tau,q) \approx {\lambda^{3 \over 4} \over 16} \bigg({ \pi \over  \beta }\bigg)^{7 \over 2} e^{ 2 \beta \lambda^{-{1 \over 2}} + \pi^2 \beta^{-1} \lambda^{-{3 \over 2}}  } {1 \over \cos^2({\pi \tau \over \beta})}\,.
\end{equation}

\section{Some more details about the quantum Lobachevsky space}
\label{app:details_qLobachevsky}

In this Appendix, we work out the action of $\mathcal{U}_{\hq}^{\mathbb{C}}$ on the algebra of functions $\mathcal{B}^{\mathbb{C}}_{\hq}$ on the $\hq$-Lobachevsky space ${\mathbb{H}}_{\hq}^3$. We will be following \cite{Olshanetsky:1993sw} and correct a few of their typos along the way.

The action of $\mathcal{U}_{\hq}^{\mathbb{C}}$ on $\mathcal{B}^{\mathbb{C}}_{\hq}$ follows naturally from  the action on the coordinate functions $\alpha,\beta, \gamma$ and $\delta$ on the quantum group. We begin by recalling the algebra $\widetilde{\mathcal{A}}^{\mathbb{C}}_{\hq}$ that we introduced before in section \ref{subsec:A_hopf_alg}, see also equations (1.9)--(1.25) of \cite{Podles:1990zq}:
\begin{equation}
    \begin{split}
        &\alpha \beta = \hq \beta \alpha, \qquad \alpha \gamma = \hq \gamma \alpha, \qquad \alpha \delta - \hq \beta \gamma = 1,\qquad [\beta , \gamma] =0,\qquad \beta \delta = \hq \delta \beta \\
        &\gamma \delta = \hq \delta \gamma,\qquad \delta \alpha - \hq^{-1} \beta \gamma = 1 \\
        &\beta \alpha^* = \hq^{-1} \alpha^* \beta + \hq^{-1} (1-\hq^2) \gamma^* \delta \\
        &\gamma \alpha^* = \hq \alpha^* \gamma , \qquad [\delta, \alpha^*] = [ \gamma, \beta^*]= 0 \\
        &\delta \beta^* = \hq \beta^* \delta - \hq(1-\hq^2) \alpha^* \gamma \\
        &\delta \gamma^* = \hq^{-1} \gamma^* \delta, \qquad \alpha \alpha^* = \alpha^* \alpha + (1-\hq^2) \gamma^* \gamma \\
        &\beta \beta^* = \beta^* \beta + (1-\hq^2) (\delta^* \delta - \alpha^* \alpha) - (1-\hq^2)^2 \gamma^* \gamma \\
        &[\gamma , \gamma^*] = 0, \qquad \delta \delta^* = \delta^* \delta -(1-\hq^2) \gamma^* \gamma \,,
     \end{split}
\end{equation}
(we have corrected typos of \cite{Olshanetsky:1993sw}). We are interested in the action of the generators of  $\mathcal{U}_{\hq}^{\mathbb{C}}$. The action on the coordinate functions on quantum group was already given in \eqref{eq:AlgActGrp}, and it can be extended to the full $\widetilde{\mathcal{A}}^{\mathbb{C}}_{\hq}$ by using appropriate coproduct rules. Since the coordinates on ${\mathbb{H}}_{\hq}^3$ are just quadratic combinations of $\alpha, \beta, \gamma, \delta$, the action of generators of  $\mathcal{U}_{\hq}^{\mathbb{C}}$ can be computed in a straightforward (but tedious) manner. We turn to the details of this calculation below.

The $\hq$-Lobachevsky coordinates $H,z, z^*$ are defined by the same formula as in \eqref{eq:LubPlan1}:
\begin{equation*}
    H \colonequals \alpha^* \alpha + \gamma^* \gamma , \qquad Hz \colonequals \alpha^* \beta + \gamma^* \delta \,.
\end{equation*}
One can readily check the commutation relations \eqref{eq:OR1 Lobachevsky commutations}. To find the action of generators on Lobachevsky coordinates, it is first useful to define a so called normal ordering of operators in $\widetilde{\mathcal{A}}^{\mathbb{C}}_{\hq}$. This is unambiguous thanks to the PBW-theorem for the algebra and takes the form
\begin{equation}\label{eq:def normal ordering on A_q}
    : a : = \sum_k c_k\  a^*_{1,k}\   a_{2,k}, \quad \forall a \in \widetilde{\mathcal{A}}^{\mathbb{C}}_{\hq}\,,
\end{equation}
where $a_{1,k}^*$ is a monomial in $\alpha^*,\beta^*,\gamma^*,\delta^*$, and, similarly, $a_{2,k}$ is a monomial in $\alpha,\beta,\gamma,\delta$. If we define
\begin{align}
\xi \colonequals \alpha^{-1} \beta ,\qquad \zeta \colonequals  \gamma^* H^{-1} \alpha^{-1}, \qquad y \colonequals \gamma \alpha^{-1}\,,
\end{align}
then one can check that the powers of $\hq$-Lobachevsky coordinates $z, H$ can be normal-ordered as follows:
\begin{equation}\begin{split}
    z^n & = \sum_{l=0}^n {\hq}^{l(n-1)} \binom{n}{ l }_{\hq^2} \zeta^l \xi^{n-l},\qquad \zeta^l = \hq^{-{l(l-1) \over 2}} {\gamma^*}^l H^{-l} \alpha^{-l} \\
    H^l  &= \sum_{k=0}^\infty  \hq^{(l-1)k} \binom{l}{k}_{\hq^2}   {\alpha^*}^l {y^*}^k y^k \alpha^l\,.
    \end{split}
\end{equation}

\subsection{Action of generators}
With the definition of normal ordering as in \eqref{eq:def normal ordering on A_q}, the right action of any generator $u \in \mathcal{U}_{\hq}^{\mathbb{C}}$ on an element $a$ of $\widetilde{\mathcal{A}}^{\mathbb{C}}_{\hq}$ is then given by
\begin{equation}
   u \cdot  a   \colonequals \sum_k c_k\  a_{1,k}^* \left(u \cdot  a_{2,k}  \right) \,.
\end{equation}
Note that the action $ a_{2,k} \cdot u$ is the usual left action of $\mathcal{U}_{\hq}^{\mathbb{C}}$ on $\widetilde{\mathcal{A}}^{\mathbb{C}}_{\hq}$, as defined in subsection \ref{subsec:pairing}.

We are interested in finding the action of generators\footnote{In what follows, we will not display the action of $D$ since it is the same as $A^{-1}$.}  $A,B,C$ on the polynomials made out of coordinate functions on the $\hq$-Lobachevsky space. In other words, they are built as linear combinations of the ordered monomials ${z^*}^m H^r z^n$. As an intermediate step, it will be useful to compute first the action of the generators on a slightly different monomial:  
\begin{equation}
    \begin{split}
        w^{k,r,l} & \colonequals {\zeta^*}^k H^r \zeta^l \,.
    \end{split}
\end{equation}
A straightforward (but long) calculation shows that
\begin{equation}    \begin{split}
    A \cdot w^{k,r,l} &= \hq^{{r \over 2} - l}w^{k,r,l} \\
    B \cdot w^{k,r,l} & =  0  \\
    C \cdot w^{k,r,l}
        & = -\hq^{1-r+2l \over 2} [2l-r]_{\hq^2}   w^{k,r,l} \xi - [l-r]_{\hq^2} \hq^{1-r + 4l  \over 2} w^{k,r,l+1} \\
        & \hspace{20mm} + \hq^{1-2k + 2l + r \over 2}  [k]_{\hq^2}  w^{k-1,r-2,l}\,.
   \end{split}
\end{equation}
The object of interest on which we want to calculate the action of generators is
\begin{equation}\begin{split}
    v^{m,r,n} & \equiv (\hq^{1\over 2}z^*)^{m} H^r( \hq^{1\over 2} z)^{n}\,.
\end{split} \end{equation}  
It can be easily related to the $ w^{k,r,l}$ we defined above, via
\begin{equation}\begin{split}
        v^{m,r,n} = \hq^{m + n\over 2} \sum_{k=0}^m \sum_{l=0}^{n} \hq^{k(m-1) + l(n-1)} \binom{m}{k}_{\hq^2} \binom{n}{l}_{\hq^2} {\xi^*}^{m-k} w^{k,r,l} \xi^{n-l}\,,
\end{split} \end{equation}          
(again some typos corrected relative to equation (45) of \cite{Olshanetsky:1993sw} and in the action of $B$ on $v^{m,r,n}$ given there), which brings us to the final answer \eqref{eq:OSAdS2}.    

\bibliography{DRAFT_BIB}
\bibliographystyle{JHEP}

\end{document}